\documentclass[aps,pra,epsf,superscriptaddress,amsmath,amssymb,amsfonts,twocolumn,showpacs,nofootinbib]{revtex4-1}

\usepackage{amssymb}
\usepackage{amsmath}
\usepackage{amssymb}
\usepackage{subfigure}
\usepackage{color}
\usepackage{mathrsfs} 
\usepackage{float}
\usepackage{comment}
\usepackage{soul}

\usepackage{hyperref}

\usepackage{epsfig}
\usepackage{dcolumn}
\usepackage{bm}
\usepackage{braket}
\usepackage{mathtools}
\usepackage{graphicx,color,xcolor}
\usepackage{hyperref}
\newcommand{\abs}[1]{\left| #1 \right|} 
\usepackage{multirow}
\DeclareMathOperator{\sech}{sech} 

\usepackage[normalem]{ulem} 

\begin{document}

\title{Solitary waves in a quantum droplet-bearing system}

\author{G. C. Katsimiga}
\affiliation{Department of Mathematics and Statistics, University of Massachusetts,Amherst, MA 01003-4515, USA}
\author{S. I. Mistakidis}
\affiliation{ITAMP,  Center for Astrophysics $|$ Harvard $\&$ Smithsonian, Cambridge, MA 02138 USA}
\affiliation{ Department of Physics, Harvard University, Cambridge, Massachusetts 02138, USA}
\author{G. N. Koutsokostas}
\affiliation{Department of Physics, National and Kapodistrian University of Athens,
Panepistimiopolis, Zografos, Athens 15784, Greece}
\author{\\ D. J. Frantzeskakis}
\affiliation{Department of Physics, National and Kapodistrian University of Athens,
Panepistimiopolis, Zografos, Athens 15784, Greece}
\author{R. Carretero-Gonz{\'a}lez}
\affiliation{Nonlinear Dynamical Systems Group, Computational Sciences Research Center,
and Department of Mathematics and Statistics, San Diego State University, San Diego, California 92182-7720, USA}
\author{P.~G.\ Kevrekidis}
\affiliation{Department of Mathematics and Statistics, University of Massachusetts,Amherst, MA 01003-4515, USA}

\date{\today}

\begin{abstract} 

We unravel the existence and stability properties of dark soliton solutions 
as they extend from the regime of trapped quantum droplets towards the Thomas-Fermi 
limit in homonuclear symmetric Bose mixtures. 
Leveraging a phase-plane analysis, we identify the regimes of existence of
different types of quantum droplets and subsequently examine the possibility of
black and gray solitons and kink-type structures in this system. 
Moreover, we employ the Landau dynamics approach to extract an analytical estimate of the 
oscillation frequency of a single dark soliton in the relevant 
extended Gross-Pitaevskii model. 
Within this  framework, we also find that the single soliton 
immersed in a droplet is stable, while multisoliton configurations exhibit 
parametric windows of oscillatory instabilities. 
Our results pave the way for studying dynamical features of nonlinear multisoliton 
excitations in a droplet environment in contemporary experimental settings.

\end{abstract}

\maketitle

\section{Introduction}
\label{intro}

Multicomponent cold atom macroscopic systems provide the possibility to assess quantum fluctuation phenomena within the weakly interacting regime~\cite{petrov2015quantum,luo2021new,malomed2020family,bottcher2020new}. 
A recent intriguing manifestation constitutes the formation of self-bound quantum droplets owing to the presence of the first-order quantum Lee-Huang-Yang (LHY) correction term~\cite{lee1957eigenvalues} acting repulsively in higher than one dimension to prevent the collapse potentially favored by the mean-field interactions. 
Such states of matter have been originally observed in dipolar gases~\cite{chomaz2022dipolar} and afterwards in relevant mixtures~\cite{bottcher2020new}. 
Recently, they were also realized in short-range attractively interacting two-component, 
both homonuclear~\cite{cheiney2018bright,semeghini2018self,cabrera2018quantum} and heteronuclear~\cite{d2019observation,burchianti2020dual} bosonic mixtures. 
Other proposals also suggest the existence of these states in Bose-Fermi mixtures~\cite{cui2018spin,rakshit2019quantum} under the presence of three-body interactions~\cite{sekino2018quantum,morera2021quantum,xu2022three}, 
optical lattices~\cite{malomedOL} and spin-orbit coupling~\cite{gangwar2022dynamics,li2017two}. Besides atomic platforms, droplet configurations can also be generated, e.g., in photonic systems~\cite{wilson2018observation}, vapors~\cite{holyst2013evaporation,feder1966homogeneous} and liquid Helium~\cite{barranco2006helium,toennies2004superfluid}, further promoting their broader relevance.

The theoretical modeling of droplets in atomic systems is achieved through an extended Gross-Pitaevskii (eGPE) framework~\cite{petrov2015quantum,petrov2016ultradilute} which has been utilized to probe a plethora of their properties. 
In short-range bosonic mixtures that we investigate herein, these include, but are not limited to, their inelastic collisions~\cite{ferioli2019collisions,astrakharchik2018dynamics}, structural deformations from a Gaussian-shape to a flat-top
profile~\cite{astrakharchik2018dynamics}, the behavior of their collective modes~\cite{astrakharchik2018dynamics,sturmer2021breathing,tylutki2020collective,cappellaro2018collective,hu2020collective,tengstrand2022droplet}, the triggering of modulational instability events~\cite{mithun2020modulational}
and their statistical 
mechanics~\cite{mithun2021statistical}, as well as the effects of thermal instabilities~\cite{de2021thermal,Wang_2020,boudjemaa2021many}. 
Beyond LHY correlation effects have also been discussed by considering self-consistently higher-order corrections~\cite{parisi2019liquid,parisi2020quantum,mistakidis2021formation,ota2020beyond,mistakidis2022cold} revealing, for instance, slight alterations in the breathing frequency or an enhancement of the expansion velocity. 

The focus of the vast majority of the above investigations was on the ground state properties, collective excitations and dynamical response of droplets.
Yet, it is natural, motivated also from corresponding analysis of
simpler, mean-field-driven  dynamical settings~\cite{kevrekidis2015defocusing},
to examine excited states and their dynamical stability in such mixtures.
This concerns in particular, the existence of nonlinearity-driven coherent 
structures, in the form of dark solitons~\cite{edmonds2022dark,gangwar2022dynamics} and vortices~\cite{kartashov2022spinor,kartashov2018three,li2018two,tengstrand2019rotating,examilioti2020ground,caldara2022vortices}, 
as well as vortex rings~\cite{sheikh} embedded in these self-bound states,
a topic which has been touched upon only very recently. 
For instance, in the one-dimensional (1D) geometry which is the central 
focus of the present study, the crossover from dark solitons at weak repulsive 
couplings to dark quantum droplets at attractive interactions in free space was demonstrated~\cite{edmonds2022dark}. 
However, it remains elusive under which conditions dark soliton states persist in the presence of the external trap, a common feature of relevant 
experiments~\cite{cheiney2018bright,d2019observation,cabrera2018quantum}, and importantly whether additional solutions, e.g., gray 
solitons~\cite{shukla2021kink,kartashov2022spinor} occur as it was argued recently for dipolar as well as in spin-orbit coupled~\cite{kaur2022supersolid} condensates in the presence of the LHY correction~\cite{kopycinski2022ultrawide}. 
Additionally, an analytical prediction regarding the dark soliton in-trap oscillation frequency similar to the one known in repulsive condensates~\cite{busch2001dark,frantzeskakis2010dark} constitutes a central question. 

Another intriguing aspect, that we tackle herein, concerns the existence, 
stability and dynamics of relevant multisoliton configurations especially so due 
to the presence of beyond mean-field nonlinearities in the mixture setting. 
On a different note, we remark that recently there is renewed interest on the 
experimental realization of a diverse array of solitonic structures in repulsive 
condensates~\cite{RDS,mossman2022dense,bakkali2021realization,chai2021magnetic,fritsch2020creating} 
which is expected to lead to interesting extensions for self-bound droplet environments.

To address the questions posed above we leverage a variety of theoretical 
and numerical tools. 
These include (i) a phase-plane analysis of the ensuing dynamical system,
(ii) a computational analysis of its excitation spectrum deploying the eGPE framework 
exploring the regime from the quantum droplet, low-density limit, to the 
Thomas-Fermi (TF), large-density limit, and 
(iii) direct numerical simulations in the context of dynamically unstable solutions. 
Specifically, we consider a homonuclear symmetric bosonic mixture showcasing the existence of various localized solutions in free space. 
We go beyond the well-studied
conventional bright droplets~\cite{luo2021new,malomed2020family} and the largely unexplored dark quantum droplets ({referred to as bubbles hereafter})~\cite{edmonds2022dark}, examining
black~\cite{edmonds2022dark} and gray solitons~\cite{kartashov2022spinor}. Notice that while the existence of a fraction of these nonlinear structures has been recently discussed in Refs.~\cite{edmonds2022dark,kartashov2022spinor}, here we unravel their origin from a completely different perspective. 
Namely, we rely on solutions of the dynamical system but also expose their persistence in the presence of the trap, as well as investigate their spectrum. The coexistence of multiple of the aforementioned waveforms is 
elucidated, as are their dynamical properties and the regimes where 
quantum fluctuations are central to their persistence.

\begin{figure*}[ht]
\includegraphics[width=0.95\textwidth]{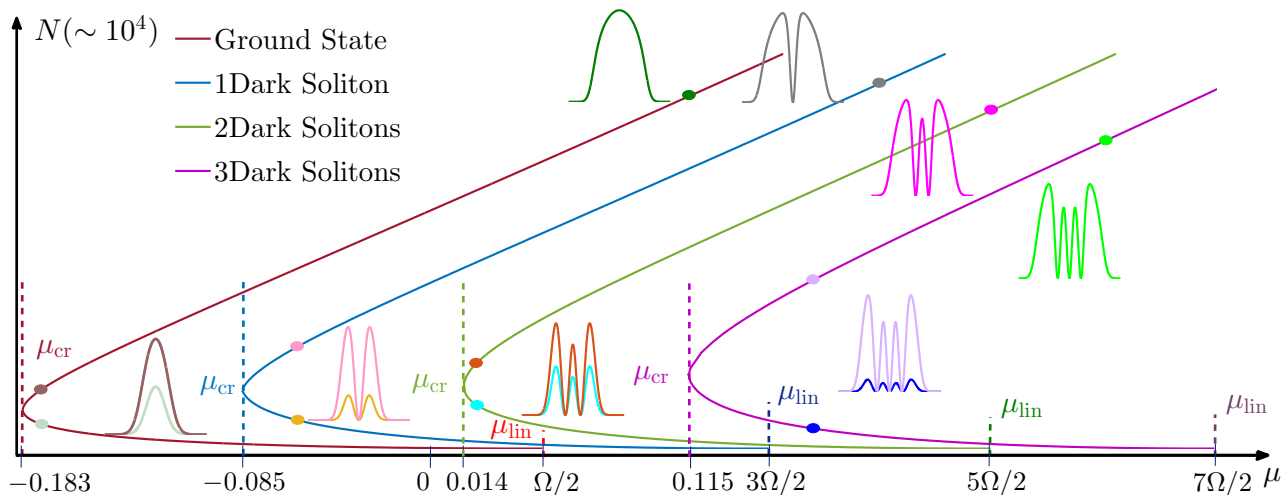}
\caption{Schematic representation of the underlying bifurcation diagram
of the particle number $N(\mu)$ as a function of the chemical potential $\mu$ 
for the droplet ground state as well as 
the single, two and three droplet dark solitons for $\delta=1$. 
Vertical dashed lines indicate the turning 
point, $\mu_{\rm cr}$, and the corresponding linear limit, 
$\mu_{\rm lin}=(n+1/2)\Omega$ of the adimensionalized model,
where $\Omega=0.1$ denotes the trap frequency.
The solution (Hermite-Gauss state) departs from the linear limit 
and for decreasing chemical potential acquires a droplet background 
since the LHY dominates until it reaches $\mu_{\rm cr}$. 
Afterwards, the configuration gradually deforms from a Gaussian-type backdrop 
towards a TF-type one bearing one or more strongly localized dark soliton(s) for 
increasing $\mu$ where the cubic nonlinearity prevails. 
Characteristic density profiles of the ensuing localized waveforms at 
specific chemical potentials (see ellipse markers) are also provided. 
The latter correspond to 
$\mu=-0.175$ and $0.115$ for the ground states, 
$\mu=-0.06$ and $0.2$ for the single, 
$\mu=0.02$ and $0.25$ for the two, and 
$\mu=0.17$ and $0.3$ for the three dark 
soliton solutions respectively. Note that for visualization purposes 
the curves are slightly shifted upwards from $N=0$. Namely, all
the lower curves asymptote to $N=0$ as $\mu\rightarrow\infty$.
}
\label{bif_diag} 
\end{figure*}

Moreover, utilizing the Landau dynamics approach~\cite{konotop2004landau}, the LHY-dependent oscillation frequency for a single
dark soliton is extracted recovering the standard mean-field prediction in the 
appropriate large density limit~\cite{kevrekidis2008emergent,pethick2008bose,PitaevskiiStringari2016}. Exploring the underlying excitation spectrum reveals that single dark solitons are stable structures for varying chemical potential, trap and LHY strengths. Strikingly, the interplay between the attractive in 1D LHY term (dominating for small chemical potentials) and the repulsive cubic nonlinearity (prevailing for large chemical potentials) 
is central for understanding the behavior of solitonic branches. For instance, a dark 
soliton in this setting bifurcates from the linear limit (either from the band edge 
in the absence of the trap, or from the first excited state in its presence) and tends 
to the soliton-inside-a-droplet for decreasing chemical potentials before
turning around so as to acquire a TF-type background for larger chemical potentials. 
Along these transitions, the dark soliton experiences modifications in its intensity and thus its core size. A summary of the
bifurcation structure for the multisoliton configurations is depicted in Fig.~\ref{bif_diag}. It is found that the aforementioned turning point which is a direct imprint of quantum fluctuations depends on the strengths of the LHY term and the trap but also on the soliton number. Turning to the largely unexplored, in this context, multisoliton solutions, we showcase that their turning points are shifted and, importantly, these states experience parametric windows of oscillatory instabilities. These manifest through the amplification of their out-of-phase vibration accompanied by a breathing of the entire configuration. 

The content of this work unfolds as follows. 
Section~\ref{model_phaseplane} introduces the model under investigation and the 
relevant theory framework based on the eGPE and its hydrodynamic form. 
Section~\ref{phase_plane} unveils  
the phase-plane analysis of the underlying dynamical system enabling,
among others, to identify their in-trap oscillation frequency. 
In Sec.~\ref{single_sol} we discuss in detail the excitation spectrum, existence and stability properties of the single soliton with respect to variations of the chemical potential, the strength of the LHY and the trap frequency. 
Section~\ref{multisol} elaborates on the persistence of multisoliton configurations and unveils their parametric windows of instability. 
Finally, in Sec.~\ref{conclusions} we summarize our results and offer future perspectives and challenges.

\section{Attractive Bose mixture and its hydrodynamic formulation}
\label{model_phaseplane}

We consider a harmonically trapped homonuclear Bose-Bose mixture with equal masses ($m_1=m_2\equiv m$) and intra-species repulsive interactions ($g_{11}=g_{22}\equiv g$).  
The inter-component couplings lie in the attractive regime ($g_{12}<0$), such that droplet configurations are generated depending on the value of the chemical potential. 
Due to these symmetry considerations it has been argued that the description of the genuine two-component system reduces to an effective single-component one~\cite{petrov2016ultradilute,petrov2015quantum,astrakharchik2018dynamics}, so that the two components are described by the same wave function, i.e., $\psi_1=\psi_2=\psi$. 
Such a setting has been suggested as accessible to experimental 
realizations via the utilization of two hyperfine states, $\ket{F,m_F}$, of $^{39}$K, e.g. $\ket{1,-1}$ and $\ket{1,0}$ as it was done in Refs.~\cite{cheiney2018bright,cabrera2018quantum,semeghini2018self}.  
Nevertheless, we should mention that, although this setting has been 
considered experimentally in the above works for 
three-dimensional quantum droplets, we are not
presently aware of a genuine 1D experimental realization in the realm
of the above model. This remains an important outstanding challenge
for the current experimental state-of-the-art.

The corresponding dimensional effective eGPE describing droplet configurations and including the first-order LHY quantum correction~\cite{petrov2016ultradilute} has the following form~\cite{tylutki2020collective}   
\begin{equation}
i\hbar \psi_t= -\frac{\hbar^2}{2m}\psi_{xx} + \frac{\delta g}{2}|\psi|^2 \psi - \frac{\sqrt{m}}{\pi \hbar} g^{3/2} |\psi| \psi +V_{\rm tr}(x)\psi,\label{eGPE_ORIG}
\end{equation}
where the subscripts denote partial derivatives, $\delta g=g_{12}+g$ quantifies the deviation from the mean-field balance point $\delta g=0$ (see the Supplemental
material of Ref.~\cite{cheiney2018bright} for the dependence and controllability 
of $\delta g$ based on external magnetic fields), and $V_{\rm tr}(x)$ is
the (usually parabolic) external trapping potential.
The results to be presented in all the figures below are provided in dimensional 
units in order to be directly comparable with current state-of-the-art experiments.
However, for ease of exposition, all the theoretical analysis 
will be provided using the corresponding adimensional 
1D eGPE~\cite{petrov2016ultradilute,kartashov2022spinor}  
\begin{equation}
i\psi_t= -\frac{1}{2}\psi_{xx} + |\psi|^2 \psi - \delta |\psi| \psi +V_{\rm tr}(x)\psi.
\label{eGPE}
\end{equation}
Here, energy, time, space, and interaction strengths are respectively measured in units of $\mathcal{E}_0\equiv \hbar \omega_{\perp}$, $\omega_{\perp}^{-1}$, $\sqrt{\hbar/(m \omega_{\perp})}$, and $\sqrt{\hbar^3 \omega_{\perp}/m}$. 
The atomic mass is $m$ and $\omega_{\perp}$ refers to the confinement frequency in the transversal direction which can be experimentally tuned with the aid of confinement induced resonances~\cite{olshanii1998atomic}. 
Importantly, the parameter $\delta$ (with $\delta>0$) describes the ``strength" of the 
LHY contribution~\cite{petrov2016ultradilute,kartashov2022spinor} and it is given by 
\begin{equation}
\delta = \frac{1}{\pi}\left[ \frac{2g^3}
{\mathcal{E}_0 \left( g_{12}+ g\right)} \right]^{1/2}.
\label{GP_notused}
\end{equation}
It is evident that $\delta$ depends on the involved interaction strengths (which can be routinely adjusted in the experiment through Fano-Feshbach resonances~\cite{cheiney2018bright,semeghini2018self,chin2010feshbach}) and also on the transverse confinement via $\mathcal{E}_0$. 
It is a main focus of this work to expose the impact of the LHY ``strength" on the solitonic solutions, see for instance Sec.~\ref{single_sol}. 
Notice that for $\delta=0$ the eGPE reduces to the common Gross-Pitaevskii equation~\cite{PitaevskiiStringari2016,pethick2008bose}. 
Moreover, we chose a standard parabolic external trapping potential $V_{\rm tr}(x)=(1/2)\Omega^2 x^2$, with $\Omega= \hbar \omega_0/\mathcal{E}_0$ and $\omega_0$ is the trap frequency in the longitudinal direction. 
To restrict the atomic motion in a 1D geometry, where transversal excitations do not play any role, we employ parametric variations in the interval $0 \leq \Omega \leq 0.1$.

To begin our analysis, we first employ the Madelung 
transformation~\cite{madelung1927quantentheorie,pethick2008bose}, $\psi=\sqrt{\rho}\,\exp(i\phi)$, where $\rho(x,t)$ and $\phi(x,t)$ denote, respectively, the 1D density and phase of the gas. 
This way, Eq.~(\ref{eGPE}) is expressed in the following 
hydrodynamic form
\begin{subequations}
\label{h1h2} 
\begin{eqnarray}
&&\rho_t+ (\rho \phi_x)_x=0, 
\label{h1} \\
&&\phi_t +\frac{1}{2}\phi_x^2 + \rho- \delta |\rho^{1/2}| 
- \frac{1}{2}\rho^{-1/2} (\rho^{1/2})_{xx}+V_{\rm tr}(x)=0.\nonumber \\ 
\label{h2}
\end{eqnarray}
\end{subequations}
The above coupled system of equations can be used for the derivation of stationary states of the system in free space [$V_{\rm tr}(x)=0$], as well as its ground state in the presence of the trap [$V_{\rm tr}(x) \neq 0$]. 

\section{Phase-plane analysis}
\label{phase_plane}

\subsection{Stationary states in free space}
\label{sec:C1=0}

Assuming that the potential $V_{\rm tr}(x)$ can be ignored in order to assess
the model properties in free space, we may seek stationary solutions of the 
form $\rho=\rho(x)$ and $ \phi(x,t)=\phi(x)$ for the system of Eqs.~(\ref{h1h2}). 
In such a case, integrating Eq.~(\ref{h1}) leads to $\phi_x=C_1/\rho$, 
where $C_1$ is a to-be-determined constant. 
It is important to note at this stage that the Madelung transformation
allows us to identify the gradient of the phase, $\phi_x$, as the
velocity of the traveling stationary states under consideration.
Therefore, configurations with $C_1=0$ correspond to fixed (non-traveling) 
steady states while the case $C_1\not=0$ corresponds to co-traveling
configurations.

The equation for $\phi_x$ in conjunction with Eq.~(\ref{h2}) 
leads to the following expression for the phase 
\begin{equation}
\phi(x,t)=\int \frac{C_1}{\rho(x)}dx -\mu t + \theta_0,
\label{ff}
\end{equation}
where $\mu$ plays the role of the chemical potential and $\theta_0$ is 
an integration constant. 
From this expression it becomes evident that the phase of the solution can be determined by the form of $\rho(x)$ (and 
the constants $C_1$, $\mu$, and $\theta_0$). 
The density $\rho(x)$, on the other hand, can be 
found from Eq.~(\ref{h2}), upon substituting the expression~(\ref{ff}). 
Indeed, introducing the auxiliary field $q=\rho^{1/2}$, 
Eq.~(\ref{h2}) takes the form\footnote{This substitution seems to suggest that $q>0$. 
Yet, a careful inspection of the original model leads to the conclusion 
that $q$ simply needs to be assumed to be real.} 
\begin{equation}
q_{xx}+2\mu q +2\delta |q|q -2q^3 -\frac{C_1^2}{q^3}=0, 
\label{feqq}
\end{equation} 
which can be viewed as the equation of motion of a unit mass particle 
in the presence of the effective potential
\begin{equation}
V(q)=\mu q^2 +\frac{2\delta}{3}q^2|q| -\frac{1}{2}q^4 +\frac{C_1^2}{2q^2}.
\label{epot}
\end{equation}
Notice that integration of Eq.~(\ref{feqq}) leads to the equation 
\begin{equation}
\frac{1}{2}q_x^2 +V(q)=E,
\nonumber
\end{equation}
where $E$ is the total energy of the system. This equation can be readily integrated, resulting in the implicit solution 
\begin{equation}
\int \frac{dq}{\sqrt{2[E-V(q)]}} = x-x_0,
\nonumber
\end{equation} 
where $x_0$ is an integration constant. 
In what follows, we provide a systematic study of the structure of the phase plane 
$(q,q_x)$ associated with the dynamical system of Eq.~(\ref{feqq}) 
focusing, in particular, on localized waveforms $q(x)$ corresponding to homoclinic 
[i.e., tending asymptotically to the same steady state
from the Greek ``$\kappa \lambda \iota \nu \omega$'' (to tend) and 
``$o \mu o \iota o$'' (same)]
or heteroclinic orbits [from ``$\kappa \lambda \iota \nu \omega$'' 
and ``$\epsilon \tau \epsilon \rho o$'' (different), i.e., tending 
asymptotically to different steady states] in the $(q,q_x)$ plane.  

Let us initially, for simplicity, consider the case of non-traveling configurations, 
$C_1=0$. The co-traveling case for $C_1\not=0$ is considered for completeness in 
Appendix~\ref{sec:append}.
For $C_1=0$, the extrema of the potential, that set the fixed points of the
system, are determined by $dV/dq=0$, i.e., by $q(-q^2 +\delta |q| +\mu) =0$. 
To solve this equation, it is necessary to 
distinguish the cases with $q>0$ and $q<0$, and with $\mu<0$ and $\mu \geq 0$. In particular, 
for $q \geq 0$ and $\mu \leq 0$, the fixed points are given by
\begin{equation}
q=0, \quad q_{\pm}= \frac{1}{2}\left(\delta \pm \sqrt{\delta^2 -4|\mu|} \right).
\label{fps}
\end{equation}   
This implies that there exist 
\begin{itemize}
\item[(a)] one fixed point, at $q=0$, for $\delta^2<4|\mu|$,
\item[(b)] two fixed points, at $q=0$ and $q=\delta/2$, for $\delta^2=4|\mu|$,
\item[(c)] three fixed points, at $q=0$ and $q=\frac{1}{2}\left(\delta \pm \sqrt{\delta^2 -4|\mu|} \right)$, for $\delta^2>4|\mu|$,
\end{itemize}
while for $q \geq 0$ and $\mu \geq 0$ there exist two fixed points corresponding to 
\begin{equation}
q=0, \quad q= \frac{1}{2}\left(\delta + \sqrt{\delta^2 +4\mu} \right).
\nonumber
\end{equation}   
Finally, in the case of $q<0$, and due to the fact that the potential is an even function 
of $q$, the fixed points are mirror symmetric to the ones above, namely a fixed point 
$q_{\rm fp}$ for $q>0$ maps to $-q_{\rm fp}$ for $q<0$.

\begin{figure*}[ht]
\includegraphics[height=8.7cm]{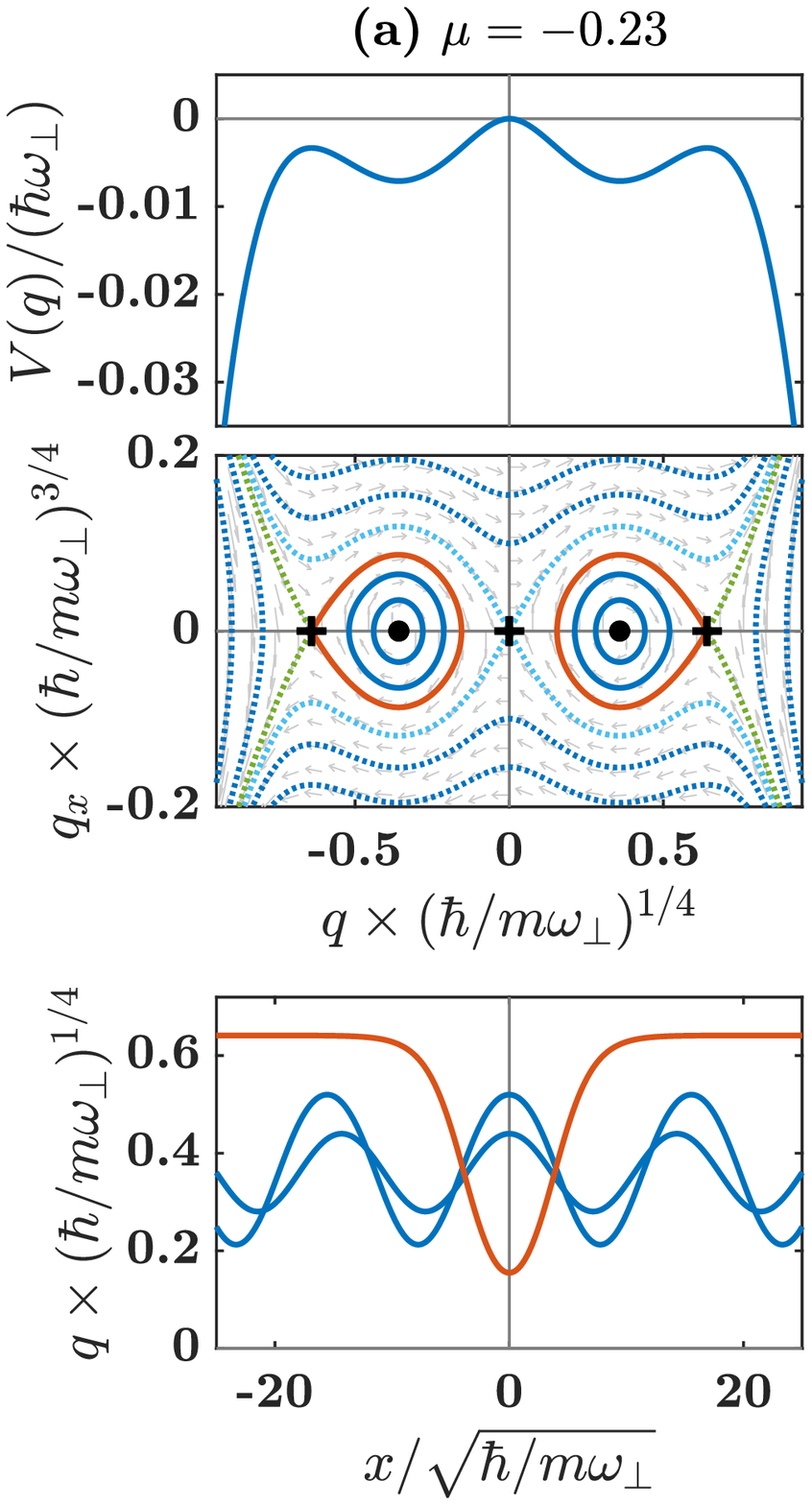}~
\includegraphics[height=8.7cm]{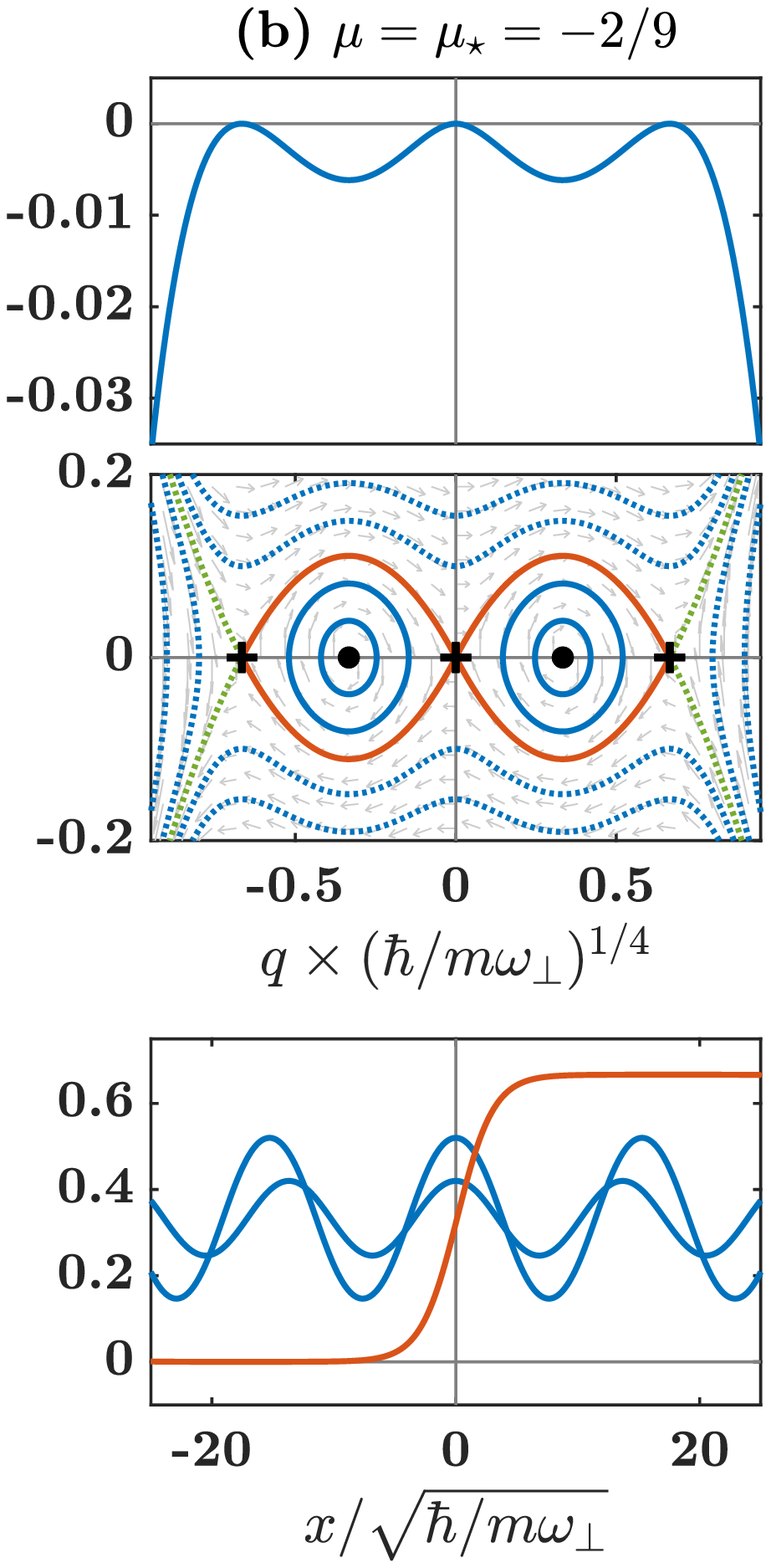}~
\includegraphics[height=8.7cm]{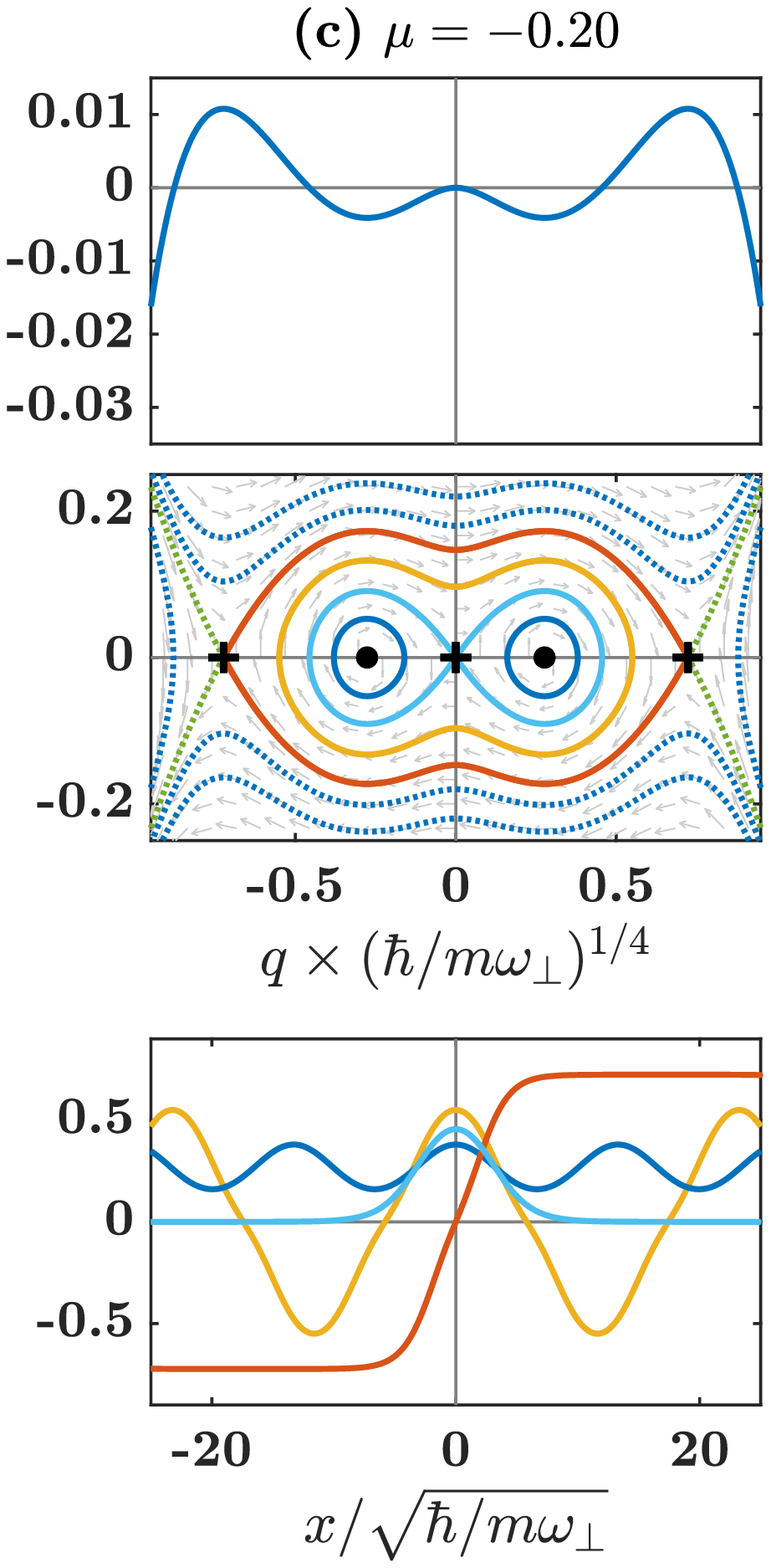}~~
\includegraphics[height=8.7cm]{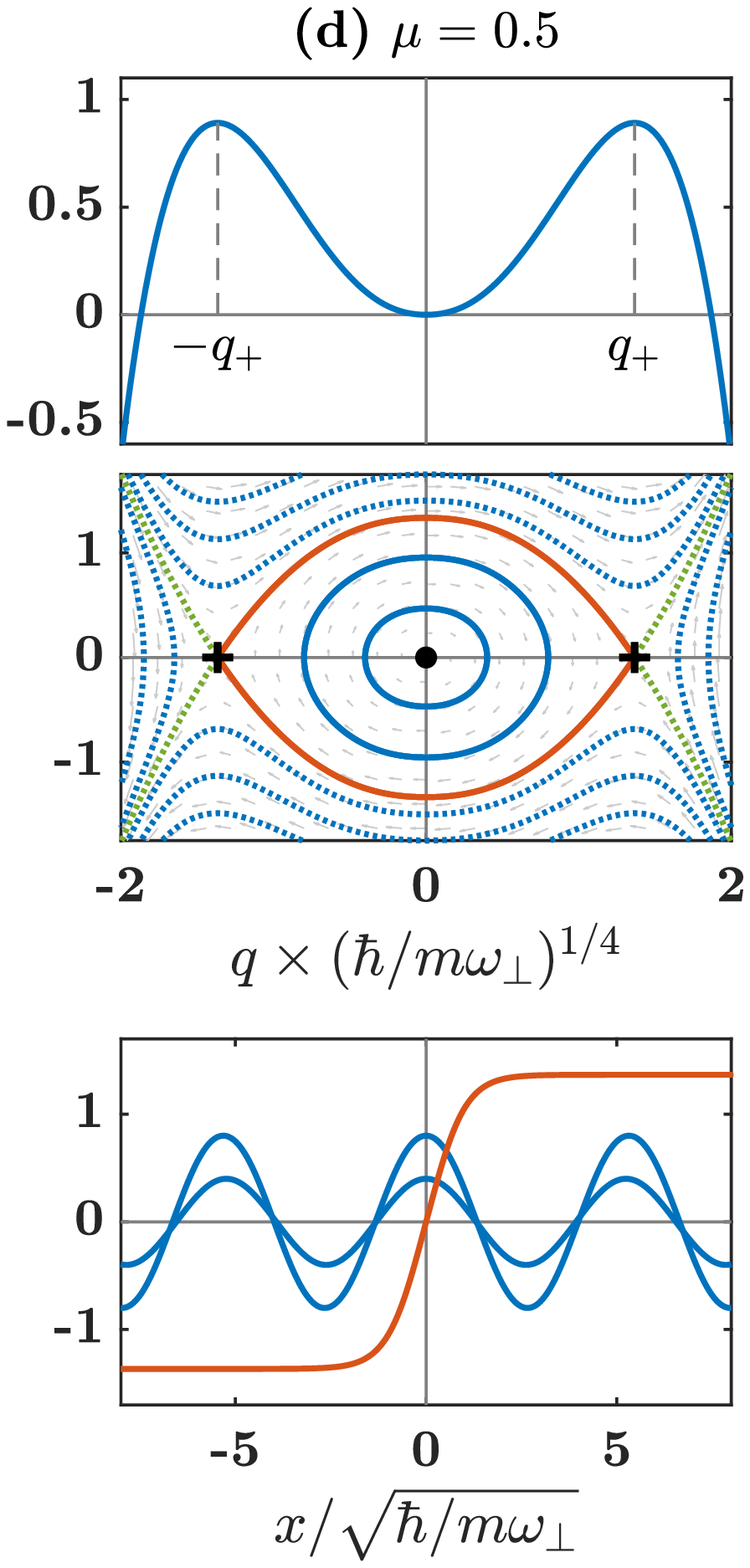}
\caption{Potential $V(q)$ (top panels), respective phase planes (middle panels) 
and bounded orbits (bottom panels) for $C_1=0$ and $\delta=1$
for which $\mu_{\star}=-2/9$.
The different columns of panels correspond to representative cases, for the values 
of $\mu$ indicated, for all the qualitative different scenarios that produce 
bounded orbits. 
Bounded (unbounded) orbits are depicted with solid (dashed) lines in the middle panels.
For clarity of exposition, when symmetric ($q \leftrightarrow -q$) solutions exist, 
only positive bounded 
orbits are shown in the lower panels.
Case (a) corresponds to $-\delta^2/4<\mu<\mu_{\star}$ which gives rise to homoclinic orbits
supported by a non-zero background.
Case (b) corresponds to $\mu=\mu_{\star}$ where the two homoclinic orbits of case (a) 
collide and give rise to heteroclinic orbits connecting zero and non-zero backgrounds.
Case (c) corresponds to $\mu_{\star}<\mu<0$ where the heteroclinic orbits
of case (b) merge into heteroclinic orbits connecting non-zero backgrounds.
Case (d) corresponds to $\mu>0$ where the homoclinic orbits of case (c) 
disappear while the heteroclinic orbits connecting non-zero backgrounds
are preserved.
Note that in order to better relate our results to experimental conditions,
we opt to display in this figure (as it is the case for all subsequent figures) 
all of the relevant quantities in full dimensional units as indicated in 
the labels.
}
\label{mns} 
\end{figure*} 

\begin{figure*}[ht]
\includegraphics[width=1.0\textwidth]{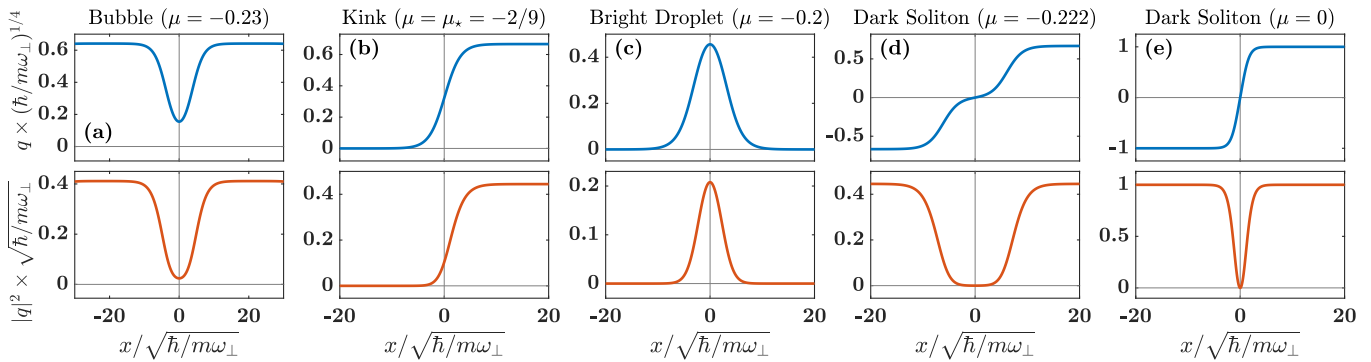}
\caption{{``Zoo" of nonlinear waves in the crossover from the droplet to the TF regime. 
Top and bottom sets of panels depict, respectively, the wave function, $q$,
and the relevant density profiles, $|q|^2$.
(a) Bubble wave function for $\mu=-0.23$ referring to a homoclinic solution
anchored at $q_+$. Bubbles are found within the interval 
$-\delta^2/4<\mu<\mu_{\star}$
where $q_+$ exists and $V(0)>V(\pm q_+)$, see also Fig.~\ref{mns}(a).
(b) Kink solution corresponding to a heteroclinic connection 
between $0$ and $q_+$ (or $-q_+$), occurring solely 
for $\mu=\mu_{\star}=-2\delta^2/9$. Here, the maxima of the effective potential have the same 
height (energy), i.e., $V(0)=V(\pm q_+)$ [see Fig.~\ref{mns}(b)].
(c) Bright droplet for $\mu=-0.2$. These states form for $\mu_{\star}<\mu<0$
when $V(0)<V(\pm q_+)$ 
and are homoclinic solutions at $q=0$ [see also Fig.~\ref{mns}(c)].
(d) [(e)] Dark soliton configuration for $\mu=-0.222$ [$\mu=0$] being
a heteroclinic orbit connecting $-q_+$ and $q_+$.
These structures arise when  $\mu>\mu_{\star}$ and, as $\mu\rightarrow\mu_{\star}^+$,
they become wider due to the bottleneck induced by the potential maximum at $q=0$
as evidenced in panel (d).
In all cases, $C_1=0$ and $\delta=1$.} 
}
\label{solnm} 
\end{figure*}

The structure of the effective potential and its associated phase plane 
become particularly interesting when $\mu=\mu_{\star}=-2\delta^2/9$.
For this value, $\delta^2>4|\mu_{\star}|$, and hence 
there exist five fixed points, namely three saddle points (associated with 
potential energy maxima) and two centers (associated with potential energy minima)
at the ordinary differential equation 
level. The effective potential in this case factorizes as 
\begin{equation}
V(q)=-\frac{1}{2}q^2\left(|q|-\frac{2\delta}{3}\right)^2,
\nonumber
\end{equation}
and the fixed points share the same energy, $E=0$. 
The relevant form of the potential (exhibiting two wells at $q=\pm \delta/3$) and 
the associated phase plane are depicted in Fig.~\ref{mns}(b) for $\delta=1$. 
Therefore, for $\mu=\mu_{\star}$ there exists a quartet of heteroclinic orbits, 
which correspond to \textit{kink}-type shapes of $q(x)$; 
see the relevant wave function depicted in red in the
bottom panel of Fig.~\ref{mns}(b) and in Fig.~\ref{solnm}(b). 
All these heteroclinic orbits can be found by direct integration, in an 
explicit analytical form (see, e.g., 
also Refs.~\cite{astrakharchik2018dynamics,mithun2020modulational}), namely
\begin{equation}
q_{\rm \,kink}(x)= \pm \frac{\delta}{3} \left[1 \pm \tanh\left( \frac{\delta}{3} x\right) \right].
\label{kinks}
\end{equation}
Notice that, inside the eight-shaped pattern that is formed by the quartet of the 
heteroclinic orbits,
periodic solutions are present (that can be expressed in terms of the Jacobi 
elliptic functions~\cite{mithun2020modulational}), while outside of this 
region all trajectories are unbounded. 
We remark that in the case of $\mu< -\delta^2/4$ for our setting of $\delta=1$, 
a straightforward analysis shows that all emergent trajectories,   
$q(x)$, are unbounded for $|x|\rightarrow \infty$, 
since there is only a potential energy maximum at the origin. 
For $-\delta^2/4<\mu< \mu_{\star}$, the situation is akin to the one presented in Fig.~\ref{mns}(a).
In this case, that merits separate investigation, there exist homoclinic solutions. 
However, rather than these being homoclinic to the vanishing background, as the well-studied
case of bright droplets, these are homoclinic to the potential maxima at $\pm q_+$,
hence representing so-called bubble solutions~\cite{baras1989}.
Furthermore, even though both bubbles and dark solitons exist on top of a finite 
background, bubbles do not feature
a phase jump (and an accompanying sign change) as dark solitons do.
In fact, bubbles are states which commence and return to the same
nontrivial equilibrium state (either $q_+$ or $-q_+$) and as such are homoclinic;
see red curve in Fig.~\ref{mns}(a) and also Fig.~\ref{solnm}(a).
On the other hand, dark solitons commence from one of the nontrivial fixed
points (say $q_+$) and end on the other (in this example, $-q_+$,
or vice-versa starting at $-q_+$ and ending at $q_+$) and, thus,
correspond to heteroclinic orbits;
see red curves in Fig.~\ref{mns}(c)-(d) and also Figs.~\ref{solnm}(d)-(e).
Bubble configurations and their (in)stability will be the subject of a separate study. Indeed,
here, as concerns droplets, we will examine solely the bright ones present for 
$\mu_{\star}<\mu<0$ and the dark solitonic excitations potentially present therein.

Turning to $\mu>\mu_{\star}$, the value of the two maxima of the potential on either side of $q=0$  
increases as $\mu$ increases while the central peak at $q=0$ remains unchanged. 
Precisely at $\mu=\mu_{\star}$ the two side peaks have the same height as the central peak creating the
quartet of heteroclinic orbits mentioned above [see Fig.~\ref{mns}(b)].
For larger values of $\mu$, the central peak becomes shallower than the side peaks resulting in the 
emergence of a pair of homoclinic orbits to the origin, i.e., the well-recognized
bright  quantum droplets~\cite{astrakharchik2018dynamics,mithun2020modulational,tylutki2020collective} 
[see light blue curves in Fig.~\ref{mns}(c)]. 
These can be found as exact analytical solutions:
\begin{equation}
q_{\rm \,droplet}(x)=\mp \frac{3\mu}{\delta} \frac{1}{1+\sqrt{1+\frac{9\mu}{2\delta^2}}\cosh(\sqrt{-2\mu}x)},
\label{droplet}
\end{equation} 
with the $-$ and $+$ signs corresponding to the homoclinic orbit with $q>0$ and $q<0$, 
in line with what was reported, e.g., in
Refs.~\cite{astrakharchik2018dynamics,tylutki2020collective,mithun2020modulational} for $\delta=1$. 
Notice that, as mentioned above, bright droplets 
exist within the interval $\mu_{\star}=-(2/9)\delta^2<\mu <0$. 
A characteristic example of the bright droplet solution [Eq.~(\ref{droplet})]
is depicted in Fig.~\ref{solnm}(c) [see also the orbit depicted in light blue
in Fig.~\ref{mns}(c)].
Obviously, the presence of the LHY term in our extended mean-field model is 
essential for the existence of such bright droplet waveforms.

Finally, let us focus on the dark soliton solutions. 
In general, for $\mu_{\star}<\mu$ (still for $C_1=0$), 
the potential $V(q)$ features two outer extrema $\pm q_+$ associated with a pair of 
saddle fixed points on either side of $q=0$ [top panels in Figs.~\ref{mns}(c) and (d)]. 
These outermost saddle fixed points are connected by a pair of heteroclinic orbits producing
dark soliton structures, see red curves in the middle and bottom panels of 
Figs.~\ref{mns}(c) and (d). However, as long as $\mu<0$ there exists also a saddle point at the origin. Due to the presence of this saddle point at $q=0$,
the relevant pair of heteroclinic orbits exhibits a local minimum in $q_x$.
As a result, the corresponding dark solitons do not 
have a tanh-shaped profile as is the case of the usual dark solitons of the 
defocusing nonlinear Schr\"odinger equation (NLS)~\cite{frantzeskakis2010dark},
but change their slope in the vicinity of the origin.
In fact, due to the bottleneck induced by the effective potential maximum
at $q=0$, the width of the dark soliton may be rendered arbitrarily large
in the limit $\mu\rightarrow\mu_{\star}^+$~\cite{edmonds2022dark}.
{A dark soliton profile with an enlarged core is depicted in Fig.~\ref{solnm}(d) for $\mu=-0.222$,
which should be contrasted with the case when the maximum of the effective potential 
at $q=0$ is absent, namely for $\mu\geq 0$.
In this latter case [see, e.g., Fig.~\ref{solnm}(e) for $\mu=0$], the dark (black) 
soliton has a profile closer to the familiar 
tanh-shaped one (with minimum density equal to zero) yet not precisely the same due 
to the presence of the LHY correction term; see Eq.~(\ref{DS}).}

Even though the profiles of these dark soliton states are far more complex than the 
standard NLS dark solitons~\cite{frantzeskakis2010dark}, they can still 
be obtained analytically using the following formula
\begin{eqnarray}
\label{DS}
q_{\rm \,dark}(x)=q_++\frac{-{\cal B}(\mu) + \sqrt{{\cal B}^2(\mu)-4 {\cal A}(\mu) {\cal C}(\mu)}}{2 {\cal A}(\mu)},\qquad
\end{eqnarray}
where 
${\cal A}(\mu)=B^2-4 A \tanh^2(\sqrt{A} (x))$, 
${\cal B}(\mu)=4 A B \sech^2(\sqrt{A} (x))$, and
${\cal C}(\mu)=4 A^2 \sech^2(\sqrt{A} (x))$, 
with $A=4 \mu + (1+\sqrt{1+4 \mu})$ and
$B=2 (\frac{1}{3}+\sqrt{1+4 \mu})$.
The expression of Eq.~(\ref{DS}) 
is valid for $x$ such that $q(x)>0$, while the other ``half'' of the
solution (when $q(x)<0$) is obtained by anti-symmetrizing the wave's profile past
the point of its zero-crossing.
In the above expression, $q_+$ is given by Eq.~(\ref{fps}), while
it should be noted that for all the solutions discussed in the absence of the trap
[cf.~also the kinks of Eq.~(\ref{kinks}) and droplets of Eq.~(\ref{droplet})], 
their invariance with respect to
translation allows to center them at any position $x_0$, even though in the above 
analytical expressions we have implicitly assumed that they are centered at $x_0=0$.
Note also, that despite the similarity between the kink structure of Fig.~\ref{solnm}(b) 
and the dark soliton of Fig.~\ref{solnm}(e) at the wave function level, the former 
solution exists only for $\mu=\mu_{\star}$ and, additionally, the two states have 
well distinguished asymptotics.

Summarizing, the systematic analysis of the behavior of the dynamical system 
associated with the stationary solutions of the eGPE of Eq.~(\ref{eGPE}) 
for $V_{\rm tr}(x)=0$ reveals a wealth of localized stationary states.  
These include bubbles, kinks, bright quantum droplets, dark (black) solitons, 
and (as discussed in the Appendix~\ref{sec:append}) gray solitons. 
The kink-like structures, bubbles, as well as the bright quantum droplets 
can only be supported in the system due to the presence of the LHY quadratic 
nonlinearity term. 
Notice that periodic nonlinear waves, such as the sn-,  cn- and dn-type Jacobi 
elliptic functions, also exist but are not the focus of this work
(see oscillatory, periodic, solutions in the bottom panels of Fig.~\ref{mns}). 
Moreover, it is particularly relevant to investigate interactions among the 
above-discussed entities in order to understand their character and whether 
they bear similarities with interacting solitons. This is discussed in detail 
in a forthcoming work~\cite{multidroplets}.

\subsection{The ground state of the trapped system}
\label{GS}

Let us now consider the ground state of the system in the presence
of the external trapping potential $V_{\rm tr}(x) \neq 0$.  
Within the large density TF limit, we seek solutions of Eqs.~(\ref{h1h2})
having the form $\rho=\rho_0(x)$, $\phi_t=-\mu$ and $\phi_x=0$. 
In this regime, the quantum pressure term $(1/2) \rho^{-1/2} (\rho^{1/2})_{xx}$ 
can be neglected, and the pertinent ground state is  obtained upon solving the 
reduced Eq.~(\ref{h2}) for $\rho_0^{1/2}$. Namely  
\begin{equation}
\left|\rho_0^{1/2}(x)\right|=\frac{\delta}{2} + \left[\left(\frac{\delta}{2}\right)^2+\mu-V_{\rm tr}(x) \right]^{1/2}, 
\label{TF}
\end{equation}
which is the only relevant solution since $\mu-V_{\rm tr}(x)>0$.  
Notice that in the absence of the LHY term ($\delta=0$), the well-known 
form of the TF cloud~\cite{pethick2008bose,PitaevskiiStringari2016} is retrieved. 
A careful inspection of the solution~(\ref{TF}) for the ground state in the large 
density limit unveils that it does not vanish for any $x$ [as in the case with 
$\delta=0$, where $\rho_0(\pm \sqrt{2\mu}/\Omega)=0$], but it rather takes the 
constant value $\rho_0=\delta/2$ for $|x|\geq \sqrt{\delta^2/2+2\mu}/\Omega$. 
An intriguing mathematical question concerns the decay of the trapped state
for large $\mu>0$ towards zero. To address this question, one needs
to carefully consider the turning point region where $\mu-V_{\rm tr}(x) \approx 0$.
A linear approximation thereof is expected to lead to a variation of the
Painlev{\'e}-II equation~\cite{gallo} that should be leveraged in order to identify 
the decay towards vanishing amplitudes. While this is an avenue that we do not
pursue herein, we recognize this as a fruitful direction for further mathematical
studies associated with the relevant ground state.

Below we focus on the large density limit, investigating the stability spectrum of the 
ground state since our aim is to embed dark solitons on top of such a background. 
The stability can be checked upon considering small-amplitude perturbations on the
ground state density $\rho_0(x)$ by substituting in Eqs.~(\ref{h1h2}) the ansatz 
$\rho(x,t)=\rho_0(x)+\epsilon \rho_1(x,t)$ and $\phi(x,t)=-\mu t + \epsilon \phi_1(x,t)$
where $0<\epsilon \ll 1$ is a formally small perturbation parameter. 
In this sense, keeping terms of order $\mathcal{O}(\epsilon)$ the following linear 
equation for $\rho_1(x,t)$ is derived
\begin{equation}
\rho_{1,tt}-c^2\rho_{1,xx} +\frac{1}{4}\rho_{1,xxxx}=0.
\nonumber
\end{equation}
The above equation admits plane wave solutions $\propto \exp[i(kx-\omega t)]$, where 
the wavenumber $k$ and frequency $\omega$ obey the Bogoliubov dispersion relation 
\begin{equation}
\omega^2=c^2 k^2 +\frac{1}{4}k^4,
\label{dr}
\end{equation}
while $c^2(x)=\rho_0^{1/2}(x) \left(\rho_0^{1/2}(x) -\delta/2 \right)$ 
is the square of the local speed of sound~\cite{landau1959fluid}. 
This reveals how the local speed of sound depends on the ``strength" of the first order quantum correction. 
Note also that the relevant expression is in line with earlier ones for this 
setting; see, e.g., Eq.~(10) of Ref.~\cite{sheikh}, upon accounting for the density 
expression of Eq.~(\ref{TF}).
Therefore, by measuring the speed of sound, it is possible to quantify the presence 
of quantum fluctuations. Additionally, as dictated by Eq.~(\ref{dr}), the frequency 
is always real for every wavenumber and thus, as it should be
expected, the ground state is always stable in the TF limit. 

As a final remark regarding the pertinent trapped ground state, it is also 
of interest to consider its low-density limit in the vicinity of negative 
chemical potentials, in connection with the relevant continuation 
illustrated in the bifurcation diagram of Fig.~\ref{bif_diag}.  
Here, in line with earlier works~\cite{kartashov2022spinor}, 
we see that the trapped ground state branch does not bifurcate from $\mu=0$ 
as in the pure cubic NLS model, but rather starts, as one might expect, 
from the ground state of the quantum harmonic oscillator at $\mu=\Omega/2$. 
The effective focusing nature of the model for small/intermediate intensities 
(due to the dominance for such amplitudes of the quadratically nonlinear term) 
leads to the formation of a droplet-like state. The latter is strongly 
reminiscent of the analytically obtained one in 
Refs.~\cite{astrakharchik2018dynamics,mithun2020modulational}, yet it 
is ``compressed'' in comparison (i.e., narrower) due to the effect of 
the trapping potential.  The resulting branch of solutions accordingly 
does not have a termination point at $\mu=-2 \delta^2/9$.  Rather, the 
corresponding solutions encounter a turning point $\mu_{\rm cr}>\mu_{\star}$, 
and subsequently turn and head towards the TF limit discussed above. 
Indeed, one can identify this turning point as the parametric threshold 
where the cubic nonlinearity term takes over, eventually leading 
(for higher values of the chemical potential) to the asymptotic large 
density regime.

\subsection{Landau dynamics of dark solitons}
\label{landau}

In the homogeneous case of $V_{\rm tr}(x)=0$  and in the absence of 
the LHY correction, the setting at hand reduces to the completely 
integrable NLS equation, characterized by the energy 
\begin{equation}
H=\int_{-\infty}^{+\infty} \left[|\psi_x|^2 + (\mu-|\psi|^2)^2 \right] dx.
\label{dsen}
\end{equation} 
Furthermore, the NLS equation possesses the following exact analytical dark 
soliton solution~\cite{zakharov1973interaction}
\begin{equation}
\psi(x,t)=\left\{ \sqrt{\mu-v^2}\tanh\left[\sqrt{\mu-v^2}(x-X_0) \right] +iv \right\} 
e^{-i\mu t}, \label{ds}
\end{equation}  
where $X_0$ is the soliton center and $dX_0/dt=v$ denotes its  
velocity~\cite{frantzeskakis2010dark,kevrekidis2015defocusing}.
The energy of the dark soliton, can be found upon substituting 
Eq.~(\ref{ds}) into Eq.~(\ref{dsen}), leading to  
\begin{equation}
E_{\rm DS} = \frac{4}{3} (c^2 -v^2)^{3/2},
\label{dsen2}
\end{equation}
where $c^2=\mu$ in this limit.

Let us now consider the dynamics of the dark soliton in the TF limit,  
in the presence of both the LHY term and the external potential.
For this purpose, we treat the LHY term as a small perturbation and
we assume the potential to be slowly varying on the soliton scale. 
In such a case, we may employ the so-called Landau dynamics approach~\cite{konotop2004landau} 
(see also Ref.~\cite{kevrekidis2017adiabatic}), according to which the soliton 
energy~(\ref{dsen}) is treated as an adiabatic invariant in the presence of perturbations. 
Namely, the background density $\mu$ will be slowly varying according to 
$\mu \rightarrow \mu-V_{\rm tr}(x)$, 
while $c^2\rightarrow c^2(x)=\rho_0^{1/2}(x) \left(\rho_0^{1/2}(x) -\delta/2 \right)$ 
accounting for the LHY contribution.     

Then, assuming the adiabatic invariance of the soliton energy in Eq.~(\ref{dsen2}), 
i.e., $c^2(x)-v^2 = [(3/4)E_{\rm DS}]^{2/3} \approx {\rm const.}$, 
using $v=dX_0/dt=\dot X_0$, and considering a parabolic trap with strength $\Omega$, 
namely $V_{\rm tr}(x)=(1/2) \Omega^2 x^2$, we derive the following nonlinear evolution 
equation for the dark soliton position
\begin{equation}
\ddot X_0
+ \frac{\Omega^2}{2}\left\{1+\frac{\delta}{4} 
\left[\frac{\delta^2}{4}\!\!+\mu - \frac{1}{2}\Omega^2 X_0^2\right]^{-1/2} \right\}X_0=0.
\nonumber
\end{equation}
Supposing that the soliton motion takes place in the vicinity of the trap center, 
we extract the oscillation frequency, $\Omega_{\rm osc}$, of the dark soliton in 
the presence of quantum fluctuations as
\begin{equation}
\Omega_{\rm osc} = \frac{\Omega}{\sqrt{2}}\left[1+ \frac{\delta}
{4\sqrt{\frac{\delta^2}{4} +\mu}} \right]^{1/2}. 
\label{osc_dark}
\end{equation}
Notice that in the absence of the LHY contribution (i.e., $\delta \rightarrow 0$), 
as well as  in the large density limit (with $\mu \rightarrow +\infty$), 
the oscillation frequency of the dark soliton retrieves the well-known value 
$\Omega_{\rm osc} =\Omega/\sqrt{2}$~\cite{frantzeskakis2010dark,kevrekidis2015defocusing}. 
As we will explicate later on, this analytical estimate correctly 
captures the trend of $\Omega_{\rm osc}$ in terms of $\mu$ as predicted 
from the numerical solution of the eGPE. 
However, deviations occur especially as $\mu$ decreases and as we depart 
from the TF regime. Indeed, as expected, the relevant frequency in the low 
density, near-linear limit tends to $\Omega_{\rm osc} = {\Omega}$, with the anomalous 
mode (see the discussion below)
interpolating between these two distinctive limits of small and large $\mu$.

\section{Excitation spectrum of single dark soliton solutions}
\label{single_sol} 

Having identified through the aforementioned phase-space analysis a plethora of localized solutions that exist 
in the setup at hand, let us now study stationary single and multiple  (see Sec.~\ref{multisol}) dark soliton
solutions in the presence of 
parabolic confinement~\cite{kevrekidis2015defocusing}.
The incorporation of the external trapping potential is of relevance 
for contemporary  
experiments~\cite{cheiney2018bright,cabrera2018quantum,semeghini2018self} dealing with homonuclear BEC mixtures.
In practice, we obtain the relevant trapped solutions by solving the time-independent version of the eGPE~(\ref{eGPE}) using a  
fixed point iterative scheme of the Newton type~\cite{kelley2003solving}\footnote{For the results 
presented throughout this work we use a spatial discretization of 
$dx=10^{-4}$ with a second (crosschecked with the outcome of a fourth) 
order finite differences scheme in space and a fourth order
Runge-Kutta method for the dynamical evolution 
of the system with temporal discretization $dt=10^{-5}$.}.
Specifically, stationary states are found upon varying the chemical potential, $\mu$, 
addressing both the small and large (TF) density limit for different strengths 
of the LHY contribution, $\delta$, spanning low, intermediate, 
and comparable to the standard cubic nonlinearity interactions.  
Characteristic density profiles of the droplet ground state as well as the different  single 
and multiple dark soliton configurations are depicted as insets in Fig.~\ref{bif_diag}.

To address the stability of the above obtained solutions, the following 
ansatz is introduced in the time-independent eGPE~(\ref{eGPE})
\begin{eqnarray}
 \Psi(x,t)=\Big[\psi_0 (x) + \epsilon \bigg( a(x) e^{-i \Omega_{\rm osc} t} +  b^{*}(x) e^{i \Omega_{\rm osc} t} 
 \bigg) \Big] e^{-i \mu t},\nonumber \\ 
 \label{eq:BdGpert} 
\end{eqnarray}
where $\psi_0(x)$ is the iteratively found equilibrium solution 
and $\epsilon$ denotes a small amplitude (formal) perturbation parameter. 
Additionally, $\Omega_{\rm osc}$ are the eigenfrequencies and $\big (a(x), b^{*}(x) \big)^T$ 
the eigenfunctions of the eigenvalue problem, which to $\mathcal{O}(\epsilon)$ reads
\begin{equation}
\Omega_{\rm osc} \left[ {\begin{array}{c}
    a \\
    b \\
  \end{array} } \right] =
    \left[ {\begin{array}{rr}
   L_{11} & L_{12}   \\
    -L^*_{12} & -L_{11} \\
  \end{array} } \right] \\
  \left[ {\begin{array}{c}
    a \\
    b \\
    \end{array} } \right],    
  \end{equation}
where $(\cdot)^*$ denotes complex conjugation.
Here, the block matrix elements are~\cite{tylutki2020collective}
\begin{eqnarray}
L_{11}&=&-\frac{1}{2} \partial^2_x +V_{\rm tr}-\mu+ 2 \abs{\psi_0}^2 
         -\frac{3}{2} \delta \abs{\psi_0},
\nonumber
\\[1.0ex]
\nonumber
L_{12}&=& \abs{\psi_0}^2 - \frac{1}{2} \delta \abs{\psi_0}.
\end{eqnarray}
Notice that the latter expression for $L_{12}$ notably hinges on the 
fact that the solution $\psi_0$ is real.

Recall, that in the absence of beyond mean-field corrections (i.e., for $\delta=0$),
the dark solitons of the cubic NLS are well-known to interpolate between 
the linear limit of the first excited state~\cite{alfimov2007nonlinear,chernyavsky2017krein}, 
with energy $E=\mu=\Omega (n + 1/2)$ for $n=1$ and the 
TF limit of large density for large values of the chemical potential $\mu$. 
Similarly to the ground state of the system (see Sec.~\ref{GS}), the relevant waveform
is dynamically stable for all values of $\mu$.
However, contrary to what is the case for the ground state,
the linearization around such a waveform bears a negative
energy (negative Krein signature)~\cite{skryabin2000instabilities,katsimiga2020observation} mode that interpolates
between the linear limit of frequency $\Omega_{\rm osc}=\Omega$
and the highly nonlinear TF limit
of $\Omega_{\rm osc}=\Omega/\sqrt{2}$~\cite{coles2010excited}. 
The Krein signature for the eGPE model under consideration is defined as 
$K=\Omega_{{\rm {osc}}}\int dx \left(|a|^2 - |b|^2\right)$.
It constitutes a key quantity of the Bogoliubov de Gennes stability analysis since it identifies the energy contribution of each mode to the unperturbed system. 
Specifically, depending on the eigenvectors ($a$, $b$) such a mode can have a positive frequency $\Omega_{{\rm osc }}$ but a negative energy $K<0$ or negative Krein signature.
It is these modes that, in what follows, are called anomalous modes (alias negative energy modes).

\begin{figure*}[ht]
\includegraphics[width=0.80\textwidth]{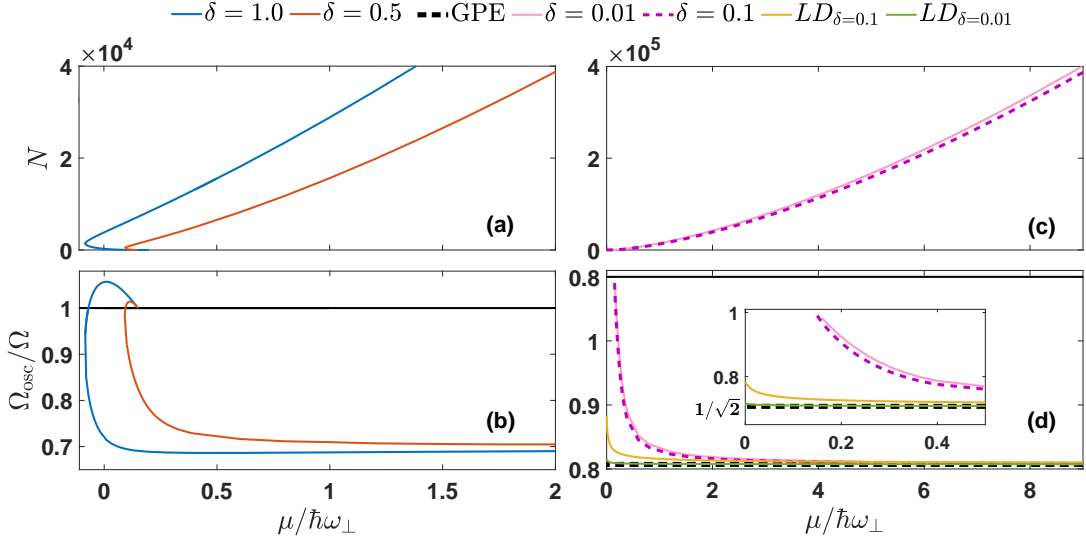}
\caption{(a), (c) Bifurcation diagram of the particle number, $N$, for
the single dark soliton solutions as a function of the chemical potential, 
$\mu$, for different values of the LHY correction as indicated.
(b), (d) Corresponding dependence of the internal (anomalous) mode of eigenfrequency 
$\Omega_{\rm osc}/\Omega$ vs.~$\mu$. 
Notice, than in addition to the dipolar mode of frequency $\Omega_{\rm osc}/\Omega=1$ 
(see the horizontal black line), the anomalous mode starts from the linear limit 
$\mu_{\rm lin}=(3/2)\Omega$. As $N$ is increased, for small enough $\delta$ values (see 
cases $\delta=0.01$ and $\delta=0.1$), the anomalous mode decreases and then
asymptotes to $\Omega_{\rm osc}/\Omega=1/\sqrt{2}$ for large $\mu$ values 
[see the horizontal dashed black line in (d)].
The inset in panel (d) corresponds to a magnification at small chemical potentials 
where also the difference from the prediction of Eq.~(\ref{osc_dark}) is maximal.
In contrast, for large enough values of $\delta$ (see cases $\delta=0.5$ and $\delta=1$),
the anomalous mode frequency 
first increases and then decreases as $N$ increases while the 
chemical potential also reaches a turning point for a minimal critical value of
$\mu$ at $\mu_{\rm cr}$ ($\mu_{\rm cr}=-0.085$ for $\delta=1$ and $\mu_{\rm cr}=0.091$
for $\delta=0.5$).
The trapping strength for all cases is fixed at $\Omega=0.1$.
}
\label{fig:1dark} 
\end{figure*} 
    
A natural question here concerns how the relevant bifurcation structure may change 
in the presence of the quadratic nonlinearity of the LHY term. 
Indeed, Figs.~\ref{fig:1dark}(a) and~(b) illustrate that the picture 
is drastically different when the LHY strength is present.
Contrary to the defocusing (repulsive) cubic nonlinearity case, which induces a bifurcation 
towards values of $\mu>E$ (i.e., to the right of the linear eigenvalue of the quantum harmonic
oscillator)~\cite{kevrekidis2015defocusing}, in the 1D geometry at hand, the focusing (attractive) 
nature of the LHY~\cite{luo2021new,mistakidis2022cold} correction is the dominant one for 
small intensities. The corresponding bifurcation diagram of the particle number, 
$N=\int |\psi(x)|^2 dx$, as a function of $\mu$ is depicted in Fig.~\ref{fig:1dark}(a)
for a pair of values of the LHY correction and a fixed trapping strength $\Omega=0.1$. 
As it can be seen, for increasing $N$, eventually, the defocusing nature of the 
large amplitude solutions ``takes over'' and leads to the emergence of a turning point,
for $\delta=1$, at $\mu=\mu_{\rm cr}=-0.085$. After this turning point, the dependence 
of $N$ on $\mu$ becomes monotonically increasing, as is representative 
of a defocusing nonlinearity~\cite{kevrekidis2015defocusing,chernyavsky2017krein}. 
Interestingly, the relevant critical point when concerning the ground state of the 
droplet is shifted to more negative values of $\mu$ as compared to the dark droplet
soliton solution [see Fig.~\ref{bif_diag}]. Naturally,
as the trap strength tends to vanishing, accordingly
both the starting point of the branch and its 
$\mu_{\rm cr}$ shift to the left approaching the homogeneous
limit values. 
Therefore, when lowering $\Omega$, the region of existence of the droplet background 
becomes larger, a statement that holds independently of the number of embedded solitons.

\begin{figure*}[ht]
\includegraphics[width=0.80\textwidth]{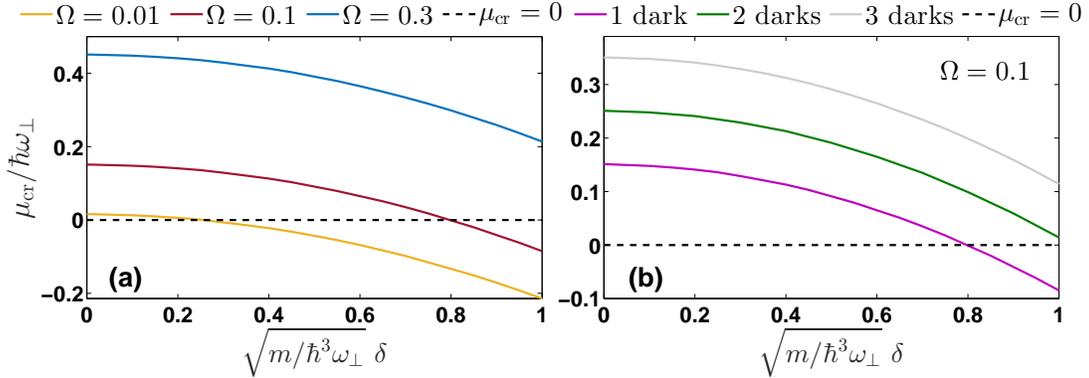}
\caption{
Critical value of the chemical potential, $\mu_{\rm cr}$, versus the LHY strength, $\delta$, 
for (a) single dark soliton solutions and different trap frequencies $\Omega$ (see legend) 
and (b) distinct dark soliton complexes ranging from one to three (see legend) 
and for $\Omega=0.1$. 
Evidently, for fixed $\delta$ the turning point gets shifted to lower chemical 
potentials, either by decreasing the trap frequency or the soliton number. 
In, all cases larger $\delta$ reduces $\mu_{\rm cr}$. 
The dashed horizontal line at $\mu_{\rm cr}=0$ is added as a guide to the eye.}
\label{fig:critical} 
\end{figure*} 

It is also relevant to follow the dependence of the corresponding ``anomalous'' 
(negative energy or negative Krein sign) spectral mode; see Fig.~\ref{fig:1dark}$\rm{(b)}$. 
The relevant (rescaled) eigenfrequency, at the linear limit bifurcates 
from the value of $\Omega_{\rm osc }/\Omega=1$ 
when $\mu=\mu_{\rm lin }=(3/2) \Omega$.
Initially, once again, the effective focusing nature of the nonlinearity 
leads to larger values of $\Omega_{\rm osc}/\Omega$, but eventually the 
defocusing character of the large amplitude (high nonlinearity) limit
leads $\Omega_{\rm osc}/\Omega$ to turn around and start 
decreasing as shown in Fig.~\ref{fig:1dark}$\rm{(b)}$. 
In the large $\mu$ (TF) limit, the corresponding eigenfrequency 
asymptotes towards the constant value $\Omega_{\rm osc}/\Omega=1/\sqrt{2}$.
In this latter case, the solution is largely reminiscent of the corresponding cubic 
problem, while near the linear limit of low density,
the solution resembles a first-excited Hermite-Gauss eigenmode; 
see the relevant profiles in Fig.~\ref{bif_diag}. 
Indeed, residing on the upper (lower) branch, the soliton core widens 
for smaller (larger) $\mu$ since the atom number is significantly reduced. 
The same overall phenomenology occurs also for intermediate strengths of 
the LHY term as depicted also in Figs.~\ref{fig:1dark}(a)
and~(b) for $\delta=0.5$. 
Notice however, that for smaller values of the parameter $\delta$, the 
turning point, appearing at $\mu_{\rm cr}=0.091$ for $\delta=0.5$, 
is shifted towards more positive values.

\begin{figure*}[ht]
\includegraphics[width=0.98\textwidth]{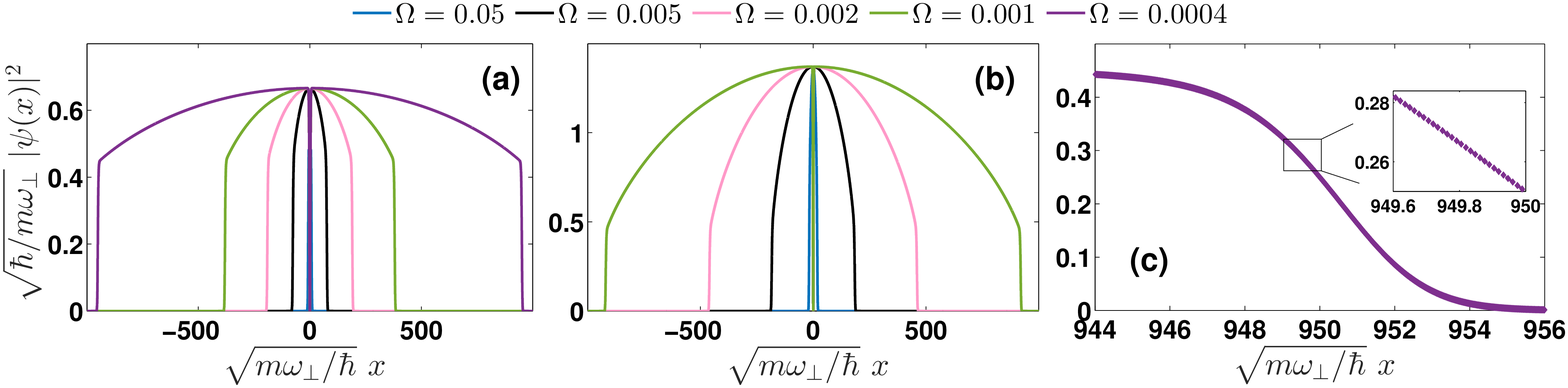}
\caption{
Selected density profiles of the droplet dark soliton for (a) $\mu=-0.15$ and 
(b) $\mu=0.2$ and distinct $\Omega$ values (see legends) in the case of $\delta=1$. 
Reducing the trap strength leads to a flat-top droplet dark configuration. 
Panel (c) and its inset depict different magnifications of the density profile of panel (a) for 
$\Omega=0.0004$ illustrating the spatial resolution used.
}
\label{fig:trap_var} 
\end{figure*} 

\begin{figure*}[ht]
\includegraphics[width=0.95\textwidth]{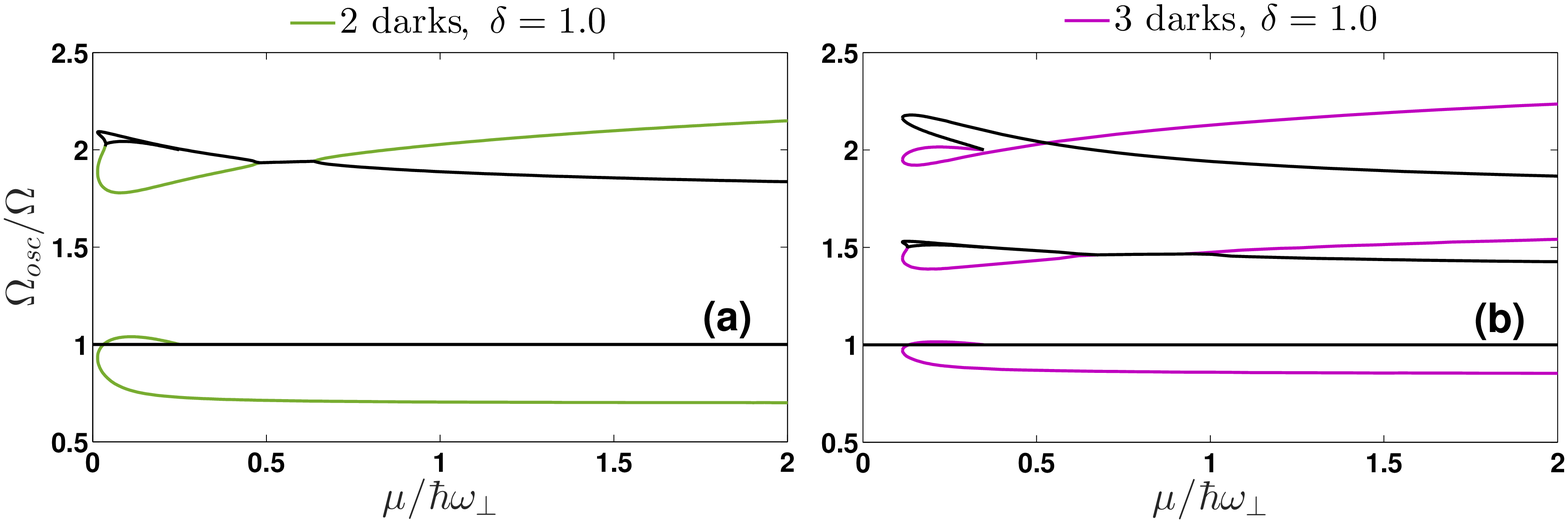}
\caption{
Dependence of the anomalous mode, $\Omega_{\rm osc}/\Omega$, for (a) two 
and (b) three dark soliton solutions as a function of $\mu$ for $\delta=1$. 
The relevant turning points, being shifted to more positive $\mu$ values 
for higher soliton complexes, appear at $\mu_{\rm c}=0.014$ and 
$\mu_{\rm cr}=0.115$.
In both cases the collision of the second anomalous mode (green and purple lines) with 
the background ones (black lines) signals the occurrence of an oscillatory instability.
}
\label{fig:3darks} 
\end{figure*} 

The significance of the LHY contribution on the stability properties of a 
single dark soliton is further 
elucidated by treating this term as a (weak) perturbative one. 
Namely, when considering significantly lower values of the relevant interaction 
coefficient such as $\delta=0.1$ and $0.01$; see Figs.~\ref{fig:1dark}(c) and~(d).
Evidently, since repulsion prevails in both cases, $N(\mu)$ is a monotonically 
increasing function as $\mu$ increases from the linear limit of $\mu=3 \Omega/2$; 
see Fig.~\ref{fig:1dark}(c).
This monotonicity for small $\delta$ 
is also present in the relevant dependence of the anomalous mode 
for each of the distinct values of $\delta$; see Fig.~\ref{fig:1dark}(d). 
In cases of higher $\delta$, as shown in Fig.~\ref{fig:1dark} [see, e.g., panel (b)], 
the asymptotics of the relevant anomalous mode still approach the same (scaled) limit 
of $1/\sqrt{2}$ [as is also predicted by Eq.~(\ref{osc_dark})] for large $\mu$.
Yet, they do so through a multivalued $\mu$-dependence, reflecting in this
way the corresponding $N(\mu)$ curve.
Finally, Fig.~\ref{fig:1dark}(d) also depicts the results for the prediction of
the dark soliton oscillation frequency using the Landau dynamics approach
as per Eq.~(\ref{osc_dark}). As the figure suggests, the Landau prediction correctly 
captures the qualitative tendency of the oscillation frequency for increasingly larger 
values of $\mu$ albeit with a noticeable discrepancy for small $\mu$ values as 
depicted in the inset. 
In that vein, it is relevant to recall that the relevant prediction is expected
to be asymptotically valid in the limit of large chemical potential.

Next, we aim to determine the impact of the LHY strength on the location of the critical 
point of the bifurcation diagram $N(\mu)$ for distinct trap frequencies and soliton numbers.
To this end, we initially obtain dark soliton solutions upon varying $\delta$ in the interval $[0, 1]$
covering this way the small, intermediate and large density limits 
and identify the relevant turning point $\mu=\mu_{\rm cr}$.
Focusing on the single soliton case, the behavior of $\mu_{\rm cr}$ in terms of $\delta$
is depicted in Fig.~\ref{fig:critical}$\rm{(a)}$ not only for $\Omega=0.1$ that is typically 
utilized herein, but also upon varying the trapping frequency. 
It becomes apparent that, irrespective of the trap strength,
$\mu_{\rm cr}$ decreases for increasing $\delta$.
Additionally, for fixed $\Omega$, $\mu_{\rm cr}$ is shifted to more positive values 
as $\delta \rightarrow 0$ whilst for looser traps $\mu_{\rm cr} \rightarrow -0.2$ 
for $\delta \rightarrow 1$, approaching this way the value of $\mu_{\star}=-2/9$  
found in our previous phase-plane analysis in free space. 
Furthermore, it is possible to infer from which strength of the LHY contribution
onward, $N(\mu)$ features one instead of two (i.e.,~lower and upper) branches.
For instance, when $\Omega=0.1$ and $\delta\approx 0.2$, $\mu_{\rm cr}\approx \mu_{\rm lin}=0.15$, 
and thus for $\delta \lesssim 0.2$ the bifurcation diagram consists only of the upper branch. 
A similar behavior of $\mu_{\rm cr}(\delta)$ takes place for higher soliton numbers, see in 
particular Fig.~\ref{fig:critical}$\rm{(b)}$. 
As can also be seen, $\mu_{\rm cr}$ is larger for fixed ($\Omega$, $\delta$) when 
increasing the number of solitons; see also the relevant discussion
of such solutions in the next section. 
We remark that in the absence of a soliton the respective critical point of the emergent 
droplet occurs at lower $\mu$, e.g., when $\delta=1$ then $\mu_{\rm cr}\approx-0.183$; 
see leftmost curve in Fig.~\ref{bif_diag}.

Since dark solitons and bright quantum droplets coexist, in what follows 
we explicitly depict in Figs.~\ref{fig:trap_var}(a) and~(b) representative 
droplet-dark soliton density profiles for different $\mu$ values upon
considering variations of the trapping frequency for $\delta=1$. 
In all cases, we have verified the spectral stability of these solutions 
as the homogeneous limit is approached. 
In line with our previous findings, solutions bearing 
smaller $\mu$ values, such as $\mu=-0.15$, persist for trap frequencies 
$\sim 10^{-4}$ but they cease to exist for $\Omega \sim 10^{-2}$.  
While traversing the relevant branch, the density profiles 
gradually alter their shape from a configuration proximal to the 
first-excited Hermite-Gauss eigenstate to a progressively wider dark droplet
configuration; see Figs.~\ref{fig:trap_var}(a) and~(b). 
Furthermore, looser traps lead to droplet-dark configurations with a wider flat-top
portion, being more pronounced for smaller values of $\mu$. 
Notice that the droplet background is modified [see Figs.~\ref{fig:trap_var}(a) and~(b)] 
when compared to the homogeneous case~\cite{tylutki2020collective,edmonds2022dark} 
due to the presence of the trap as has been also observed for 
dipolar bosons in Ref.~\cite{PhysRevB.106.014503}.
Furthermore, a sharp decay of the density, with respect to the size of the condensate, 
can be observed at the edge of the cloud. This sharp transition, when compared to the 
entire domain, forces the numerical characterization of the corresponding solutions to 
contain a large number of mesh points in order to keep a fine enough discretization to
resolve the solution over the entire domain as depicted in the zoomed panel 
of Fig.~\ref{fig:trap_var}(c).
It is exactly in this turning point region where an analysis tantamount to that 
of Ref.~\cite{gallo} is relevant to perform, a topic of interest for future 
mathematical studies.

\begin{figure*}[ht]
\includegraphics[width=1.0\textwidth]{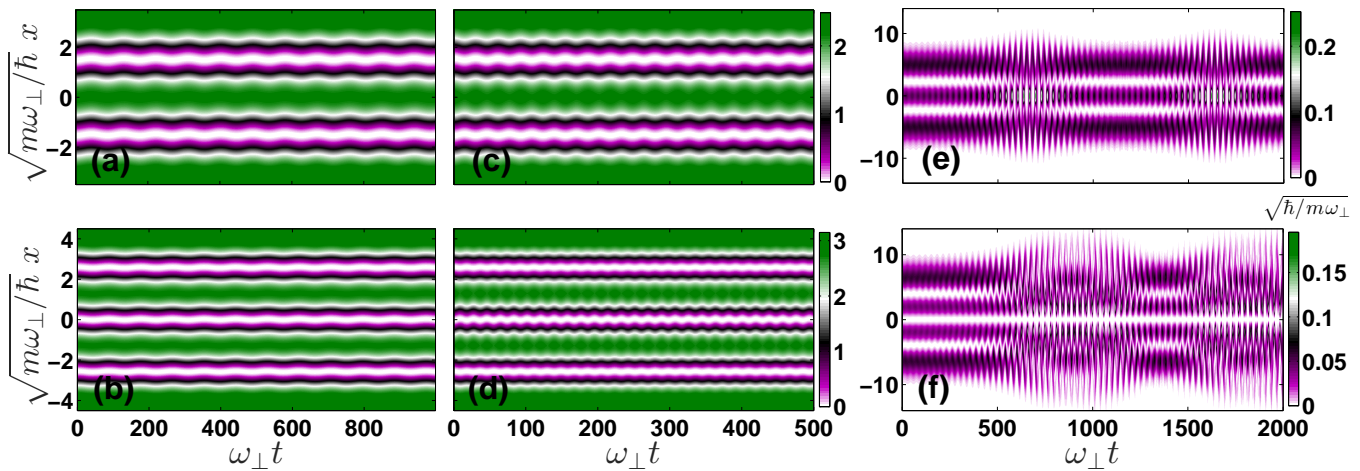}
\caption{Density evolution of anomalous modes $|\Psi(x,t)_{{\rm AM}_i}|^2$.
In each case, the steady state is perturbed with the corresponding anomalous modes 
initialized with an amplitude of $\epsilon=0.3$ and then allowed to evolve in time.
Panels (a)--(d) correspond to stable modes while (e) and (f) pertain to unstable modes for 
the two-soliton [(a) and (c)] and the three-soliton [(b), (d)--(f)] 
configurations.
The anomalous modes in panels (a) and (b) represent in-phase oscillations
while in panels (c) and (d) correspond to out-of-phase oscillations.
The unstable cases depicted in panels (e) and (f) for
two and three solitons, respectively, showcase the destabilization of the second 
anomalous mode (${\rm AM}_2$) through a complex eigenfrequency quartet.
It is observed that the dynamical evolution of this instability is
manifested as amplified out-of-phase vibrations
that eventually saturate and are subsequently
subject to recurrence.
}  
\label{fig:dynamics} 
\end{figure*} 

\section{Multisoliton solutions and their dynamics}
\label{multisol}

Let us now consider a homonuclear harmonically confined BEC mixture, 
in which the LHY contribution 
is taken on equal footing with the cubic nonlinearity (i.e., $\delta=1$),
and offer a generalization of our findings of the preceding section 
to multisoliton complexes; see Fig.~\ref{bif_diag}.
Note here that, although we restrict ourselves to configurations 
consisting of two and three dark solitons, our results can be 
generalized to $\mathcal{N}$-dark soliton states. 
Starting with a general qualitative 
remark, we note that 
higher-order dark solitons (2-solitons, 3-solitons, etc.)
are progressively higher excited states of the system; see the relevant
discussion of Ref.~\cite{Chernyavsky2018} for the cubic nonlinearity case.
As such,
each of them bears a progressively higher number of 
negative Krein-sign eigenvalues. As has been proved in the work
of Ref.~\cite{kapitula}, the number of such 
negative Krein-sign modes is equal to the number of the solitons
in the configuration. While this does not, a priori, render
the state dynamically unstable, it renders it far more prone to
so-called oscillatory instabilities arising from the collisions
of such modes with the rest of the excitation spectrum. The latter, 
asymptotically, resembles that of
the ground state configuration. In short, an $\mathcal{N}$-dark-soliton
state bears $\mathcal{N}$ ``excitation modes'' (pertaining structurally
to in-phase, out-of-phase, and mixed-phase motions of the
dark solitons) which may contribute to its potential instability.

The stability analysis outcome of the identified multisoliton 
solutions is showcased in Fig.~\ref{fig:3darks}.
Similarly to the bifurcation diagram of Fig.~\ref{fig:1dark}, also here,
the dominant attractive role of the LHY term for small $\mu$ values and the 
repulsive nature of the cubic term for large ones lead to the appearance 
of a turning point in $N(\mu)$ both for the two- and the 
three-dark soliton configurations. 
This turning point appears at $\mu_{\rm cr}=0.014$ for the two-soliton states 
while it occurs at $\mu_{\rm cr}=0.115$ for the three-soliton complexes.
However, in contrast to the single dark soliton case, the corresponding 
spectrum of these higher soliton complexes is more involved as can be 
seen by inspecting the behavior of the anomalous modes shown in 
Fig.~\ref{fig:3darks}.
Here, the lowest lying anomalous mode (designated by ${\rm AM}_1$) for these
multisoliton configurations follows a similar trend as the one found 
in the single dark scenario.
This is natural to expect as this mode pertains to the in-phase motion 
of the dark solitons (see, e.g., also Ref.~\cite{weller2}). In this case, 
the distance between the solitons does not change and hence the oscillation 
frequency is tantamount to that of the motion of a single soliton in the 
trap, i.e., effectively of the center of mass.
The lowest lying spectral modes depart from the linear limit 
of $\Omega_{\rm osc}/\Omega=1$ when $\mu=\mu_{\rm lin}=5 \Omega/2$
and $\mu=\mu_{\rm lin}=7 \Omega/2$ for the two- and three-dark 
soliton solutions respectively.
Additionally, the two and all three anomalous modes turn simultaneously 
at each of the aforementioned critical points;  
notice the relevant ``loops'' 
present for the second (${\rm AM}_2$) and the second and the third (${\rm AM}_3$) 
modes for the two- and the three-dark solitons respectively.
In particular, the higher modes for both two- and three-soliton
configurations, bearing negative Krein signature ($K<0$),
undergo collisions with the background modes
that are characterized by a positive Krein signature ($K>0$).
Such collision events give rise to complex eigenfrequency quartets signaling 
the presence of an oscillatory instability~\cite{katsimiga2020observation} 
for the ensuing configuration;
see also the discussion 
in Refs.~\cite{weller2,kevrekidis2015defocusing,Chernyavsky2018}.
The relevant instability windows for the two-soliton states appear for 
$\mu \in [0.037, 0.249]$ and 
$\mu \in [0.482, 0.636]$ whereas they occur for 
$\mu \in [0.129, 0.349]$ and 
$\mu \in [0.650, 0.925]$ for the three-soliton ones. 

Dynamical evolution of the steady states perturbed by the anomalous modes
is showcased in Fig.~\ref{fig:dynamics}. Specifically, the system is
initialized with a perturbed density $|\Psi(x,t)_{{\rm AM}_i}|^2$ where,
as per Eq.~(\ref{eq:BdGpert}), the initial condition is given by
$\Psi(x,0)_{{\rm AM}_i}=\psi_0 (x)+\epsilon [a_i(x)+b_i^{*}(x)]$
with the subscript $i$ indicating the anomalous mode number and 
$\epsilon=0.3$ is chosen so as to observe the ensuing oscillations 
already at an early stage within the dynamical simulation.
The dynamics of the ensuing anomalous mode perturbations for
$i=1,2,3$  of the different multisoliton configurations depicted 
in Fig.~\ref{fig:dynamics} confirms the above stability analysis 
results, while exposing the internal type of motion that 
each of the modes induces.
Particularly, for both two- and three-soliton states, the lowest 
lying anomalous eigenfrequency (${\rm AM}_1$) leads, once activated, 
i.e., upon adding to the stationary solution the eigenvector 
associated with it, to the in-phase vibration of these entities.
This is illustrated in Fig.~\ref{fig:dynamics}(a) [Fig.~\ref{fig:dynamics}(b)] when   
$\mu=1$ [$\mu=1.5$] and $\delta=1$ for the two- [three-]soliton configuration.
On the other hand, ${\rm AM}_2$ triggers the out-of-phase  oscillation of the two soliton 
solution [Fig.~\ref{fig:dynamics}(c)].
The same type of motion but in a pairwise fashion is activated when exciting the 
three-soliton state through ${\rm AM}_3$ [Fig.~\ref{fig:dynamics}(d)]. 
The remaining anomalous mode, ${\rm AM}_2$, in the three-soliton case produces an 
out-of-phase motion of the outermost dark solitons while the central 
one is unaffected throughout the evolution (results not shown here).
Note that for the stable configurations in Fig.~\ref{fig:dynamics},
we verified that the evolution remained coherent up to times
$t\sim 3\times 10^3$. 

Finally, let us briefly describe the typical dynamics ensuing from the
instabilities of multisoliton configurations. Specifically, for parameter 
values residing in the above-discussed instability windows, 
we find that, irrespectively of which mode is added to the initially 
stationary two- and three-soliton configuration, the resonant (unstable)
second mode is eventually excited.
The activation of this mode is depicted in Figs.~\ref{fig:dynamics}(e)
and~(f), respectively, for two and three dark soliton
solutions and for chemical potentials corresponding to the maximum 
instability growth rate. 
Namely, for $\mu=0.08$ and $\mu=0.2$ having an imaginary part 
$\rm{Im}(\Omega_{\rm osc}/\Omega)\approx 0.013$ and 
$\rm{Im}(\Omega_{\rm osc}/\Omega)\approx 0.048$, respectively.
In both cases, the amplification of the out-of-phase vibration is 
triggered and progressively leads to an overall breathing of the entire configuration. 
{This overall breathing pattern of the background is more transparent in the three dark 
soliton state, when compared to the two soliton one, with the central wave remaining nearly unaffected.
The impact of this breathing is also reflected on the occurrence of ``beats" present, for instance, at the contraction intervals around $t\approx 700$ or $t=1650$ in Fig.~\ref{fig:dynamics}(e).
These beats stem from resonances between the anomalous modes of dark solitons and those associated with the background state (on which these solitons are placed). 
This is a fundamental feature
pertaining to such oscillatory instabilities~\cite{kevrekidis2015defocusing}.}

\begin{figure*}[ht]
\includegraphics[height=9.0cm]{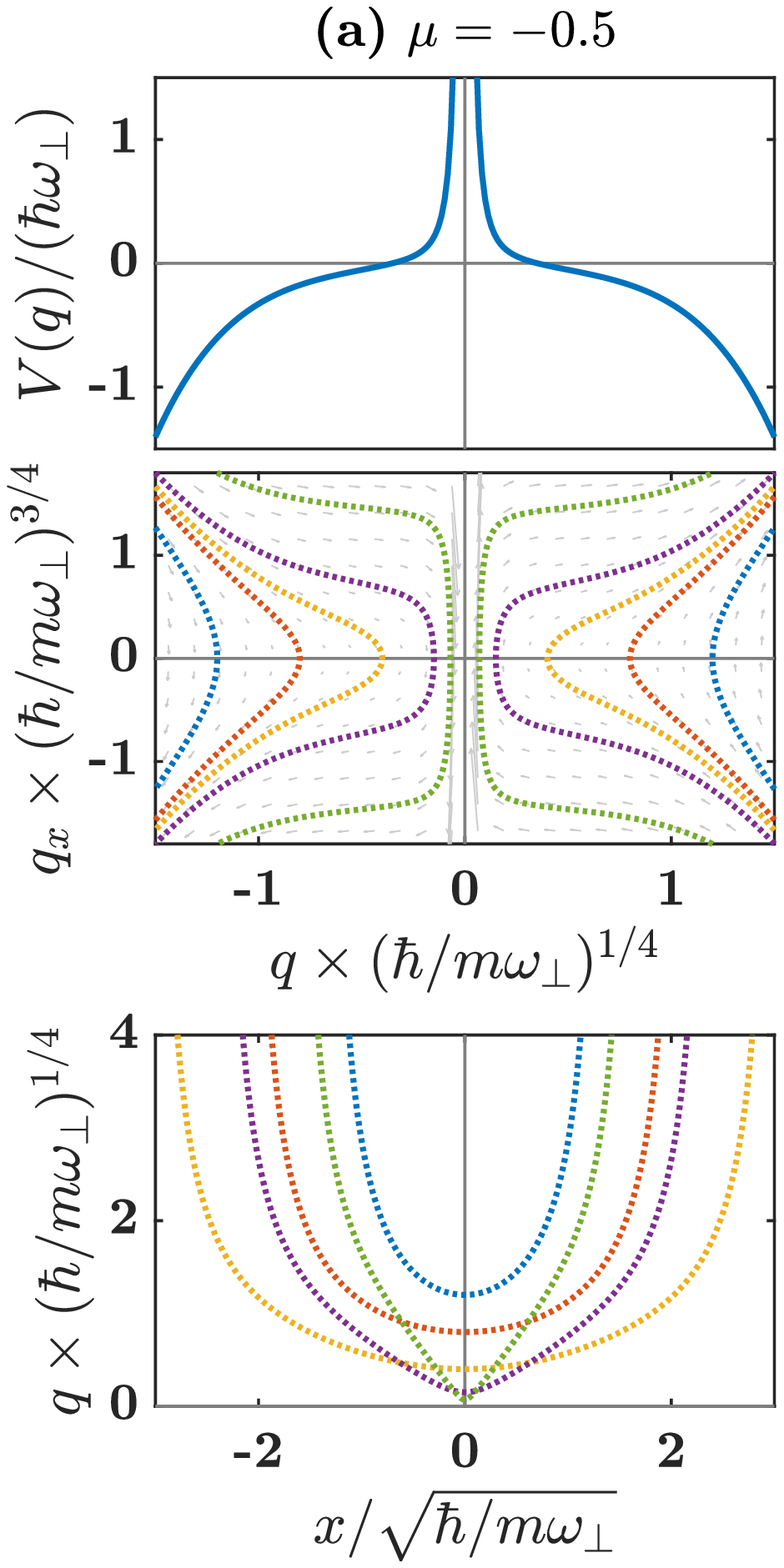}~~~
\includegraphics[height=9.0cm]{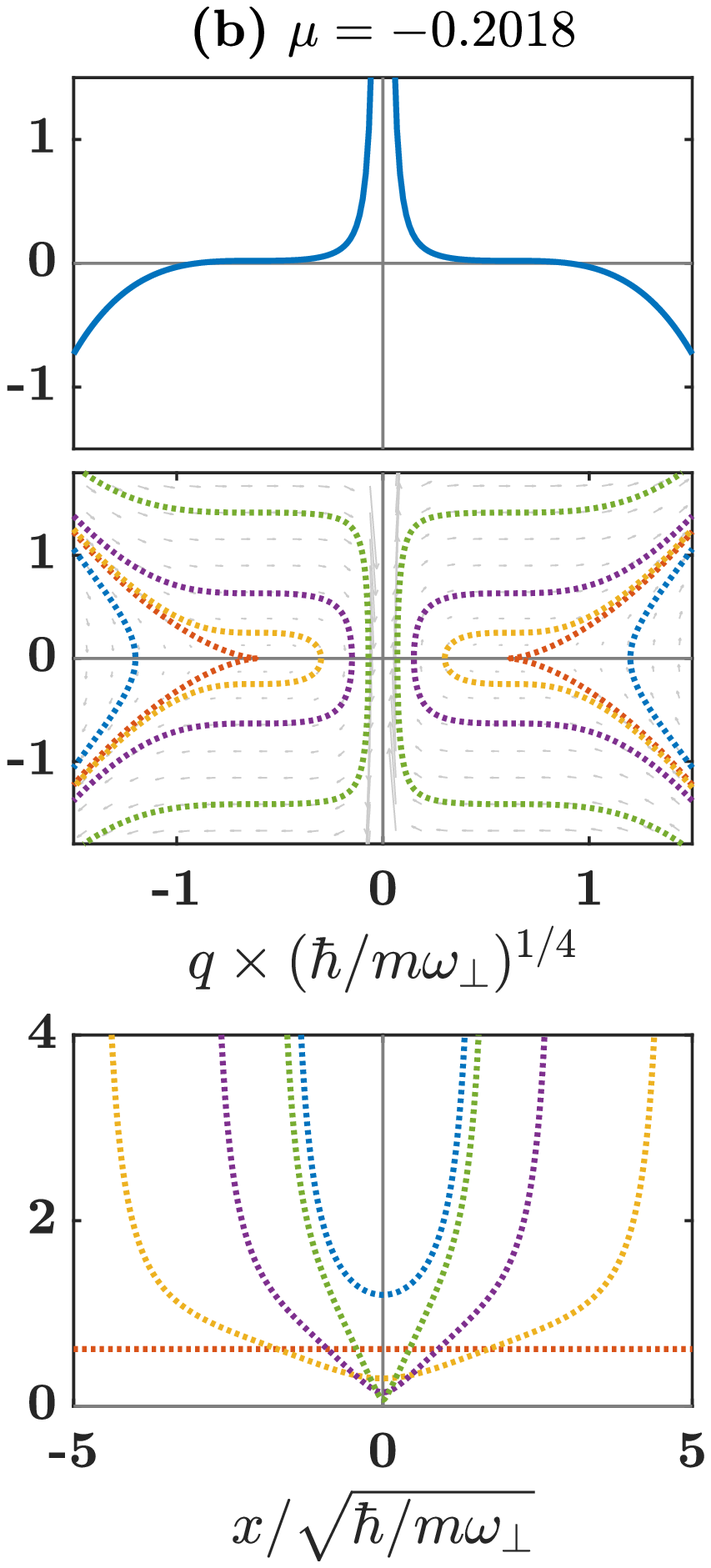}~~~
\includegraphics[height=9.0cm]{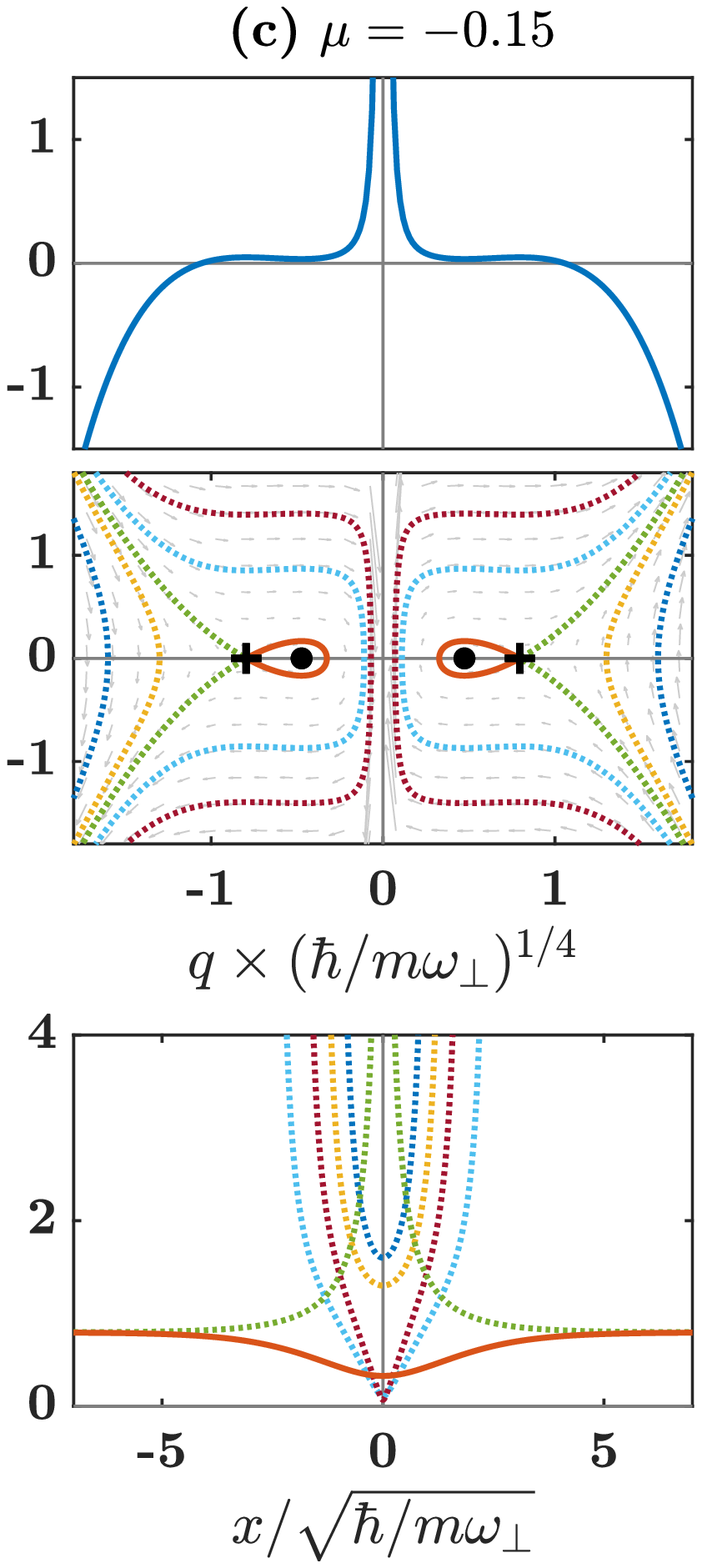}~~~
\includegraphics[height=9.0cm]{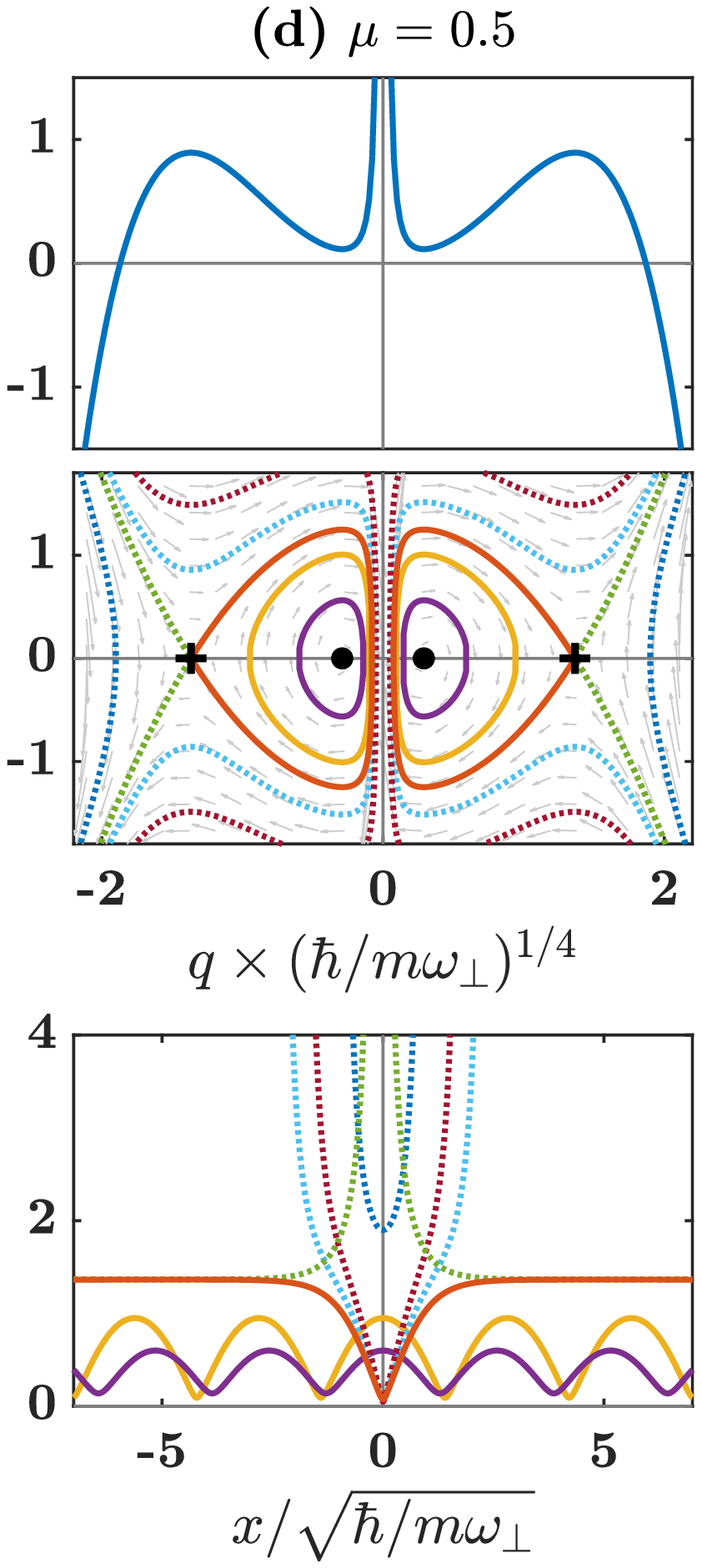}
\caption{The potential $V(q)$ (top panels) and respective phase planes (middle panels) 
and orbits (bottom panels) for $C_1=0.1$ and $\delta=1$.
Same layout and notation as in Fig.~\ref{mns}.
For (a) $\mu=-0.5$ and (b) $\mu=-0.2018$, all trajectories are unbounded. 
For values of $\mu$ larger than case (b) ($\mu=-0.2018$) there exist
bounded periodic and homoclinic orbits.
For example, for (c) $\mu=-0.15$ and (d) $\mu=0.5$ there exist homoclinic orbits 
corresponding, respectively, to a shallow and a deep (almost black) gray soliton
solution; see the relevant solid red curves in each case.
}
\label{mnu2} 
\end{figure*} 

\section{Conclusions and perspectives}
\label{conclusions} 

We have investigated the existence and stability properties of 
single- and multi-solitonic configurations in the presence of quantum 
fluctuations as captured by the LHY contribution both in free space 
and under the influence of a harmonic trap in one spatial dimension. 
Our analysis covers the regime from the quantum droplet (low density) to 
the TF (large density) limit. 
The hydrodynamic formulation of the eGPE describing a homonuclear 
symmetric bosonic mixture is utilized delivering insights on the 
existence of localized solutions of the underlying dynamical system. 

In free space, the respective phase-plane analysis 
reveals a variety of localized waveforms such as black 
and gray solitons, kink-type structures, as well as 
bright droplets, and the most recently found bubbles 
{(alias dark quantum droplets).}
These are identified in the relevant phase portraits as either 
homoclinic or heteroclinic orbits and analytical solutions for the 
kinks, the bright droplets, and the dark solitons are discussed. 
It should be emphasized that kink, bright quantum droplet, and bubble 
configurations arise exclusively due to the presence of the LHY term.  

For harmonically trapped configurations, 
the so-called Landau dynamics approach which treats the soliton energy 
as an adiabatic invariant, is employed to obtain a generalized, 
LHY-dependent, internal oscillation frequency of the dark soliton. 
This altered frequency bears the imprint  of 
quantum fluctuations, and can be traced back to the already modified local 
speed of sound, which can be used to diagnose beyond mean-field effects. 

The validity of the aforementioned theory is compared with the 
numerical evaluation of the corresponding eGPE excitation spectrum.  
In particular, the stability analysis unveils that single dark solitons 
are stable configurations upon chemical potential and trap variations 
for different strengths of the LHY interaction parameter. 
Importantly, the interplay of the LHY (dominating for small chemical potentials) 
and the cubic nonlinearity (prevailing for large chemical potentials) leads 
to the existence of a turning point as the number of particles is increased. 
This turning point is associated with the structural deformation of the 
configuration from a Hermite-Gauss linear state to a TF-like dark soliton solution,
and depends on both the LHY and the trap strength, as well as the soliton number.
As such, the presence of the LHY term alters the structure of the dark soliton spectrum 
as compared to the mean-field outcome providing another imprint of quantum fluctuations. 
Furthermore, spectral stability is retained for different chemical potentials 
independently of the trap strength. 
Multisoliton solutions feature a similar phenomenology as the single
dark soliton ones, but with their respective 
turning points shifted towards more positive chemical potentials for 
increasing number of solitons in the complexes. 
Interestingly, multisoliton configurations experience parametric windows 
of oscillatory instabilities displaying a gradually amplified out-of-phase 
motion of the individual entities leading to a periodic overall breathing 
of the entire entity.

There is a plethora of future research directions which can be pursued 
based on our findings.  
For instance, a direct extension would be to study in further detail 
the involvement and influence of dark-droplet states in more
practical applications as, e.g., in scattering problems 
since they are expected to act as material barriers/absorbers. 
A further understanding of the 
difference between the analytical prediction 
of the dark soliton oscillation frequency compared to the eGPE numerical 
prediction might be fruitful using other perturbation/analytical schemes. 
It still remains an open question whether the presence of the trap impacts 
the Bogoliubov modes and thus the LHY term or to what extent higher-order 
quantum corrections will play a crucial role as argued in 
Ref~\cite{englezos2023correlated}; see also the review of Ref.~\cite{mistakidis2022cold}. 
Also, it is particularly relevant to extend the present considerations 
to higher-dimensional settings, e.g., for configurations
bearing solitonic stripes where transverse excitations can play a 
crucial role and induce further instabilities. 
Certainly, the investigation of soliton and vortex structures
and interactions thereof in higher 
dimensions is of interest. 
Moreover, the generalization of our results to heteronuclear mixtures 
where more complex nonlinear structures such as dark-bright and dark-antidark 
solitons can be formed is another intriguing aspect. 

\acknowledgements

The work of GNK was supported by the Hellenic Foundation for Research 
and Innovation (HFRI) under the HFRI PhD Fellowship grant (Fellowship No.~5860). 
This material is based upon work supported by the U.S.\ National Science 
Foundation under the awards PHY-2110038 (RCG) and PHY-2110030 and DMS-2204702 (PGK).
SIM gratefully acknowledges financial support
from the NSF through a grant for ITAMP at Harvard University.
PGK gratefully acknowledges iterations with Mithun Thudiyangal at the early 
stages of the project.

\appendix
\section{Traveling configurations in free space}
\label{sec:append}

To complement the study of non-traveling configurations for $C_1=0$ 
in Sec.~\ref{sec:C1=0}, here we turn our attention to the case of
traveling configurations corresponding to $C_1 \neq 0$
(still in free space, i.e.,~$V_{\rm tr} = 0$). 
In this case, the fixed point at $q=0$ disappears due to the emergent 
singularity: in particular, $V(q)$ diverges in the 
vicinity of $q=0$; see the top panels in Fig.~\ref{mnu2}.
In this parameter regime, the dynamical system can have zero, two, or four 
fixed points, depending on the values of $\mu$, $\delta$ and $C_1$. 

Since analytical solutions are not 
straightforward to obtain, we will proceed by providing 
representative examples corresponding to different values of $\mu$, for 
fixed $\delta=1$ and $C_1=0.1$. 
We have checked that other choices of the aforementioned parameters lead 
to qualitatively similar results. 
For these parameter values, we find that in the interval 
$\mu \lesssim -0.2018$,
all trajectories are unbounded for $|x| \rightarrow \infty$;  
a pertinent example is given in Fig.~\ref{mnu2}(a) for $\mu=-0.5$ while the 
threshold case for $\mu =-0.2018$ is depicted in Fig.~\ref{mnu2}(b). 
On the other hand, for larger values of $\mu$, see for instance
panels (c) and (d) (corresponding, respectively, to $\mu=-0.2$ and $\mu=0.5$),
center and saddle fixed points on either side of $q=0$ emerge.

Importantly, in these cases, such an arrangement of fixed points allows 
for the appearance of homoclinic orbits, with the absolute maximum of 
$q(x)$ occurring closer to the origin as $\mu$ is increased. The relevant 
forms of $q(x)$ resemble gray solitons of the usual defocusing NLS equation, 
which are characterized by a nonzero density minimum.
Accordingly, the gray soliton corresponding to 
the homoclinic orbit for $\mu=-0.15$ depicted in panel (c) of 
Fig.~\ref{mnu2} is a shallow one, while the soliton corresponding 
to $\mu=0.5$, depicted in panel (d) of Fig.~\ref{mnu2}, is almost black.
A detailed study of these gray solitons, including their stability, 
falls outside of the scope of this work and will be presented elsewhere.


\begin{thebibliography}{85}%
\makeatletter
\providecommand \@ifxundefined [1]{%
 \@ifx{#1\undefined}
}%
\providecommand \@ifnum [1]{%
 \ifnum #1\expandafter \@firstoftwo
 \else \expandafter \@secondoftwo
 \fi
}%
\providecommand \@ifx [1]{%
 \ifx #1\expandafter \@firstoftwo
 \else \expandafter \@secondoftwo
 \fi
}%
\providecommand \natexlab [1]{#1}%
\providecommand \enquote  [1]{``#1''}%
\providecommand \bibnamefont  [1]{#1}%
\providecommand \bibfnamefont [1]{#1}%
\providecommand \citenamefont [1]{#1}%
\providecommand \href@noop [0]{\@secondoftwo}%
\providecommand \href [0]{\begingroup \@sanitize@url \@href}%
\providecommand \@href[1]{\@@startlink{#1}\@@href}%
\providecommand \@@href[1]{\endgroup#1\@@endlink}%
\providecommand \@sanitize@url [0]{\catcode `\\12\catcode `\$12\catcode
  `\&12\catcode `\#12\catcode `\^12\catcode `\_12\catcode `\%12\relax}%
\providecommand \@@startlink[1]{}%
\providecommand \@@endlink[0]{}%
\providecommand \url  [0]{\begingroup\@sanitize@url \@url }%
\providecommand \@url [1]{\endgroup\@href {#1}{\urlprefix }}%
\providecommand \urlprefix  [0]{URL }%
\providecommand \Eprint [0]{\href }%
\providecommand \doibase [0]{http://dx.doi.org/}%
\providecommand \selectlanguage [0]{\@gobble}%
\providecommand \bibinfo  [0]{\@secondoftwo}%
\providecommand \bibfield  [0]{\@secondoftwo}%
\providecommand \translation [1]{[#1]}%
\providecommand \BibitemOpen [0]{}%
\providecommand \bibitemStop [0]{}%
\providecommand \bibitemNoStop [0]{.\EOS\space}%
\providecommand \EOS [0]{\spacefactor3000\relax}%
\providecommand \BibitemShut  [1]{\csname bibitem#1\endcsname}%
\let\auto@bib@innerbib\@empty
\bibitem [{\citenamefont {Petrov}(2015)}]{petrov2015quantum}%
  \BibitemOpen
  \bibfield  {author} {\bibinfo {author} {\bibfnamefont {D.~S.}\ \bibnamefont
  {Petrov}},\ }\href@noop {} {\bibfield  {journal} {\bibinfo  {journal} {Phys.
  Rev. Lett.}\ }\textbf {\bibinfo {volume} {115}},\ \bibinfo {pages} {155302}
  (\bibinfo {year} {2015})}\BibitemShut {NoStop}%
\bibitem [{\citenamefont {Luo}\ \emph {et~al.}(2021)\citenamefont {Luo},
  \citenamefont {Pang}, \citenamefont {Liu}, \citenamefont {Li},\ and\
  \citenamefont {Malomed}}]{luo2021new}%
  \BibitemOpen
  \bibfield  {author} {\bibinfo {author} {\bibfnamefont {Z.-H.}\ \bibnamefont
  {Luo}}, \bibinfo {author} {\bibfnamefont {W.}~\bibnamefont {Pang}}, \bibinfo
  {author} {\bibfnamefont {B.}~\bibnamefont {Liu}}, \bibinfo {author}
  {\bibfnamefont {Y.-Y.}\ \bibnamefont {Li}}, \ and\ \bibinfo {author}
  {\bibfnamefont {B.~A.}\ \bibnamefont {Malomed}},\ }\href@noop {} {\bibfield
  {journal} {\bibinfo  {journal} {Frontiers of Physics}\ }\textbf {\bibinfo
  {volume} {16}},\ \bibinfo {pages} {1} (\bibinfo {year} {2021})}\BibitemShut
  {NoStop}%
\bibitem [{\citenamefont {Malomed}(2021)}]{malomed2020family}%
  \BibitemOpen
  \bibfield  {author} {\bibinfo {author} {\bibfnamefont {B.~A.}\ \bibnamefont
  {Malomed}},\ }\href@noop {} {\bibfield  {journal} {\bibinfo  {journal}
  {Frontiers of Physics}\ }\textbf {\bibinfo {volume} {16}},\ \bibinfo {pages}
  {22504} (\bibinfo {year} {2021})}\BibitemShut {NoStop}%
\bibitem [{\citenamefont {B{\"o}ttcher}\ \emph {et~al.}(2020)\citenamefont
  {B{\"o}ttcher}, \citenamefont {Schmidt}, \citenamefont {Hertkorn},
  \citenamefont {N~g}, \citenamefont {Graham}, \citenamefont {Guo},
  \citenamefont {Langen},\ and\ \citenamefont {Pfau}}]{bottcher2020new}%
  \BibitemOpen
  \bibfield  {author} {\bibinfo {author} {\bibfnamefont {F.}~\bibnamefont
  {B{\"o}ttcher}}, \bibinfo {author} {\bibfnamefont {J.-N.}\ \bibnamefont
  {Schmidt}}, \bibinfo {author} {\bibfnamefont {J.}~\bibnamefont {Hertkorn}},
  \bibinfo {author} {\bibfnamefont {K.~S.~H.}\ \bibnamefont {N~g}}, \bibinfo
  {author} {\bibfnamefont {S.~D.}\ \bibnamefont {Graham}}, \bibinfo {author}
  {\bibfnamefont {M.}~\bibnamefont {Guo}}, \bibinfo {author} {\bibfnamefont
  {T.}~\bibnamefont {Langen}}, \ and\ \bibinfo {author} {\bibfnamefont
  {T.}~\bibnamefont {Pfau}},\ }\href@noop {} {\bibfield  {journal} {\bibinfo
  {journal} {Rep. Progr. Phys.}\ }\textbf {\bibinfo {volume} {84}},\ \bibinfo
  {pages} {012403} (\bibinfo {year} {2020})}\BibitemShut {NoStop}%
\bibitem [{\citenamefont {Lee}\ \emph {et~al.}(1957)\citenamefont {Lee},
  \citenamefont {Huang},\ and\ \citenamefont {Yang}}]{lee1957eigenvalues}%
  \BibitemOpen
  \bibfield  {author} {\bibinfo {author} {\bibfnamefont {T.~D.}\ \bibnamefont
  {Lee}}, \bibinfo {author} {\bibfnamefont {K.}~\bibnamefont {Huang}}, \ and\
  \bibinfo {author} {\bibfnamefont {C.~N.}\ \bibnamefont {Yang}},\ }\href@noop
  {} {\bibfield  {journal} {\bibinfo  {journal} {Phys. Rev.}\ }\textbf
  {\bibinfo {volume} {106}},\ \bibinfo {pages} {1135} (\bibinfo {year}
  {1957})}\BibitemShut {NoStop}%
\bibitem [{\citenamefont {Chomaz}\ \emph {et~al.}(2023)\citenamefont {Chomaz},
  \citenamefont {Ferrier-Barbut}, \citenamefont {Ferlaino}, \citenamefont
  {Laburthe-Tolra}, \citenamefont {Lev},\ and\ \citenamefont
  {Pfau}}]{chomaz2022dipolar}%
  \BibitemOpen
  \bibfield  {author} {\bibinfo {author} {\bibfnamefont {L.}~\bibnamefont
  {Chomaz}}, \bibinfo {author} {\bibfnamefont {I.}~\bibnamefont
  {Ferrier-Barbut}}, \bibinfo {author} {\bibfnamefont {F.}~\bibnamefont
  {Ferlaino}}, \bibinfo {author} {\bibfnamefont {B.}~\bibnamefont
  {Laburthe-Tolra}}, \bibinfo {author} {\bibfnamefont {B.~L.}\ \bibnamefont
  {Lev}}, \ and\ \bibinfo {author} {\bibfnamefont {T.}~\bibnamefont {Pfau}},\
  }\href@noop {} {\bibfield  {journal} {\bibinfo  {journal} {Rep. Prog. Phys.}\
  }\textbf {\bibinfo {volume} {86}},\ \bibinfo {pages} {026401} (\bibinfo
  {year} {2023})}\BibitemShut {NoStop}%
\bibitem [{\citenamefont {Cheiney}\ \emph {et~al.}(2018)\citenamefont
  {Cheiney}, \citenamefont {Cabrera}, \citenamefont {Sanz}, \citenamefont
  {Naylor}, \citenamefont {Tanzi},\ and\ \citenamefont
  {Tarruell}}]{cheiney2018bright}%
  \BibitemOpen
  \bibfield  {author} {\bibinfo {author} {\bibfnamefont {P.}~\bibnamefont
  {Cheiney}}, \bibinfo {author} {\bibfnamefont {C.~R.}\ \bibnamefont
  {Cabrera}}, \bibinfo {author} {\bibfnamefont {J.}~\bibnamefont {Sanz}},
  \bibinfo {author} {\bibfnamefont {B.}~\bibnamefont {Naylor}}, \bibinfo
  {author} {\bibfnamefont {L.}~\bibnamefont {Tanzi}}, \ and\ \bibinfo {author}
  {\bibfnamefont {L.}~\bibnamefont {Tarruell}},\ }\href@noop {} {\bibfield
  {journal} {\bibinfo  {journal} {Phys. Rev. Lett.}\ }\textbf {\bibinfo
  {volume} {120}},\ \bibinfo {pages} {135301} (\bibinfo {year}
  {2018})}\BibitemShut {NoStop}%
\bibitem [{\citenamefont {Semeghini}\ \emph {et~al.}(2018)\citenamefont
  {Semeghini}, \citenamefont {Ferioli}, \citenamefont {Masi}, \citenamefont
  {Mazzinghi}, \citenamefont {Wolswijk}, \citenamefont {Minardi}, \citenamefont
  {Modugno}, \citenamefont {Modugno}, \citenamefont {Inguscio},\ and\
  \citenamefont {Fattori}}]{semeghini2018self}%
  \BibitemOpen
  \bibfield  {author} {\bibinfo {author} {\bibfnamefont {G.}~\bibnamefont
  {Semeghini}}, \bibinfo {author} {\bibfnamefont {G.}~\bibnamefont {Ferioli}},
  \bibinfo {author} {\bibfnamefont {L.}~\bibnamefont {Masi}}, \bibinfo {author}
  {\bibfnamefont {C.}~\bibnamefont {Mazzinghi}}, \bibinfo {author}
  {\bibfnamefont {L.}~\bibnamefont {Wolswijk}}, \bibinfo {author}
  {\bibfnamefont {F.}~\bibnamefont {Minardi}}, \bibinfo {author} {\bibfnamefont
  {M.}~\bibnamefont {Modugno}}, \bibinfo {author} {\bibfnamefont
  {G.}~\bibnamefont {Modugno}}, \bibinfo {author} {\bibfnamefont
  {M.}~\bibnamefont {Inguscio}}, \ and\ \bibinfo {author} {\bibfnamefont
  {M.}~\bibnamefont {Fattori}},\ }\href@noop {} {\bibfield  {journal} {\bibinfo
   {journal} {Phys. Rev. Lett.}\ }\textbf {\bibinfo {volume} {120}},\ \bibinfo
  {pages} {235301} (\bibinfo {year} {2018})}\BibitemShut {NoStop}%
\bibitem [{\citenamefont {Cabrera}\ \emph {et~al.}(2018)\citenamefont
  {Cabrera}, \citenamefont {Tanzi}, \citenamefont {Sanz}, \citenamefont
  {Naylor}, \citenamefont {Thomas}, \citenamefont {Cheiney},\ and\
  \citenamefont {Tarruell}}]{cabrera2018quantum}%
  \BibitemOpen
  \bibfield  {author} {\bibinfo {author} {\bibfnamefont {C.~R.}\ \bibnamefont
  {Cabrera}}, \bibinfo {author} {\bibfnamefont {L.}~\bibnamefont {Tanzi}},
  \bibinfo {author} {\bibfnamefont {J.}~\bibnamefont {Sanz}}, \bibinfo {author}
  {\bibfnamefont {B.}~\bibnamefont {Naylor}}, \bibinfo {author} {\bibfnamefont
  {P.}~\bibnamefont {Thomas}}, \bibinfo {author} {\bibfnamefont
  {P.}~\bibnamefont {Cheiney}}, \ and\ \bibinfo {author} {\bibfnamefont
  {L.}~\bibnamefont {Tarruell}},\ }\href@noop {} {\bibfield  {journal}
  {\bibinfo  {journal} {Science}\ }\textbf {\bibinfo {volume} {359}},\ \bibinfo
  {pages} {301} (\bibinfo {year} {2018})}\BibitemShut {NoStop}%
\bibitem [{\citenamefont {D'Errico}\ \emph {et~al.}(2019)\citenamefont
  {D'Errico}, \citenamefont {Burchianti}, \citenamefont {Prevedelli},
  \citenamefont {Salasnich}, \citenamefont {Ancilotto}, \citenamefont
  {Modugno}, \citenamefont {Minardi},\ and\ \citenamefont
  {Fort}}]{d2019observation}%
  \BibitemOpen
  \bibfield  {author} {\bibinfo {author} {\bibfnamefont {C.}~\bibnamefont
  {D'Errico}}, \bibinfo {author} {\bibfnamefont {A.}~\bibnamefont
  {Burchianti}}, \bibinfo {author} {\bibfnamefont {M.}~\bibnamefont
  {Prevedelli}}, \bibinfo {author} {\bibfnamefont {L.}~\bibnamefont
  {Salasnich}}, \bibinfo {author} {\bibfnamefont {F.}~\bibnamefont
  {Ancilotto}}, \bibinfo {author} {\bibfnamefont {M.}~\bibnamefont {Modugno}},
  \bibinfo {author} {\bibfnamefont {F.}~\bibnamefont {Minardi}}, \ and\
  \bibinfo {author} {\bibfnamefont {C.}~\bibnamefont {Fort}},\ }\href@noop {}
  {\bibfield  {journal} {\bibinfo  {journal} {Phys. Rev. Research}\ }\textbf
  {\bibinfo {volume} {1}},\ \bibinfo {pages} {033155} (\bibinfo {year}
  {2019})}\BibitemShut {NoStop}%
\bibitem [{\citenamefont {Burchianti}\ \emph {et~al.}(2020)\citenamefont
  {Burchianti}, \citenamefont {D'Errico}, \citenamefont {Prevedelli},
  \citenamefont {Salasnich}, \citenamefont {Ancilotto}, \citenamefont
  {Modugno}, \citenamefont {Minardi},\ and\ \citenamefont
  {Fort}}]{burchianti2020dual}%
  \BibitemOpen
  \bibfield  {author} {\bibinfo {author} {\bibfnamefont {A.}~\bibnamefont
  {Burchianti}}, \bibinfo {author} {\bibfnamefont {C.}~\bibnamefont
  {D'Errico}}, \bibinfo {author} {\bibfnamefont {M.}~\bibnamefont
  {Prevedelli}}, \bibinfo {author} {\bibfnamefont {L.}~\bibnamefont
  {Salasnich}}, \bibinfo {author} {\bibfnamefont {F.}~\bibnamefont
  {Ancilotto}}, \bibinfo {author} {\bibfnamefont {M.}~\bibnamefont {Modugno}},
  \bibinfo {author} {\bibfnamefont {F.}~\bibnamefont {Minardi}}, \ and\
  \bibinfo {author} {\bibfnamefont {C.}~\bibnamefont {Fort}},\ }\href@noop {}
  {\bibfield  {journal} {\bibinfo  {journal} {Cond. Matt.}\ }\textbf {\bibinfo
  {volume} {5}},\ \bibinfo {pages} {21} (\bibinfo {year} {2020})}\BibitemShut
  {NoStop}%
\bibitem [{\citenamefont {Cui}(2018)}]{cui2018spin}%
  \BibitemOpen
  \bibfield  {author} {\bibinfo {author} {\bibfnamefont {X.}~\bibnamefont
  {Cui}},\ }\href@noop {} {\bibfield  {journal} {\bibinfo  {journal} {Phys.
  Rev. A}\ }\textbf {\bibinfo {volume} {98}},\ \bibinfo {pages} {023630}
  (\bibinfo {year} {2018})}\BibitemShut {NoStop}%
\bibitem [{\citenamefont {Rakshit}\ \emph {et~al.}(2019)\citenamefont
  {Rakshit}, \citenamefont {Karpiuk}, \citenamefont {Brewczyk},\ and\
  \citenamefont {Gajda}}]{rakshit2019quantum}%
  \BibitemOpen
  \bibfield  {author} {\bibinfo {author} {\bibfnamefont {D.}~\bibnamefont
  {Rakshit}}, \bibinfo {author} {\bibfnamefont {T.}~\bibnamefont {Karpiuk}},
  \bibinfo {author} {\bibfnamefont {M.}~\bibnamefont {Brewczyk}}, \ and\
  \bibinfo {author} {\bibfnamefont {M.}~\bibnamefont {Gajda}},\ }\href@noop {}
  {\bibfield  {journal} {\bibinfo  {journal} {SciPost Phys.}\ }\textbf
  {\bibinfo {volume} {6}},\ \bibinfo {pages} {079} (\bibinfo {year}
  {2019})}\BibitemShut {NoStop}%
\bibitem [{\citenamefont {Sekino}\ and\ \citenamefont
  {Nishida}(2018)}]{sekino2018quantum}%
  \BibitemOpen
  \bibfield  {author} {\bibinfo {author} {\bibfnamefont {Y.}~\bibnamefont
  {Sekino}}\ and\ \bibinfo {author} {\bibfnamefont {Y.}~\bibnamefont
  {Nishida}},\ }\href@noop {} {\bibfield  {journal} {\bibinfo  {journal} {Phys.
  Rev. A}\ }\textbf {\bibinfo {volume} {97}},\ \bibinfo {pages} {011602}
  (\bibinfo {year} {2018})}\BibitemShut {NoStop}%
\bibitem [{\citenamefont {Morera}\ \emph {et~al.}(2021)\citenamefont {Morera},
  \citenamefont {Juli{\'a}-D{\'\i}az},\ and\ \citenamefont
  {Valiente}}]{morera2021quantum}%
  \BibitemOpen
  \bibfield  {author} {\bibinfo {author} {\bibfnamefont {I.}~\bibnamefont
  {Morera}}, \bibinfo {author} {\bibfnamefont {B.}~\bibnamefont
  {Juli{\'a}-D{\'\i}az}}, \ and\ \bibinfo {author} {\bibfnamefont
  {M.}~\bibnamefont {Valiente}},\ }\href@noop {} {\bibfield  {journal}
  {\bibinfo  {journal} {arXiv:2103.16499}\ } (\bibinfo {year} {2021})},\
  \bibinfo {note} {{Q}uantum liquids and droplets with low-energy interactions
  in one dimension}\BibitemShut {NoStop}%
\bibitem [{\citenamefont {Xu}\ \emph {et~al.}(2022)\citenamefont {Xu},
  \citenamefont {Lei}, \citenamefont {Du}, \citenamefont {Zhao}, \citenamefont
  {Hua},\ and\ \citenamefont {Zeng}}]{xu2022three}%
  \BibitemOpen
  \bibfield  {author} {\bibinfo {author} {\bibfnamefont {S.-L.}\ \bibnamefont
  {Xu}}, \bibinfo {author} {\bibfnamefont {Y.-B.}\ \bibnamefont {Lei}},
  \bibinfo {author} {\bibfnamefont {J.-T.}\ \bibnamefont {Du}}, \bibinfo
  {author} {\bibfnamefont {Y.}~\bibnamefont {Zhao}}, \bibinfo {author}
  {\bibfnamefont {R.}~\bibnamefont {Hua}}, \ and\ \bibinfo {author}
  {\bibfnamefont {J.-H.}\ \bibnamefont {Zeng}},\ }\href@noop {} {\bibfield
  {journal} {\bibinfo  {journal} {Chaos, Solitons \& Fractals}\ }\textbf
  {\bibinfo {volume} {164}},\ \bibinfo {pages} {112665} (\bibinfo {year}
  {2022})}\BibitemShut {NoStop}%
\bibitem [{\citenamefont {Zhang}\ \emph {et~al.}(2019)\citenamefont {Zhang},
  \citenamefont {Xu}, \citenamefont {Zheng}, \citenamefont {Chen},
  \citenamefont {Liu}, \citenamefont {Huang}, \citenamefont {Malomed},\ and\
  \citenamefont {Li}}]{malomedOL}%
  \BibitemOpen
  \bibfield  {author} {\bibinfo {author} {\bibfnamefont {X.}~\bibnamefont
  {Zhang}}, \bibinfo {author} {\bibfnamefont {X.}~\bibnamefont {Xu}}, \bibinfo
  {author} {\bibfnamefont {Y.}~\bibnamefont {Zheng}}, \bibinfo {author}
  {\bibfnamefont {Z.}~\bibnamefont {Chen}}, \bibinfo {author} {\bibfnamefont
  {B.}~\bibnamefont {Liu}}, \bibinfo {author} {\bibfnamefont {C.}~\bibnamefont
  {Huang}}, \bibinfo {author} {\bibfnamefont {B.~A.}\ \bibnamefont {Malomed}},
  \ and\ \bibinfo {author} {\bibfnamefont {Y.}~\bibnamefont {Li}},\ }\href@noop
  {} {\bibfield  {journal} {\bibinfo  {journal} {Phys. Rev. Lett.}\ }\textbf
  {\bibinfo {volume} {123}},\ \bibinfo {pages} {133901} (\bibinfo {year}
  {2019})}\BibitemShut {NoStop}%
\bibitem [{\citenamefont {Gangwar}\ \emph {et~al.}(2022)\citenamefont
  {Gangwar}, \citenamefont {Ravisankar}, \citenamefont {Muruganandam},\ and\
  \citenamefont {Mishra}}]{gangwar2022dynamics}%
  \BibitemOpen
  \bibfield  {author} {\bibinfo {author} {\bibfnamefont {S.}~\bibnamefont
  {Gangwar}}, \bibinfo {author} {\bibfnamefont {R.}~\bibnamefont {Ravisankar}},
  \bibinfo {author} {\bibfnamefont {P.}~\bibnamefont {Muruganandam}}, \ and\
  \bibinfo {author} {\bibfnamefont {P.~K.}\ \bibnamefont {Mishra}},\
  }\href@noop {} {\bibfield  {journal} {\bibinfo  {journal} {Phys. Rev. A}\
  }\textbf {\bibinfo {volume} {106}},\ \bibinfo {pages} {063315} (\bibinfo
  {year} {2022})}\BibitemShut {NoStop}%
\bibitem [{\citenamefont {Li}\ \emph {et~al.}(2017)\citenamefont {Li},
  \citenamefont {Luo}, \citenamefont {Liu}, \citenamefont {Chen}, \citenamefont
  {Huang}, \citenamefont {Fu}, \citenamefont {Tan},\ and\ \citenamefont
  {Malomed}}]{li2017two}%
  \BibitemOpen
  \bibfield  {author} {\bibinfo {author} {\bibfnamefont {Y.}~\bibnamefont
  {Li}}, \bibinfo {author} {\bibfnamefont {Z.}~\bibnamefont {Luo}}, \bibinfo
  {author} {\bibfnamefont {Y.}~\bibnamefont {Liu}}, \bibinfo {author}
  {\bibfnamefont {Z.}~\bibnamefont {Chen}}, \bibinfo {author} {\bibfnamefont
  {C.}~\bibnamefont {Huang}}, \bibinfo {author} {\bibfnamefont
  {S.}~\bibnamefont {Fu}}, \bibinfo {author} {\bibfnamefont {H.}~\bibnamefont
  {Tan}}, \ and\ \bibinfo {author} {\bibfnamefont {B.~A.}\ \bibnamefont
  {Malomed}},\ }\href@noop {} {\bibfield  {journal} {\bibinfo  {journal} {New
  J. Phys.}\ }\textbf {\bibinfo {volume} {19}},\ \bibinfo {pages} {113043}
  (\bibinfo {year} {2017})}\BibitemShut {NoStop}%
\bibitem [{\citenamefont {Wilson}\ \emph {et~al.}(2018)\citenamefont {Wilson},
  \citenamefont {Westerberg}, \citenamefont {Valiente}, \citenamefont {Duncan},
  \citenamefont {Wright}, \citenamefont {{\"O}hberg},\ and\ \citenamefont
  {Faccio}}]{wilson2018observation}%
  \BibitemOpen
  \bibfield  {author} {\bibinfo {author} {\bibfnamefont {K.~E.}\ \bibnamefont
  {Wilson}}, \bibinfo {author} {\bibfnamefont {N.}~\bibnamefont {Westerberg}},
  \bibinfo {author} {\bibfnamefont {M.}~\bibnamefont {Valiente}}, \bibinfo
  {author} {\bibfnamefont {C.~W.}\ \bibnamefont {Duncan}}, \bibinfo {author}
  {\bibfnamefont {E.~M.}\ \bibnamefont {Wright}}, \bibinfo {author}
  {\bibfnamefont {P.}~\bibnamefont {{\"O}hberg}}, \ and\ \bibinfo {author}
  {\bibfnamefont {D.}~\bibnamefont {Faccio}},\ }\href@noop {} {\bibfield
  {journal} {\bibinfo  {journal} {Phys. Rev. Lett.}\ }\textbf {\bibinfo
  {volume} {121}},\ \bibinfo {pages} {133903} (\bibinfo {year}
  {2018})}\BibitemShut {NoStop}%
\bibitem [{\citenamefont {Ho{\l}yst}\ \emph {et~al.}(2013)\citenamefont
  {Ho{\l}yst}, \citenamefont {Litniewski}, \citenamefont {Jakubczyk},
  \citenamefont {Kolwas}, \citenamefont {Kolwas}, \citenamefont {Kowalski},
  \citenamefont {Migacz}, \citenamefont {Palesa},\ and\ \citenamefont
  {Zientara}}]{holyst2013evaporation}%
  \BibitemOpen
  \bibfield  {author} {\bibinfo {author} {\bibfnamefont {R.}~\bibnamefont
  {Ho{\l}yst}}, \bibinfo {author} {\bibfnamefont {M.}~\bibnamefont
  {Litniewski}}, \bibinfo {author} {\bibfnamefont {D.}~\bibnamefont
  {Jakubczyk}}, \bibinfo {author} {\bibfnamefont {K.}~\bibnamefont {Kolwas}},
  \bibinfo {author} {\bibfnamefont {M.}~\bibnamefont {Kolwas}}, \bibinfo
  {author} {\bibfnamefont {K.}~\bibnamefont {Kowalski}}, \bibinfo {author}
  {\bibfnamefont {S.}~\bibnamefont {Migacz}}, \bibinfo {author} {\bibfnamefont
  {S.}~\bibnamefont {Palesa}}, \ and\ \bibinfo {author} {\bibfnamefont
  {M.}~\bibnamefont {Zientara}},\ }\href@noop {} {\bibfield  {journal}
  {\bibinfo  {journal} {Rep. Progr. Phys.}\ }\textbf {\bibinfo {volume} {76}},\
  \bibinfo {pages} {034601} (\bibinfo {year} {2013})}\BibitemShut {NoStop}%
\bibitem [{\citenamefont {Feder}\ \emph {et~al.}(1966)\citenamefont {Feder},
  \citenamefont {Russell}, \citenamefont {Lothe},\ and\ \citenamefont
  {Pound}}]{feder1966homogeneous}%
  \BibitemOpen
  \bibfield  {author} {\bibinfo {author} {\bibfnamefont {J.}~\bibnamefont
  {Feder}}, \bibinfo {author} {\bibfnamefont {K.}~\bibnamefont {Russell}},
  \bibinfo {author} {\bibfnamefont {J.}~\bibnamefont {Lothe}}, \ and\ \bibinfo
  {author} {\bibfnamefont {G.}~\bibnamefont {Pound}},\ }\href@noop {}
  {\bibfield  {journal} {\bibinfo  {journal} {Advances in Physics}\ }\textbf
  {\bibinfo {volume} {15}},\ \bibinfo {pages} {111} (\bibinfo {year}
  {1966})}\BibitemShut {NoStop}%
\bibitem [{\citenamefont {Barranco}\ \emph {et~al.}(2006)\citenamefont
  {Barranco}, \citenamefont {Guardiola}, \citenamefont {Hern{\'a}ndez},
  \citenamefont {Mayol}, \citenamefont {Navarro},\ and\ \citenamefont
  {Pi}}]{barranco2006helium}%
  \BibitemOpen
  \bibfield  {author} {\bibinfo {author} {\bibfnamefont {M.}~\bibnamefont
  {Barranco}}, \bibinfo {author} {\bibfnamefont {R.}~\bibnamefont {Guardiola}},
  \bibinfo {author} {\bibfnamefont {S.}~\bibnamefont {Hern{\'a}ndez}}, \bibinfo
  {author} {\bibfnamefont {R.}~\bibnamefont {Mayol}}, \bibinfo {author}
  {\bibfnamefont {J.}~\bibnamefont {Navarro}}, \ and\ \bibinfo {author}
  {\bibfnamefont {M.}~\bibnamefont {Pi}},\ }\href@noop {} {\bibfield  {journal}
  {\bibinfo  {journal} {J. Low Temp. Phys.}\ }\textbf {\bibinfo {volume}
  {142}},\ \bibinfo {pages} {1} (\bibinfo {year} {2006})}\BibitemShut {NoStop}%
\bibitem [{\citenamefont {Toennies}\ and\ \citenamefont
  {Vilesov}(2004)}]{toennies2004superfluid}%
  \BibitemOpen
  \bibfield  {author} {\bibinfo {author} {\bibfnamefont {J.~P.}\ \bibnamefont
  {Toennies}}\ and\ \bibinfo {author} {\bibfnamefont {A.~F.}\ \bibnamefont
  {Vilesov}},\ }\href@noop {} {\bibfield  {journal} {\bibinfo  {journal}
  {Angewandte Chemie International Edition}\ }\textbf {\bibinfo {volume}
  {43}},\ \bibinfo {pages} {2622} (\bibinfo {year} {2004})}\BibitemShut
  {NoStop}%
\bibitem [{\citenamefont {Petrov}\ and\ \citenamefont
  {Astrakharchik}(2016)}]{petrov2016ultradilute}%
  \BibitemOpen
  \bibfield  {author} {\bibinfo {author} {\bibfnamefont {D.~S.}\ \bibnamefont
  {Petrov}}\ and\ \bibinfo {author} {\bibfnamefont {G.~E.}\ \bibnamefont
  {Astrakharchik}},\ }\href@noop {} {\bibfield  {journal} {\bibinfo  {journal}
  {Phys. Rev. Lett.}\ }\textbf {\bibinfo {volume} {117}},\ \bibinfo {pages}
  {100401} (\bibinfo {year} {2016})}\BibitemShut {NoStop}%
\bibitem [{\citenamefont {Ferioli}\ \emph {et~al.}(2019)\citenamefont
  {Ferioli}, \citenamefont {Semeghini}, \citenamefont {Masi}, \citenamefont
  {Giusti}, \citenamefont {Modugno}, \citenamefont {Inguscio}, \citenamefont
  {Gallem{\'\i}}, \citenamefont {Recati},\ and\ \citenamefont
  {Fattori}}]{ferioli2019collisions}%
  \BibitemOpen
  \bibfield  {author} {\bibinfo {author} {\bibfnamefont {G.}~\bibnamefont
  {Ferioli}}, \bibinfo {author} {\bibfnamefont {G.}~\bibnamefont {Semeghini}},
  \bibinfo {author} {\bibfnamefont {L.}~\bibnamefont {Masi}}, \bibinfo {author}
  {\bibfnamefont {G.}~\bibnamefont {Giusti}}, \bibinfo {author} {\bibfnamefont
  {G.}~\bibnamefont {Modugno}}, \bibinfo {author} {\bibfnamefont
  {M.}~\bibnamefont {Inguscio}}, \bibinfo {author} {\bibfnamefont
  {A.}~\bibnamefont {Gallem{\'\i}}}, \bibinfo {author} {\bibfnamefont
  {A.}~\bibnamefont {Recati}}, \ and\ \bibinfo {author} {\bibfnamefont
  {M.}~\bibnamefont {Fattori}},\ }\href@noop {} {\bibfield  {journal} {\bibinfo
   {journal} {Phys. Rev. Lett.}\ }\textbf {\bibinfo {volume} {122}},\ \bibinfo
  {pages} {090401} (\bibinfo {year} {2019})}\BibitemShut {NoStop}%
\bibitem [{\citenamefont {Astrakharchik}\ and\ \citenamefont
  {Malomed}(2018)}]{astrakharchik2018dynamics}%
  \BibitemOpen
  \bibfield  {author} {\bibinfo {author} {\bibfnamefont {G.~E.}\ \bibnamefont
  {Astrakharchik}}\ and\ \bibinfo {author} {\bibfnamefont {B.~A.}\ \bibnamefont
  {Malomed}},\ }\href@noop {} {\bibfield  {journal} {\bibinfo  {journal} {Phys.
  Rev. A}\ }\textbf {\bibinfo {volume} {98}},\ \bibinfo {pages} {013631}
  (\bibinfo {year} {2018})}\BibitemShut {NoStop}%
\bibitem [{\citenamefont {St{\"u}rmer}\ \emph {et~al.}(2021)\citenamefont
  {St{\"u}rmer}, \citenamefont {Tengstrand}, \citenamefont {Sachdeva},\ and\
  \citenamefont {Reimann}}]{sturmer2021breathing}%
  \BibitemOpen
  \bibfield  {author} {\bibinfo {author} {\bibfnamefont {P.}~\bibnamefont
  {St{\"u}rmer}}, \bibinfo {author} {\bibfnamefont {M.~N.}\ \bibnamefont
  {Tengstrand}}, \bibinfo {author} {\bibfnamefont {R.}~\bibnamefont
  {Sachdeva}}, \ and\ \bibinfo {author} {\bibfnamefont {S.~M.}\ \bibnamefont
  {Reimann}},\ }\href@noop {} {\bibfield  {journal} {\bibinfo  {journal} {Phys.
  Rev. A}\ }\textbf {\bibinfo {volume} {103}},\ \bibinfo {pages} {053302}
  (\bibinfo {year} {2021})}\BibitemShut {NoStop}%
\bibitem [{\citenamefont {Tylutki}\ \emph {et~al.}(2020)\citenamefont
  {Tylutki}, \citenamefont {Astrakharchik}, \citenamefont {Malomed},\ and\
  \citenamefont {Petrov}}]{tylutki2020collective}%
  \BibitemOpen
  \bibfield  {author} {\bibinfo {author} {\bibfnamefont {M.}~\bibnamefont
  {Tylutki}}, \bibinfo {author} {\bibfnamefont {G.~E.}\ \bibnamefont
  {Astrakharchik}}, \bibinfo {author} {\bibfnamefont {B.~A.}\ \bibnamefont
  {Malomed}}, \ and\ \bibinfo {author} {\bibfnamefont {D.~S.}\ \bibnamefont
  {Petrov}},\ }\href@noop {} {\bibfield  {journal} {\bibinfo  {journal} {Phys.
  Rev. A}\ }\textbf {\bibinfo {volume} {101}},\ \bibinfo {pages} {051601}
  (\bibinfo {year} {2020})}\BibitemShut {NoStop}%
\bibitem [{\citenamefont {Cappellaro}\ \emph {et~al.}(2018)\citenamefont
  {Cappellaro}, \citenamefont {Macr{\`\i}},\ and\ \citenamefont
  {Salasnich}}]{cappellaro2018collective}%
  \BibitemOpen
  \bibfield  {author} {\bibinfo {author} {\bibfnamefont {A.}~\bibnamefont
  {Cappellaro}}, \bibinfo {author} {\bibfnamefont {T.}~\bibnamefont
  {Macr{\`\i}}}, \ and\ \bibinfo {author} {\bibfnamefont {L.}~\bibnamefont
  {Salasnich}},\ }\href@noop {} {\bibfield  {journal} {\bibinfo  {journal}
  {Phys. Rev. A}\ }\textbf {\bibinfo {volume} {97}},\ \bibinfo {pages} {053623}
  (\bibinfo {year} {2018})}\BibitemShut {NoStop}%
\bibitem [{\citenamefont {Hu}\ and\ \citenamefont
  {Liu}(2020)}]{hu2020collective}%
  \BibitemOpen
  \bibfield  {author} {\bibinfo {author} {\bibfnamefont {H.}~\bibnamefont
  {Hu}}\ and\ \bibinfo {author} {\bibfnamefont {X.-J.}\ \bibnamefont {Liu}},\
  }\href@noop {} {\bibfield  {journal} {\bibinfo  {journal} {Phys. Rev. A}\
  }\textbf {\bibinfo {volume} {102}},\ \bibinfo {pages} {053303} (\bibinfo
  {year} {2020})}\BibitemShut {NoStop}%
\bibitem [{\citenamefont {Tengstrand}\ and\ \citenamefont
  {Reimann}(2022)}]{tengstrand2022droplet}%
  \BibitemOpen
  \bibfield  {author} {\bibinfo {author} {\bibfnamefont {M.~N.}\ \bibnamefont
  {Tengstrand}}\ and\ \bibinfo {author} {\bibfnamefont {S.~M.}\ \bibnamefont
  {Reimann}},\ }\href@noop {} {\bibfield  {journal} {\bibinfo  {journal} {Phys.
  Rev. A}\ }\textbf {\bibinfo {volume} {105}},\ \bibinfo {pages} {033319}
  (\bibinfo {year} {2022})}\BibitemShut {NoStop}%
\bibitem [{\citenamefont {Mithun}\ \emph {et~al.}(2020)\citenamefont {Mithun},
  \citenamefont {Maluckov}, \citenamefont {Kasamatsu}, \citenamefont
  {Malomed},\ and\ \citenamefont {Khare}}]{mithun2020modulational}%
  \BibitemOpen
  \bibfield  {author} {\bibinfo {author} {\bibfnamefont {T.}~\bibnamefont
  {Mithun}}, \bibinfo {author} {\bibfnamefont {A.}~\bibnamefont {Maluckov}},
  \bibinfo {author} {\bibfnamefont {K.}~\bibnamefont {Kasamatsu}}, \bibinfo
  {author} {\bibfnamefont {B.~A.}\ \bibnamefont {Malomed}}, \ and\ \bibinfo
  {author} {\bibfnamefont {A.}~\bibnamefont {Khare}},\ }\href@noop {}
  {\bibfield  {journal} {\bibinfo  {journal} {Symmetry}\ }\textbf {\bibinfo
  {volume} {12}},\ \bibinfo {pages} {174} (\bibinfo {year} {2020})}\BibitemShut
  {NoStop}%
\bibitem [{\citenamefont {Mithun}\ \emph {et~al.}(2021)\citenamefont {Mithun},
  \citenamefont {Mistakidis}, \citenamefont {Schmelcher},\ and\ \citenamefont
  {Kevrekidis}}]{mithun2021statistical}%
  \BibitemOpen
  \bibfield  {author} {\bibinfo {author} {\bibfnamefont {T.}~\bibnamefont
  {Mithun}}, \bibinfo {author} {\bibfnamefont {S.~I.}\ \bibnamefont
  {Mistakidis}}, \bibinfo {author} {\bibfnamefont {P.}~\bibnamefont
  {Schmelcher}}, \ and\ \bibinfo {author} {\bibfnamefont {P.~G.}\ \bibnamefont
  {Kevrekidis}},\ }\href@noop {} {\bibfield  {journal} {\bibinfo  {journal}
  {Phys. Rev. A}\ }\textbf {\bibinfo {volume} {104}},\ \bibinfo {pages}
  {033316} (\bibinfo {year} {2021})}\BibitemShut {NoStop}%
\bibitem [{\citenamefont {De~Rosi}\ \emph {et~al.}(2021)\citenamefont
  {De~Rosi}, \citenamefont {Astrakharchik},\ and\ \citenamefont
  {Massignan}}]{de2021thermal}%
  \BibitemOpen
  \bibfield  {author} {\bibinfo {author} {\bibfnamefont {G.}~\bibnamefont
  {De~Rosi}}, \bibinfo {author} {\bibfnamefont {G.~E.}\ \bibnamefont
  {Astrakharchik}}, \ and\ \bibinfo {author} {\bibfnamefont {P.}~\bibnamefont
  {Massignan}},\ }\href@noop {} {\bibfield  {journal} {\bibinfo  {journal}
  {Phys. Rev. A}\ }\textbf {\bibinfo {volume} {103}},\ \bibinfo {pages}
  {043316} (\bibinfo {year} {2021})}\BibitemShut {NoStop}%
\bibitem [{\citenamefont {Wang}\ \emph {et~al.}(2020)\citenamefont {Wang},
  \citenamefont {Hu},\ and\ \citenamefont {Liu}}]{Wang_2020}%
  \BibitemOpen
  \bibfield  {author} {\bibinfo {author} {\bibfnamefont {J.}~\bibnamefont
  {Wang}}, \bibinfo {author} {\bibfnamefont {H.}~\bibnamefont {Hu}}, \ and\
  \bibinfo {author} {\bibfnamefont {X.-J.}\ \bibnamefont {Liu}},\ }\href@noop
  {} {\bibfield  {journal} {\bibinfo  {journal} {New Journal of Physics}\
  }\textbf {\bibinfo {volume} {22}},\ \bibinfo {pages} {103044} (\bibinfo
  {year} {2020})}\BibitemShut {NoStop}%
\bibitem [{\citenamefont {Boudjem{\^a}a}(2021)}]{boudjemaa2021many}%
  \BibitemOpen
  \bibfield  {author} {\bibinfo {author} {\bibfnamefont {A.}~\bibnamefont
  {Boudjem{\^a}a}},\ }\href@noop {} {\bibfield  {journal} {\bibinfo  {journal}
  {Sci. Rep.}\ }\textbf {\bibinfo {volume} {11}},\ \bibinfo {pages} {1}
  (\bibinfo {year} {2021})}\BibitemShut {NoStop}%
\bibitem [{\citenamefont {Parisi}\ \emph {et~al.}(2019)\citenamefont {Parisi},
  \citenamefont {Astrakharchik},\ and\ \citenamefont
  {Giorgini}}]{parisi2019liquid}%
  \BibitemOpen
  \bibfield  {author} {\bibinfo {author} {\bibfnamefont {L.}~\bibnamefont
  {Parisi}}, \bibinfo {author} {\bibfnamefont {G.~E.}\ \bibnamefont
  {Astrakharchik}}, \ and\ \bibinfo {author} {\bibfnamefont {S.}~\bibnamefont
  {Giorgini}},\ }\href@noop {} {\bibfield  {journal} {\bibinfo  {journal}
  {Phys. Rev. Lett.}\ }\textbf {\bibinfo {volume} {122}},\ \bibinfo {pages}
  {105302} (\bibinfo {year} {2019})}\BibitemShut {NoStop}%
\bibitem [{\citenamefont {Parisi}\ and\ \citenamefont
  {Giorgini}(2020)}]{parisi2020quantum}%
  \BibitemOpen
  \bibfield  {author} {\bibinfo {author} {\bibfnamefont {L.}~\bibnamefont
  {Parisi}}\ and\ \bibinfo {author} {\bibfnamefont {S.}~\bibnamefont
  {Giorgini}},\ }\href@noop {} {\bibfield  {journal} {\bibinfo  {journal}
  {Phys. Rev. A}\ }\textbf {\bibinfo {volume} {102}},\ \bibinfo {pages}
  {023318} (\bibinfo {year} {2020})}\BibitemShut {NoStop}%
\bibitem [{\citenamefont {Mistakidis}\ \emph {et~al.}(2021)\citenamefont
  {Mistakidis}, \citenamefont {Mithun}, \citenamefont {Kevrekidis},
  \citenamefont {Sadeghpour},\ and\ \citenamefont
  {Schmelcher}}]{mistakidis2021formation}%
  \BibitemOpen
  \bibfield  {author} {\bibinfo {author} {\bibfnamefont {S.~I.}\ \bibnamefont
  {Mistakidis}}, \bibinfo {author} {\bibfnamefont {T.}~\bibnamefont {Mithun}},
  \bibinfo {author} {\bibfnamefont {P.~G.}\ \bibnamefont {Kevrekidis}},
  \bibinfo {author} {\bibfnamefont {H.~R.}\ \bibnamefont {Sadeghpour}}, \ and\
  \bibinfo {author} {\bibfnamefont {P.}~\bibnamefont {Schmelcher}},\
  }\href@noop {} {\bibfield  {journal} {\bibinfo  {journal} {Phys. Rev.
  Research}\ }\textbf {\bibinfo {volume} {3}},\ \bibinfo {pages} {043128}
  (\bibinfo {year} {2021})}\BibitemShut {NoStop}%
\bibitem [{\citenamefont {Ota}\ and\ \citenamefont
  {Astrakharchik}(2020)}]{ota2020beyond}%
  \BibitemOpen
  \bibfield  {author} {\bibinfo {author} {\bibfnamefont {M.}~\bibnamefont
  {Ota}}\ and\ \bibinfo {author} {\bibfnamefont {G.}~\bibnamefont
  {Astrakharchik}},\ }\href@noop {} {\bibfield  {journal} {\bibinfo  {journal}
  {SciPost Physics}\ }\textbf {\bibinfo {volume} {9}},\ \bibinfo {pages} {020}
  (\bibinfo {year} {2020})}\BibitemShut {NoStop}%
\bibitem [{\citenamefont {Mistakidis}\ \emph {et~al.}(2022)\citenamefont
  {Mistakidis}, \citenamefont {Volosniev}, \citenamefont {Barfknecht},
  \citenamefont {Fogarty}, \citenamefont {Busch}, \citenamefont {Foerster},
  \citenamefont {Schmelcher},\ and\ \citenamefont
  {Zinner}}]{mistakidis2022cold}%
  \BibitemOpen
  \bibfield  {author} {\bibinfo {author} {\bibfnamefont {S.~I.}\ \bibnamefont
  {Mistakidis}}, \bibinfo {author} {\bibfnamefont {A.~G.}\ \bibnamefont
  {Volosniev}}, \bibinfo {author} {\bibfnamefont {R.~E.}\ \bibnamefont
  {Barfknecht}}, \bibinfo {author} {\bibfnamefont {T.}~\bibnamefont {Fogarty}},
  \bibinfo {author} {\bibfnamefont {T.}~\bibnamefont {Busch}}, \bibinfo
  {author} {\bibfnamefont {A.}~\bibnamefont {Foerster}}, \bibinfo {author}
  {\bibfnamefont {P.}~\bibnamefont {Schmelcher}}, \ and\ \bibinfo {author}
  {\bibfnamefont {N.~T.}\ \bibnamefont {Zinner}},\ }\href@noop {} {\bibfield
  {journal} {\bibinfo  {journal} {arXiv:2202.11071}\ } (\bibinfo {year}
  {2022})},\ \bibinfo {note} {{C}old atoms in low dimensions--a laboratory for
  quantum dynamics}\BibitemShut {NoStop}%
\bibitem [{\citenamefont {Kevrekidis}\ \emph {et~al.}(2015)\citenamefont
  {Kevrekidis}, \citenamefont {Frantzeskakis},\ and\ \citenamefont
  {Carretero-Gonz{\'a}lez}}]{kevrekidis2015defocusing}%
  \BibitemOpen
  \bibfield  {author} {\bibinfo {author} {\bibfnamefont {P.~G.}\ \bibnamefont
  {Kevrekidis}}, \bibinfo {author} {\bibfnamefont {D.~J.}\ \bibnamefont
  {Frantzeskakis}}, \ and\ \bibinfo {author} {\bibfnamefont {R.}~\bibnamefont
  {Carretero-Gonz{\'a}lez}},\ }\href@noop {} {\emph {\bibinfo {title} {The
  defocusing nonlinear Schr{\"o}dinger equation: from dark solitons to vortices
  and vortex rings}}}\ (\bibinfo  {publisher} {SIAM},\ \bibinfo {year}
  {2015})\BibitemShut {NoStop}%
\bibitem [{\citenamefont {Edmonds}(2022)}]{edmonds2022dark}%
  \BibitemOpen
  \bibfield  {author} {\bibinfo {author} {\bibfnamefont {M.}~\bibnamefont
  {Edmonds}},\ }\href@noop {} {\bibfield  {journal} {\bibinfo  {journal}
  {arXiv:2209.00790}\ } (\bibinfo {year} {2022})},\ \bibinfo {note} {{D}ark
  quantum droplets in beyond-mean-field {B}ose-{E}instein condensate
  mixtures}\BibitemShut {NoStop}%
\bibitem [{\citenamefont {Kartashov}\ \emph {et~al.}(2022)\citenamefont
  {Kartashov}, \citenamefont {Lashkin}, \citenamefont {Modugno},\ and\
  \citenamefont {Torner}}]{kartashov2022spinor}%
  \BibitemOpen
  \bibfield  {author} {\bibinfo {author} {\bibfnamefont {Y.~V.}\ \bibnamefont
  {Kartashov}}, \bibinfo {author} {\bibfnamefont {V.~M.}\ \bibnamefont
  {Lashkin}}, \bibinfo {author} {\bibfnamefont {M.}~\bibnamefont {Modugno}}, \
  and\ \bibinfo {author} {\bibfnamefont {L.}~\bibnamefont {Torner}},\
  }\href@noop {} {\bibfield  {journal} {\bibinfo  {journal} {New J. Phys.}\
  }\textbf {\bibinfo {volume} {24}},\ \bibinfo {pages} {073012} (\bibinfo
  {year} {2022})}\BibitemShut {NoStop}%
\bibitem [{\citenamefont {Kartashov}\ \emph {et~al.}(2018)\citenamefont
  {Kartashov}, \citenamefont {Malomed}, \citenamefont {Tarruell},\ and\
  \citenamefont {Torner}}]{kartashov2018three}%
  \BibitemOpen
  \bibfield  {author} {\bibinfo {author} {\bibfnamefont {Y.~V.}\ \bibnamefont
  {Kartashov}}, \bibinfo {author} {\bibfnamefont {B.~A.}\ \bibnamefont
  {Malomed}}, \bibinfo {author} {\bibfnamefont {L.}~\bibnamefont {Tarruell}}, \
  and\ \bibinfo {author} {\bibfnamefont {L.}~\bibnamefont {Torner}},\
  }\href@noop {} {\bibfield  {journal} {\bibinfo  {journal} {Phys. Rev. A}\
  }\textbf {\bibinfo {volume} {98}},\ \bibinfo {pages} {013612} (\bibinfo
  {year} {2018})}\BibitemShut {NoStop}%
\bibitem [{\citenamefont {Li}\ \emph {et~al.}(2018)\citenamefont {Li},
  \citenamefont {Chen}, \citenamefont {Luo}, \citenamefont {Huang},
  \citenamefont {Tan}, \citenamefont {Pang},\ and\ \citenamefont
  {Malomed}}]{li2018two}%
  \BibitemOpen
  \bibfield  {author} {\bibinfo {author} {\bibfnamefont {Y.}~\bibnamefont
  {Li}}, \bibinfo {author} {\bibfnamefont {Z.}~\bibnamefont {Chen}}, \bibinfo
  {author} {\bibfnamefont {Z.}~\bibnamefont {Luo}}, \bibinfo {author}
  {\bibfnamefont {C.}~\bibnamefont {Huang}}, \bibinfo {author} {\bibfnamefont
  {H.}~\bibnamefont {Tan}}, \bibinfo {author} {\bibfnamefont {W.}~\bibnamefont
  {Pang}}, \ and\ \bibinfo {author} {\bibfnamefont {B.~A.}\ \bibnamefont
  {Malomed}},\ }\href@noop {} {\bibfield  {journal} {\bibinfo  {journal} {Phys.
  Rev. A}\ }\textbf {\bibinfo {volume} {98}},\ \bibinfo {pages} {063602}
  (\bibinfo {year} {2018})}\BibitemShut {NoStop}%
\bibitem [{\citenamefont {Tengstrand}\ \emph {et~al.}(2019)\citenamefont
  {Tengstrand}, \citenamefont {St{\"u}rmer}, \citenamefont {Karabulut},\ and\
  \citenamefont {Reimann}}]{tengstrand2019rotating}%
  \BibitemOpen
  \bibfield  {author} {\bibinfo {author} {\bibfnamefont {M.~N.}\ \bibnamefont
  {Tengstrand}}, \bibinfo {author} {\bibfnamefont {P.}~\bibnamefont
  {St{\"u}rmer}}, \bibinfo {author} {\bibfnamefont {E.}~\bibnamefont
  {Karabulut}}, \ and\ \bibinfo {author} {\bibfnamefont {S.~M.}\ \bibnamefont
  {Reimann}},\ }\href@noop {} {\bibfield  {journal} {\bibinfo  {journal} {Phys.
  Rev. Lett.}\ }\textbf {\bibinfo {volume} {123}},\ \bibinfo {pages} {160405}
  (\bibinfo {year} {2019})}\BibitemShut {NoStop}%
\bibitem [{\citenamefont {Examilioti}\ and\ \citenamefont
  {Kavoulakis}(2020)}]{examilioti2020ground}%
  \BibitemOpen
  \bibfield  {author} {\bibinfo {author} {\bibfnamefont {P.}~\bibnamefont
  {Examilioti}}\ and\ \bibinfo {author} {\bibfnamefont {G.~M.}\ \bibnamefont
  {Kavoulakis}},\ }\href@noop {} {\bibfield  {journal} {\bibinfo  {journal} {J.
  Phys. B: At. Mol. and Opt. Phys.}\ }\textbf {\bibinfo {volume} {53}},\
  \bibinfo {pages} {175301} (\bibinfo {year} {2020})}\BibitemShut {NoStop}%
\bibitem [{\citenamefont {Caldara}\ and\ \citenamefont
  {Ancilotto}(2022)}]{caldara2022vortices}%
  \BibitemOpen
  \bibfield  {author} {\bibinfo {author} {\bibfnamefont {M.}~\bibnamefont
  {Caldara}}\ and\ \bibinfo {author} {\bibfnamefont {F.}~\bibnamefont
  {Ancilotto}},\ }\href@noop {} {\bibfield  {journal} {\bibinfo  {journal}
  {Phys. Rev. A}\ }\textbf {\bibinfo {volume} {105}},\ \bibinfo {pages}
  {063328} (\bibinfo {year} {2022})}\BibitemShut {NoStop}%
\bibitem [{\citenamefont {Saqlain}\ \emph {et~al.}(2023)\citenamefont
  {Saqlain}, \citenamefont {Mithun}, \citenamefont {Carretero-Gonz{\'a}lez},\
  and\ \citenamefont {Kevrekidis}}]{sheikh}%
  \BibitemOpen
  \bibfield  {author} {\bibinfo {author} {\bibfnamefont {S.}~\bibnamefont
  {Saqlain}}, \bibinfo {author} {\bibfnamefont {T.}~\bibnamefont {Mithun}},
  \bibinfo {author} {\bibfnamefont {R.}~\bibnamefont {Carretero-Gonz{\'a}lez}},
  \ and\ \bibinfo {author} {\bibfnamefont {P.~G.}\ \bibnamefont {Kevrekidis}},\
  }\href@noop {} {\bibfield  {journal} {\bibinfo  {journal} {Phys. Rev. A}\
  }\textbf {\bibinfo {volume} {107}},\ \bibinfo {pages} {033310} (\bibinfo
  {year} {2023})}\BibitemShut {NoStop}%
\bibitem [{\citenamefont {Shukla}\ \emph {et~al.}(2021)\citenamefont {Shukla},
  \citenamefont {Panigrahi} \emph {et~al.}}]{shukla2021kink}%
  \BibitemOpen
  \bibfield  {author} {\bibinfo {author} {\bibfnamefont {A.}~\bibnamefont
  {Shukla}}, \bibinfo {author} {\bibfnamefont {P.~K.}\ \bibnamefont
  {Panigrahi}},  \emph {et~al.},\ }\href@noop {} {\bibfield  {journal}
  {\bibinfo  {journal} {J. Phys. B: At. Mol. and Opt. Phys.}\ }\textbf
  {\bibinfo {volume} {54}},\ \bibinfo {pages} {165301} (\bibinfo {year}
  {2021})}\BibitemShut {NoStop}%
\bibitem [{\citenamefont {Kaur}\ \emph {et~al.}(2022)\citenamefont {Kaur},
  \citenamefont {Gautam},\ and\ \citenamefont {Adhikari}}]{kaur2022supersolid}%
  \BibitemOpen
  \bibfield  {author} {\bibinfo {author} {\bibfnamefont {P.}~\bibnamefont
  {Kaur}}, \bibinfo {author} {\bibfnamefont {S.}~\bibnamefont {Gautam}}, \ and\
  \bibinfo {author} {\bibfnamefont {S.~K.}\ \bibnamefont {Adhikari}},\
  }\href@noop {} {\bibfield  {journal} {\bibinfo  {journal} {Phys. Rev. A}\
  }\textbf {\bibinfo {volume} {105}},\ \bibinfo {pages} {023303} (\bibinfo
  {year} {2022})}\BibitemShut {NoStop}%
\bibitem [{\citenamefont {Kopyci\ifmmode~\acute{n}\else \'{n}\fi{}ski}\ \emph
  {et~al.}(2023)\citenamefont {Kopyci\ifmmode~\acute{n}\else \'{n}\fi{}ski},
  \citenamefont {\L{}ebek}, \citenamefont {G\'orecki},\ and\ \citenamefont
  {Paw\l{}owski}}]{kopycinski2022ultrawide}%
  \BibitemOpen
  \bibfield  {author} {\bibinfo {author} {\bibfnamefont {J.}~\bibnamefont
  {Kopyci\ifmmode~\acute{n}\else \'{n}\fi{}ski}}, \bibinfo {author}
  {\bibfnamefont {M.}~\bibnamefont {\L{}ebek}}, \bibinfo {author}
  {\bibfnamefont {W.}~\bibnamefont {G\'orecki}}, \ and\ \bibinfo {author}
  {\bibfnamefont {K.}~\bibnamefont {Paw\l{}owski}},\ }\href@noop {} {\bibfield
  {journal} {\bibinfo  {journal} {Phys. Rev. Lett.}\ }\textbf {\bibinfo
  {volume} {130}},\ \bibinfo {pages} {043401} (\bibinfo {year}
  {2023})}\BibitemShut {NoStop}%
\bibitem [{\citenamefont {Busch}\ and\ \citenamefont
  {Anglin}(2001)}]{busch2001dark}%
  \BibitemOpen
  \bibfield  {author} {\bibinfo {author} {\bibfnamefont {T.}~\bibnamefont
  {Busch}}\ and\ \bibinfo {author} {\bibfnamefont {J.~R.}\ \bibnamefont
  {Anglin}},\ }\href@noop {} {\bibfield  {journal} {\bibinfo  {journal} {Phys.
  Rev. Lett.}\ }\textbf {\bibinfo {volume} {87}},\ \bibinfo {pages} {010401}
  (\bibinfo {year} {2001})}\BibitemShut {NoStop}%
\bibitem [{\citenamefont {Frantzeskakis}(2010)}]{frantzeskakis2010dark}%
  \BibitemOpen
  \bibfield  {author} {\bibinfo {author} {\bibfnamefont {D.~J.}\ \bibnamefont
  {Frantzeskakis}},\ }\href@noop {} {\bibfield  {journal} {\bibinfo  {journal}
  {J. Phys. A: Math. and Th.}\ }\textbf {\bibinfo {volume} {43}},\ \bibinfo
  {pages} {213001} (\bibinfo {year} {2010})}\BibitemShut {NoStop}%
\bibitem [{\citenamefont {Tamura}\ \emph {et~al.}(2022)\citenamefont {Tamura},
  \citenamefont {Chen},\ and\ \citenamefont {Hung}}]{RDS}%
  \BibitemOpen
  \bibfield  {author} {\bibinfo {author} {\bibfnamefont {H.}~\bibnamefont
  {Tamura}}, \bibinfo {author} {\bibfnamefont {C.-A.}\ \bibnamefont {Chen}}, \
  and\ \bibinfo {author} {\bibfnamefont {C.-L.}\ \bibnamefont {Hung}},\
  }\href@noop {} {\bibfield  {journal} {\bibinfo  {journal} {arXiv:2211.08575}\
  } (\bibinfo {year} {2022})},\ \bibinfo {note} {{O}bservation of
  self-patterned defect formation in atomic superfluids -- from ring dark
  solitons to vortex dipole necklaces}\BibitemShut {NoStop}%
\bibitem [{\citenamefont {Mossman}\ \emph {et~al.}(2022)\citenamefont
  {Mossman}, \citenamefont {Katsimiga}, \citenamefont {Mistakidis},
  \citenamefont {Romero-Ros}, \citenamefont {Bersano}, \citenamefont
  {Schmelcher}, \citenamefont {Kevrekidis},\ and\ \citenamefont
  {Engels}}]{mossman2022dense}%
  \BibitemOpen
  \bibfield  {author} {\bibinfo {author} {\bibfnamefont {S.}~\bibnamefont
  {Mossman}}, \bibinfo {author} {\bibfnamefont {G.~C.}\ \bibnamefont
  {Katsimiga}}, \bibinfo {author} {\bibfnamefont {S.~I.}\ \bibnamefont
  {Mistakidis}}, \bibinfo {author} {\bibfnamefont {A.}~\bibnamefont
  {Romero-Ros}}, \bibinfo {author} {\bibfnamefont {T.~M.}\ \bibnamefont
  {Bersano}}, \bibinfo {author} {\bibfnamefont {P.}~\bibnamefont {Schmelcher}},
  \bibinfo {author} {\bibfnamefont {P.~G.}\ \bibnamefont {Kevrekidis}}, \ and\
  \bibinfo {author} {\bibfnamefont {P.}~\bibnamefont {Engels}},\ }\href@noop {}
  {\bibfield  {journal} {\bibinfo  {journal} {arXiv:2208.10585}\ } (\bibinfo
  {year} {2022})},\ \bibinfo {note} {{D}ense dark-bright soliton arrays in a
  two-component {B}ose-{E}instein condensate}\BibitemShut {NoStop}%
\bibitem [{\citenamefont {Bakkali-Hassani}\ \emph {et~al.}(2021)\citenamefont
  {Bakkali-Hassani}, \citenamefont {Maury}, \citenamefont {Zou}, \citenamefont
  {Le~Cerf}, \citenamefont {Saint-Jalm}, \citenamefont {Castilho},
  \citenamefont {Nascimbene}, \citenamefont {Dalibard},\ and\ \citenamefont
  {Beugnon}}]{bakkali2021realization}%
  \BibitemOpen
  \bibfield  {author} {\bibinfo {author} {\bibfnamefont {B.}~\bibnamefont
  {Bakkali-Hassani}}, \bibinfo {author} {\bibfnamefont {C.}~\bibnamefont
  {Maury}}, \bibinfo {author} {\bibfnamefont {Y.-Q.}\ \bibnamefont {Zou}},
  \bibinfo {author} {\bibfnamefont {{\'E}.}~\bibnamefont {Le~Cerf}}, \bibinfo
  {author} {\bibfnamefont {R.}~\bibnamefont {Saint-Jalm}}, \bibinfo {author}
  {\bibfnamefont {P.~C.~M.}\ \bibnamefont {Castilho}}, \bibinfo {author}
  {\bibfnamefont {S.}~\bibnamefont {Nascimbene}}, \bibinfo {author}
  {\bibfnamefont {J.}~\bibnamefont {Dalibard}}, \ and\ \bibinfo {author}
  {\bibfnamefont {J.}~\bibnamefont {Beugnon}},\ }\href@noop {} {\bibfield
  {journal} {\bibinfo  {journal} {Phys. Rev. Lett.}\ }\textbf {\bibinfo
  {volume} {127}},\ \bibinfo {pages} {023603} (\bibinfo {year}
  {2021})}\BibitemShut {NoStop}%
\bibitem [{\citenamefont {Chai}\ \emph {et~al.}(2021)\citenamefont {Chai},
  \citenamefont {Lao}, \citenamefont {Fujimoto},\ and\ \citenamefont
  {Raman}}]{chai2021magnetic}%
  \BibitemOpen
  \bibfield  {author} {\bibinfo {author} {\bibfnamefont {X.}~\bibnamefont
  {Chai}}, \bibinfo {author} {\bibfnamefont {D.}~\bibnamefont {Lao}}, \bibinfo
  {author} {\bibfnamefont {K.}~\bibnamefont {Fujimoto}}, \ and\ \bibinfo
  {author} {\bibfnamefont {C.}~\bibnamefont {Raman}},\ }\href@noop {}
  {\bibfield  {journal} {\bibinfo  {journal} {Phys. Rev. Research}\ }\textbf
  {\bibinfo {volume} {3}},\ \bibinfo {pages} {L012003} (\bibinfo {year}
  {2021})}\BibitemShut {NoStop}%
\bibitem [{\citenamefont {Fritsch}\ \emph {et~al.}(2020)\citenamefont
  {Fritsch}, \citenamefont {Lu}, \citenamefont {Reid}, \citenamefont
  {Pi{\~n}eiro},\ and\ \citenamefont {Spielman}}]{fritsch2020creating}%
  \BibitemOpen
  \bibfield  {author} {\bibinfo {author} {\bibfnamefont {A.~R.}\ \bibnamefont
  {Fritsch}}, \bibinfo {author} {\bibfnamefont {M.}~\bibnamefont {Lu}},
  \bibinfo {author} {\bibfnamefont {G.~H.}\ \bibnamefont {Reid}}, \bibinfo
  {author} {\bibfnamefont {A.~M.}\ \bibnamefont {Pi{\~n}eiro}}, \ and\ \bibinfo
  {author} {\bibfnamefont {I.~B.}\ \bibnamefont {Spielman}},\ }\href@noop {}
  {\bibfield  {journal} {\bibinfo  {journal} {Phys. Rev. A}\ }\textbf {\bibinfo
  {volume} {101}},\ \bibinfo {pages} {053629} (\bibinfo {year}
  {2020})}\BibitemShut {NoStop}%
\bibitem [{\citenamefont {Konotop}\ and\ \citenamefont
  {Pitaevskii}(2004)}]{konotop2004landau}%
  \BibitemOpen
  \bibfield  {author} {\bibinfo {author} {\bibfnamefont {V.~V.}\ \bibnamefont
  {Konotop}}\ and\ \bibinfo {author} {\bibfnamefont {L.}~\bibnamefont
  {Pitaevskii}},\ }\href@noop {} {\bibfield  {journal} {\bibinfo  {journal}
  {Phys. Rev. Lett.}\ }\textbf {\bibinfo {volume} {93}},\ \bibinfo {pages}
  {240403} (\bibinfo {year} {2004})}\BibitemShut {NoStop}%
\bibitem [{\citenamefont {Kevrekidis}\ \emph {et~al.}(2008)\citenamefont
  {Kevrekidis}, \citenamefont {Frantzeskakis},\ and\ \citenamefont
  {Carretero-Gonz{\'a}lez}}]{kevrekidis2008emergent}%
  \BibitemOpen
  \bibfield  {author} {\bibinfo {author} {\bibfnamefont {P.~G.}\ \bibnamefont
  {Kevrekidis}}, \bibinfo {author} {\bibfnamefont {D.~J.}\ \bibnamefont
  {Frantzeskakis}}, \ and\ \bibinfo {author} {\bibfnamefont {R.}~\bibnamefont
  {Carretero-Gonz{\'a}lez}},\ }\href@noop {} {\emph {\bibinfo {title} {Emergent
  nonlinear phenomena in Bose-Einstein condensates: theory and experiment}}},\
  Vol.~\bibinfo {volume} {45}\ (\bibinfo  {publisher} {Springer},\ \bibinfo
  {year} {2008})\BibitemShut {NoStop}%
\bibitem [{\citenamefont {Pethick}\ and\ \citenamefont
  {Smith}(2008)}]{pethick2008bose}%
  \BibitemOpen
  \bibfield  {author} {\bibinfo {author} {\bibfnamefont {C.~J.}\ \bibnamefont
  {Pethick}}\ and\ \bibinfo {author} {\bibfnamefont {H.}~\bibnamefont
  {Smith}},\ }\href@noop {} {\emph {\bibinfo {title} {Bose--Einstein
  condensation in dilute gases}}}\ (\bibinfo  {publisher} {Cambridge university
  press},\ \bibinfo {year} {2008})\BibitemShut {NoStop}%
\bibitem [{\citenamefont {Pitaevskii}\ and\ \citenamefont
  {Stringari}(2016)}]{PitaevskiiStringari2016}%
  \BibitemOpen
  \bibfield  {author} {\bibinfo {author} {\bibfnamefont {L.}~\bibnamefont
  {Pitaevskii}}\ and\ \bibinfo {author} {\bibfnamefont {S.}~\bibnamefont
  {Stringari}},\ }\href@noop {} {\emph {\bibinfo {title} {{Bose-Einstein
  condensation and superfluidity}}}},\ International series of monographs on
  physics\ (\bibinfo  {publisher} {Oxford University Press},\ \bibinfo
  {address} {Oxford},\ \bibinfo {year} {2016})\BibitemShut {NoStop}%
\bibitem [{\citenamefont {Olshanii}(1998)}]{olshanii1998atomic}%
  \BibitemOpen
  \bibfield  {author} {\bibinfo {author} {\bibfnamefont {M.}~\bibnamefont
  {Olshanii}},\ }\href@noop {} {\bibfield  {journal} {\bibinfo  {journal}
  {Phys. Rev. Lett.}\ }\textbf {\bibinfo {volume} {81}},\ \bibinfo {pages}
  {938} (\bibinfo {year} {1998})}\BibitemShut {NoStop}%
\bibitem [{\citenamefont {Chin}\ \emph {et~al.}(2010)\citenamefont {Chin},
  \citenamefont {Grimm}, \citenamefont {Julienne},\ and\ \citenamefont
  {Tiesinga}}]{chin2010feshbach}%
  \BibitemOpen
  \bibfield  {author} {\bibinfo {author} {\bibfnamefont {C.}~\bibnamefont
  {Chin}}, \bibinfo {author} {\bibfnamefont {R.}~\bibnamefont {Grimm}},
  \bibinfo {author} {\bibfnamefont {P.}~\bibnamefont {Julienne}}, \ and\
  \bibinfo {author} {\bibfnamefont {E.}~\bibnamefont {Tiesinga}},\ }\href@noop
  {} {\bibfield  {journal} {\bibinfo  {journal} {Rev. Mod. Phys.}\ }\textbf
  {\bibinfo {volume} {82}},\ \bibinfo {pages} {1225} (\bibinfo {year}
  {2010})}\BibitemShut {NoStop}%
\bibitem [{\citenamefont {Madelung}(1927)}]{madelung1927quantentheorie}%
  \BibitemOpen
  \bibfield  {author} {\bibinfo {author} {\bibfnamefont {E.}~\bibnamefont
  {Madelung}},\ }\href@noop {} {\bibfield  {journal} {\bibinfo  {journal}
  {Zeitschrift f{\"u}r Physik}\ }\textbf {\bibinfo {volume} {40}},\ \bibinfo
  {pages} {322} (\bibinfo {year} {1927})}\BibitemShut {NoStop}%
\bibitem [{\citenamefont {Barashenkov}\ \emph {et~al.}(1989)\citenamefont
  {Barashenkov}, \citenamefont {Gocheva}, \citenamefont {Makhankov},\ and\
  \citenamefont {Puzynin}}]{baras1989}%
  \BibitemOpen
  \bibfield  {author} {\bibinfo {author} {\bibfnamefont {I.}~\bibnamefont
  {Barashenkov}}, \bibinfo {author} {\bibfnamefont {A.}~\bibnamefont
  {Gocheva}}, \bibinfo {author} {\bibfnamefont {V.}~\bibnamefont {Makhankov}},
  \ and\ \bibinfo {author} {\bibfnamefont {I.}~\bibnamefont {Puzynin}},\
  }\href@noop {} {\bibfield  {journal} {\bibinfo  {journal} {Physica D:
  Nonlinear Phenomena}\ }\textbf {\bibinfo {volume} {34}},\ \bibinfo {pages}
  {240} (\bibinfo {year} {1989})}\BibitemShut {NoStop}%
\bibitem [{\citenamefont {Katsimiga}\ \emph {et~al.}(2023)\citenamefont
  {Katsimiga}, \citenamefont {Mistakidis}, \citenamefont {Malomed},
  \citenamefont {Frantzeskakis}, \citenamefont {Carretero-Gonz{\'a}lez},\ and\
  \citenamefont {Kevrekidis}}]{multidroplets}%
  \BibitemOpen
  \bibfield  {author} {\bibinfo {author} {\bibfnamefont {G.~C.}\ \bibnamefont
  {Katsimiga}}, \bibinfo {author} {\bibfnamefont {S.~I.}\ \bibnamefont
  {Mistakidis}}, \bibinfo {author} {\bibfnamefont {B.~A.}\ \bibnamefont
  {Malomed}}, \bibinfo {author} {\bibfnamefont {D.~J.}\ \bibnamefont
  {Frantzeskakis}}, \bibinfo {author} {\bibfnamefont {R.}~\bibnamefont
  {Carretero-Gonz{\'a}lez}}, \ and\ \bibinfo {author} {\bibfnamefont {P.~G.}\
  \bibnamefont {Kevrekidis}},\ }\href@noop {} {\bibfield  {journal} {\bibinfo
  {journal} {in preparation}\ } (\bibinfo {year} {2023})}\BibitemShut {NoStop}%
\bibitem [{\citenamefont {Gallo}\ and\ \citenamefont
  {Pelinovsky}(2011)}]{gallo}%
  \BibitemOpen
  \bibfield  {author} {\bibinfo {author} {\bibfnamefont {C.}~\bibnamefont
  {Gallo}}\ and\ \bibinfo {author} {\bibfnamefont {D.}~\bibnamefont
  {Pelinovsky}},\ }\href@noop {} {\bibfield  {journal} {\bibinfo  {journal}
  {Asymptotic Analysis}\ }\textbf {\bibinfo {volume} {73}},\ \bibinfo {pages}
  {53} (\bibinfo {year} {2011})}\BibitemShut {NoStop}%
\bibitem [{\citenamefont {Landau}\ and\ \citenamefont
  {Lifshitz}(1959)}]{landau1959fluid}%
  \BibitemOpen
  \bibfield  {author} {\bibinfo {author} {\bibfnamefont {L.~D.}\ \bibnamefont
  {Landau}}\ and\ \bibinfo {author} {\bibfnamefont {E.~M.}\ \bibnamefont
  {Lifshitz}},\ }\href@noop {} {\bibfield  {journal} {\bibinfo  {journal} {New
  York}\ }\textbf {\bibinfo {volume} {61}},\ \bibinfo {pages} {10} (\bibinfo
  {year} {1959})}\BibitemShut {NoStop}%
\bibitem [{\citenamefont {Zakharov}\ and\ \citenamefont
  {Shabat}(1973)}]{zakharov1973interaction}%
  \BibitemOpen
  \bibfield  {author} {\bibinfo {author} {\bibfnamefont {V.~E.}\ \bibnamefont
  {Zakharov}}\ and\ \bibinfo {author} {\bibfnamefont {A.~B.}\ \bibnamefont
  {Shabat}},\ }\href@noop {} {\bibfield  {journal} {\bibinfo  {journal} {Sov.
  Phys. JETP}\ }\textbf {\bibinfo {volume} {37}},\ \bibinfo {pages} {823}
  (\bibinfo {year} {1973})}\BibitemShut {NoStop}%
\bibitem [{\citenamefont {Kevrekidis}\ \emph {et~al.}(2017)\citenamefont
  {Kevrekidis}, \citenamefont {Wang}, \citenamefont {Carretero-Gonz{\'a}lez},\
  and\ \citenamefont {Frantzeskakis}}]{kevrekidis2017adiabatic}%
  \BibitemOpen
  \bibfield  {author} {\bibinfo {author} {\bibfnamefont {P.~G.}\ \bibnamefont
  {Kevrekidis}}, \bibinfo {author} {\bibfnamefont {W.}~\bibnamefont {Wang}},
  \bibinfo {author} {\bibfnamefont {R.}~\bibnamefont {Carretero-Gonz{\'a}lez}},
  \ and\ \bibinfo {author} {\bibfnamefont {D.~J.}\ \bibnamefont
  {Frantzeskakis}},\ }\href@noop {} {\bibfield  {journal} {\bibinfo  {journal}
  {Phys. Rev. Lett.}\ }\textbf {\bibinfo {volume} {118}},\ \bibinfo {pages}
  {244101} (\bibinfo {year} {2017})}\BibitemShut {NoStop}%
\bibitem [{\citenamefont {Kelley}(2003)}]{kelley2003solving}%
  \BibitemOpen
  \bibfield  {author} {\bibinfo {author} {\bibfnamefont {C.~T.}\ \bibnamefont
  {Kelley}},\ }\href@noop {} {\emph {\bibinfo {title} {Solving nonlinear
  equations with Newton's method}}}\ (\bibinfo  {publisher} {SIAM},\ \bibinfo
  {year} {2003})\BibitemShut {NoStop}%
\bibitem [{\citenamefont {Alfimov}\ and\ \citenamefont
  {Zezyulin}(2007)}]{alfimov2007nonlinear}%
  \BibitemOpen
  \bibfield  {author} {\bibinfo {author} {\bibfnamefont {G.~L.}\ \bibnamefont
  {Alfimov}}\ and\ \bibinfo {author} {\bibfnamefont {D.~A.}\ \bibnamefont
  {Zezyulin}},\ }\href@noop {} {\bibfield  {journal} {\bibinfo  {journal}
  {Nonlinearity}\ }\textbf {\bibinfo {volume} {20}},\ \bibinfo {pages} {2075}
  (\bibinfo {year} {2007})}\BibitemShut {NoStop}%
\bibitem [{\citenamefont {Chernyavsky}\ \emph
  {et~al.}(2018{\natexlab{a}})\citenamefont {Chernyavsky}, \citenamefont
  {Kevrekidis},\ and\ \citenamefont {Pelinovsky}}]{chernyavsky2017krein}%
  \BibitemOpen
  \bibfield  {author} {\bibinfo {author} {\bibfnamefont {A.}~\bibnamefont
  {Chernyavsky}}, \bibinfo {author} {\bibfnamefont {P.~G.}\ \bibnamefont
  {Kevrekidis}}, \ and\ \bibinfo {author} {\bibfnamefont {D.~E.}\ \bibnamefont
  {Pelinovsky}},\ }in\ \href@noop {} {\emph {\bibinfo {booktitle} {Parity-time
  Symmetry and Its Applications}}}\ (\bibinfo  {publisher} {Springer},\
  \bibinfo {year} {2018})\ pp.\ \bibinfo {pages} {465--491}\BibitemShut
  {NoStop}%
\bibitem [{\citenamefont {Skryabin}(2000)}]{skryabin2000instabilities}%
  \BibitemOpen
  \bibfield  {author} {\bibinfo {author} {\bibfnamefont {D.~V.}\ \bibnamefont
  {Skryabin}},\ }\href@noop {} {\bibfield  {journal} {\bibinfo  {journal}
  {Phys. Rev. A}\ }\textbf {\bibinfo {volume} {63}},\ \bibinfo {pages} {013602}
  (\bibinfo {year} {2000})}\BibitemShut {NoStop}%
\bibitem [{\citenamefont {Katsimiga}\ \emph {et~al.}(2020)\citenamefont
  {Katsimiga}, \citenamefont {Mistakidis}, \citenamefont {Bersano},
  \citenamefont {Ome}, \citenamefont {Mossman}, \citenamefont {Mukherjee},
  \citenamefont {Schmelcher}, \citenamefont {Engels},\ and\ \citenamefont
  {Kevrekidis}}]{katsimiga2020observation}%
  \BibitemOpen
  \bibfield  {author} {\bibinfo {author} {\bibfnamefont {G.~C.}\ \bibnamefont
  {Katsimiga}}, \bibinfo {author} {\bibfnamefont {S.~I.}\ \bibnamefont
  {Mistakidis}}, \bibinfo {author} {\bibfnamefont {T.~M.}\ \bibnamefont
  {Bersano}}, \bibinfo {author} {\bibfnamefont {M.~K.~H.}\ \bibnamefont {Ome}},
  \bibinfo {author} {\bibfnamefont {S.~M.}\ \bibnamefont {Mossman}}, \bibinfo
  {author} {\bibfnamefont {K.}~\bibnamefont {Mukherjee}}, \bibinfo {author}
  {\bibfnamefont {P.}~\bibnamefont {Schmelcher}}, \bibinfo {author}
  {\bibfnamefont {P.}~\bibnamefont {Engels}}, \ and\ \bibinfo {author}
  {\bibfnamefont {P.~G.}\ \bibnamefont {Kevrekidis}},\ }\href@noop {}
  {\bibfield  {journal} {\bibinfo  {journal} {Phys. Rev. A}\ }\textbf {\bibinfo
  {volume} {102}},\ \bibinfo {pages} {023301} (\bibinfo {year}
  {2020})}\BibitemShut {NoStop}%
\bibitem [{\citenamefont {Coles}\ \emph {et~al.}(2010)\citenamefont {Coles},
  \citenamefont {Pelinovsky},\ and\ \citenamefont
  {Kevrekidis}}]{coles2010excited}%
  \BibitemOpen
  \bibfield  {author} {\bibinfo {author} {\bibfnamefont {M.~P.}\ \bibnamefont
  {Coles}}, \bibinfo {author} {\bibfnamefont {D.~E.}\ \bibnamefont
  {Pelinovsky}}, \ and\ \bibinfo {author} {\bibfnamefont {P.~G.}\ \bibnamefont
  {Kevrekidis}},\ }\href@noop {} {\bibfield  {journal} {\bibinfo  {journal}
  {Nonlinearity}\ }\textbf {\bibinfo {volume} {23}},\ \bibinfo {pages} {1753}
  (\bibinfo {year} {2010})}\BibitemShut {NoStop}%
\bibitem [{\citenamefont {De~Palo}\ \emph {et~al.}(2022)\citenamefont
  {De~Palo}, \citenamefont {Orignac},\ and\ \citenamefont
  {Citro}}]{PhysRevB.106.014503}%
  \BibitemOpen
  \bibfield  {author} {\bibinfo {author} {\bibfnamefont {S.}~\bibnamefont
  {De~Palo}}, \bibinfo {author} {\bibfnamefont {E.}~\bibnamefont {Orignac}}, \
  and\ \bibinfo {author} {\bibfnamefont {R.}~\bibnamefont {Citro}},\
  }\href@noop {} {\bibfield  {journal} {\bibinfo  {journal} {Phys. Rev. B}\
  }\textbf {\bibinfo {volume} {106}},\ \bibinfo {pages} {014503} (\bibinfo
  {year} {2022})}\BibitemShut {NoStop}%
\bibitem [{\citenamefont {Chernyavsky}\ \emph
  {et~al.}(2018{\natexlab{b}})\citenamefont {Chernyavsky}, \citenamefont
  {Kevrekidis},\ and\ \citenamefont {Pelinovsky}}]{Chernyavsky2018}%
  \BibitemOpen
  \bibfield  {author} {\bibinfo {author} {\bibfnamefont {A.}~\bibnamefont
  {Chernyavsky}}, \bibinfo {author} {\bibfnamefont {P.~G.}\ \bibnamefont
  {Kevrekidis}}, \ and\ \bibinfo {author} {\bibfnamefont {D.~E.}\ \bibnamefont
  {Pelinovsky}},\ }\enquote {\bibinfo {title} {Krein signature in {H}amiltonian
  and {PT}-symmetric systems},}\ in\ \href@noop {} {\emph {\bibinfo {booktitle}
  {Parity-time Symmetry and Its Applications}}},\ \bibinfo {editor} {edited by\
  \bibinfo {editor} {\bibfnamefont {D.}~\bibnamefont {Christodoulides}}\ and\
  \bibinfo {editor} {\bibfnamefont {J.}~\bibnamefont {Yang}}}\ (\bibinfo
  {publisher} {Springer Singapore},\ \bibinfo {address} {Singapore},\ \bibinfo
  {year} {2018})\ pp.\ \bibinfo {pages} {465--491}\BibitemShut {NoStop}%
\bibitem [{\citenamefont {Kapitula}\ and\ \citenamefont
  {Kevrekidis}(2005)}]{kapitula}%
  \BibitemOpen
  \bibfield  {author} {\bibinfo {author} {\bibfnamefont {T.}~\bibnamefont
  {Kapitula}}\ and\ \bibinfo {author} {\bibfnamefont {P.~G.}\ \bibnamefont
  {Kevrekidis}},\ }\href {\doibase 10.1063/1.1993867} {\bibfield  {journal}
  {\bibinfo  {journal} {Chaos}\ }\textbf {\bibinfo {volume} {15}},\ \bibinfo
  {pages} {037114} (\bibinfo {year} {2005})}\BibitemShut {NoStop}%
\bibitem [{\citenamefont {Theocharis}\ \emph {et~al.}(2010)\citenamefont
  {Theocharis}, \citenamefont {Weller}, \citenamefont {Ronzheimer},
  \citenamefont {Gross}, \citenamefont {Oberthaler}, \citenamefont
  {Kevrekidis},\ and\ \citenamefont {Frantzeskakis}}]{weller2}%
  \BibitemOpen
  \bibfield  {author} {\bibinfo {author} {\bibfnamefont {G.}~\bibnamefont
  {Theocharis}}, \bibinfo {author} {\bibfnamefont {A.}~\bibnamefont {Weller}},
  \bibinfo {author} {\bibfnamefont {J.~P.}\ \bibnamefont {Ronzheimer}},
  \bibinfo {author} {\bibfnamefont {C.}~\bibnamefont {Gross}}, \bibinfo
  {author} {\bibfnamefont {M.~K.}\ \bibnamefont {Oberthaler}}, \bibinfo
  {author} {\bibfnamefont {P.~G.}\ \bibnamefont {Kevrekidis}}, \ and\ \bibinfo
  {author} {\bibfnamefont {D.~J.}\ \bibnamefont {Frantzeskakis}},\ }\href@noop
  {} {\bibfield  {journal} {\bibinfo  {journal} {Phys. Rev. A}\ }\textbf
  {\bibinfo {volume} {81}},\ \bibinfo {pages} {063604} (\bibinfo {year}
  {2010})}\BibitemShut {NoStop}%
\bibitem [{\citenamefont {Englezos}\ \emph {et~al.}(2023)\citenamefont
  {Englezos}, \citenamefont {Mistakidis},\ and\ \citenamefont
  {Schmelcher}}]{englezos2023correlated}%
  \BibitemOpen
  \bibfield  {author} {\bibinfo {author} {\bibfnamefont {I.~A.}\ \bibnamefont
  {Englezos}}, \bibinfo {author} {\bibfnamefont {S.~I.}\ \bibnamefont
  {Mistakidis}}, \ and\ \bibinfo {author} {\bibfnamefont {P.}~\bibnamefont
  {Schmelcher}},\ }\href@noop {} {\bibfield  {journal} {\bibinfo  {journal}
  {Phys. Rev. A}\ }\textbf {\bibinfo {volume} {107}},\ \bibinfo {pages}
  {023320} (\bibinfo {year} {2023})}\BibitemShut {NoStop}%
\end{thebibliography}

%
\end{document}